\pdfoutput=1
\documentclass[12pt,table]{report}
\usepackage[letterpaper]{geometry}
 \usepackage[doublespacing]{setspace}
 \usepackage{tocloft}
 \usepackage[rm, tiny,center, compact]{titlesec}
 \usepackage{indentfirst}
 \usepackage{etoolbox}
\usepackage{tocvsec2}
 \usepackage[titletoc]{appendix}
 \usepackage{appendix}
 \usepackage{tamuconfig}
\usepackage{rotating}

\usepackage{regexpatch}

\usepackage[table]{xcolor}
\usepackage{transparent}
\usepackage{booktabs}

\usepackage{bibentry}

\usepackage{url}

\usepackage{amsmath}
\usepackage{amsthm}
\usepackage{amssymb}

\usepackage{IEEEtrantools}

\usepackage{array}
\usepackage{tikz}
\usepackage{pgfplots}
\pgfplotsset{compat=newest} 
\pgfplotsset{plot coordinates/math parser=false}
\usepackage{graphicx}

\usetikzlibrary{decorations.markings,positioning,calc}

\newif\iffull
\fulltrue

\newcommand{\tabularnewline}{\\}

\newcommand{\sasoglu}{\c{S}a\c{s}o\u{g}lu}
\newcommand{\arikan}{Ar{\i}kan}
\newcommand{\felstrom}{Felstr{\"o}m}
\newcommand{\naive}{na{\"i}ve}
\newcommand{\vecnot}[1]{\underline{#1}}

\newtheorem{thm}{Theorem}

\newtheorem{lem}[thm]{Lemma}

\theoremstyle{definition}
\newtheorem{defn}[thm]{Definition}

\newcommand\blfootnote[1]{%
  \begingroup
  \renewcommand\thefootnote{}\footnote{#1}%
  \addtocounter{footnote}{-1}%
  \endgroup
}

\newcommand{\SPBC}{{\rm\scriptscriptstyle  SPBC}}
\newcommand{\RBC}{{\rm\scriptscriptstyle  RBC}}
\newcommand{\B}{{\rm\scriptscriptstyle  B}}

\begin{document}

\renewcommand{\tamumanuscripttitle}{On Cyclic Polar Codes and The Burst Erasure Performance of Spatially-Coupled LDPC Codes}
\renewcommand{\tamupapertype}{Thesis}
\renewcommand{\tamufullname}{Narayanan Rengaswamy}
\renewcommand{\tamudegree}{Master of Science}
\renewcommand{\tamuchairone}{Henry D. Pfister}
\newcommand{\tamuchairtwo}{Krishna R. Narayanan}
\renewcommand{\tamumemberone}{Gregory H. Huff}
\newcommand{\tamumembertwo}{Anxiao Jiang}
\renewcommand{\tamudepthead}{Miroslav M. Begovic}
\renewcommand{\tamugradmonth}{December}
\renewcommand{\tamugradyear}{2015}
\renewcommand{\tamudepartment}{Electrical Engineering}

%
%
%



\begin{titlepage}
\begin{center}
\MakeUppercase{\tamumanuscripttitle}
\vspace{4em}

A \tamupapertype

by

\MakeUppercase{\tamufullname}

\vspace{4em}

\begin{singlespace}

Submitted to the Office of Graduate and Professional Studies of \\
Texas A\&M University \\

in partial fulfillment of the requirements for the degree of \\
\end{singlespace}

\MakeUppercase{\tamudegree}
\par\end{center}
\vspace{2em}
\begin{singlespace}
\begin{tabular}{ll}
 & \tabularnewline
& \cr
Chair of Committee, & \tamuchairone\tabularnewline
Co-Chair of Committee, & \tamuchairtwo\tabularnewline
Committee Members, & \tamumemberone\tabularnewline
 & \tamumembertwo\tabularnewline
Head of Department, & \tamudepthead\tabularnewline

\end{tabular}
\end{singlespace}
\vspace{3em}

\begin{center}
\tamugradmonth \hspace{2pt} \tamugradyear

\vspace{3em}

Major Subject: \tamudepartment \par
\vspace{3em}
Copyright \tamugradyear \hspace{.5em}\tamufullname 
\par\end{center}
\end{titlepage}
\pagebreak{}

%
%
%

\chapter*{ABSTRACT}
\addcontentsline{toc}{chapter}{ABSTRACT} 

\pagestyle{plain} 
\pagenumbering{roman} 
\setcounter{page}{2}

In this thesis, we produce our work on two of the state-of-the-art techniques in modern coding theory: polar codes and spatially-coupled LDPC codes.


\indent Polar codes were introduced in 2009 and proven to achieve the symmetric capacity of any binary-input discrete memoryless channel under low-complexity successive cancellation decoding.
Since then, finite length (non-asymptotic) performance has been the primary concern with respect to polar codes.
In this work, we construct cyclic polar codes based on a mixed-radix Cooley-Tukey decomposition of the Galois field Fourier transform.
The main results are:
we can, for the first time, construct, encode and decode polar codes that are cyclic, with their blocklength being arbitrary;
for a given target block erasure rate, we can achieve significantly higher code rates on the erasure channel than the original polar codes, at comparable blocklengths;
on the symmetric channel with only errors, we can perform much better than equivalent rate Reed-Solomon codes with the same blocklength, by using soft-decision decoding;
and, since the codes are subcodes of higher rate RS codes, a RS decoder can be used if suboptimal performance suffices for the application as a trade-off for higher decoding speed.
The programs developed for this work can be accessed at \url{https://github.com/nrenga/cyclic_polar}.



In 2010, it was shown that spatially-coupled low-density parity-check (LDPC) codes approach the capacity of binary memoryless channels, asymptotically, with belief-propagation (BP) decoding.
In our work, we are interested in the finite length average performance of randomly coupled LDPC ensembles on binary erasure channels with memory. 
The significant contributions of this work are:
tight lower bounds for the block erasure probability ($P_B$) under various scenarios for the burst pattern;
bounds focused on practical scenarios where a burst affects exactly one of the coupled codes;
expected error floor for the bit erasure probability ($P_b$) on the binary erasure channel;
and, characterization of the performance of random regular ensembles, on erasure channels, with a single vector describing distinct types of size-$2$ stopping sets.
%
All these results are verified using Monte-Carlo simulations.  
Further, we show that increasing variable node degree combined with expurgation can improve $P_B$ by several orders of magnitude in the number of bits per coupled code.

\pagebreak{}

%
%
%

\addcontentsline{toc}{chapter}{DEDICATION}  

\vspace*{\stretch{2}}

\begin{center}

\indent I dedicate this thesis to my wonderful parents, \emph{Rengaswamy Purushothaman} and \emph{Sudha Rengasami}, for their unconditional love and support.

\end{center}

\vspace{\stretch{3}}

\pagebreak{}

%
%
%

\chapter*{ACKNOWLEDGEMENTS}
\addcontentsline{toc}{chapter}{ACKNOWLEDGEMENTS}  

\indent I thank all my committee members for taking the time and effort to conduct my defense. 
Specifically, I extend my heartfelt thanks to Dr. Anxiao Jiang for humbly adjusting his schedule to fit the common time slot when my advisory committee can meet to conduct the defense.

My advisor, Dr. Henry Pfister, has been instrumental in motivating me throughout my thesis. 
Despite the amazing, and complex, associations between statistical physics and coding theory that he has developed in his mind, he always communicated clearly, precisely and promptly while resolving the doubts and concerns of the novice in me. 
I extend my gratitude to him for encouraging me to submit a paper to the prestigious International Symposium on Information Theory (ISIT), of the Institute of Electrical and Electronics Engineers (IEEE), and to present the work in the conference myself. 
The experience was truly international and exposed me to a lot of different researchers and research problems in the broad area of information theory. 
I also deeply appreciate his effort to recommend me for an internship in the optical communications group of Alcatel-Lucent Bell Labs at Stuttgart, Germany. 
The exposure was immensely useful, both professionally and personally, without which the work on spatially-coupled LDPC codes would not have been possible. 
Due to his encouragement, I have been able to work on two of the state-of-the-art techniques in modern coding theory: polar codes and spatially-coupled LDPC codes. 

I would like to thank Dr. Krishna Narayanan for useful discussions during my thesis work and for providing important professional advise that has enhanced my graduate experience. 
In addition to that, I thank him for his financial support for my conference trip to Hong Kong. 

I appreciate the efforts of Dr. Gregory Huff and Dr. Jean-Fran{\c c}ois Chamberland for designing a projects-based-learning course for both undergraduate and graduate students. 
The course is a great initiative in providing hands-on exposure on building systems and I feel that such a course bridges the gap between theory and practice, and also between college and graduate school. 
I sincerely thank them for appreciating my efforts during the course, having me as a research assistant in their lab and later recommending me as a teaching assistant for assisting undergraduate students in their senior design projects. 
Especially, I thank Dr. Chamberland for mentoring and guiding me many a time.
I would also like to thank Dr. Samuel Villareal for recruiting me as a teaching assistant despite his preference for doctoral candidates. 
He always keeps himself professional, lively and motivational with his assertive, confident nature and his wonderful sense of humor. 

I thank all my professors for their courses and the productive discussions both in class and in person.
Especially, the graduate level courses on Wireless Communications, by Prof. Scott Miller, and on Computer and Communication Networks, by Prof. Narasimha Reddy, were instrumental in understanding the field of communications and networking. 
The homework assignments were challenging and complemented the class lectures in providing a complete, informative experience.
They have helped me immensely in fitting my research into the big picture.

The work on SC-LDPC codes was done under the supervision of Dr. Laurent Schmalen and Dr. Vahid Aref during the summer of 2015, while I was a research intern in Alcatel-Lucent Bell Labs, Stuttgart, Germany funded by the DAAD-RisePro scholarship program.
The internship was so productive due to the professional guidance of my supervisors.
They were well versed with the literature on spatially-coupled LDPC codes and hence were able to keep me well focused during the course of my internship. 
My brain storming sessions with them have proven to be instrumental in getting good results in the short period of three months. 
I extend my sincere thanks to both of them for providing good exposure on industrial research and development.

My thesis work combined with the internship experience, the conference trip, the exposure in Dr. Huff's lab and the ongoing teaching assistantship have, as a whole, given me a lot of confidence in performing theoretical research, presenting in a professional setting, building good prototypes and helping students with their projects. I view this as an important mixture of exposures towards building my career in academia.

A graduate experience without great friends is incomplete. 
I have been able to make great friends, involve in productive discussions and build wonderful memories throughout my graduate school experience. 
Specifically, I have had my best times with Nagaraj Janakiraman, Adithyaram Narayan and Kalluri Raja Sreeram, who have been great roommates for more than two years,
and with Karthik Kalyanaraman, Balakumar Jayaraman, Karthick Sudhan, Srinivasa Varadhan, Nani Anudeep, Siddharth Agarwal, Sangeeta Panigrahy and Kartic Bhargav who always kept the fun alive.
I thank Santosh Emmadi, Avinash Vem and Santhosh Kumar for my great times in our research group. 
Santhosh Kumar also shared a lot of his graduate school wisdom during our trip to the conference. 
I found wonderful friends in Michael Bass, Brian Bass, Dipanjan Saha, Shuli Li, Yayun Lau, Zachary Partal and Desmond Uzor during our project for the course with Dr. Huff and Dr. Chamberland. 
The internship experience helped me get to know fellow interns Lei Zhang, Foad Sohrabi and Nazanin Rastegardoost with whom I have been able to strike good friendship through our chats and our trips in Europe. 
I have had some productive discussions with Lei Zhang on spatially-coupled LDPC codes and staircase codes. 
I would also like to thank my fellow teaching assistants Neal Hollingsworth, Geha Chadi, Sean Goldberger, Ahmad Bashaireh, Ahmed Morsy and Mandel Oats for making the job a fun and informative experience. 
The enthusiasm of the undergraduate students, with whom I have got a chance to interact, has always been refreshing and motivating to learn more.

I would like to stress that my graduate school experience would not have been so smooth without the support of our department's academic advisors Tammy Carda and Jeanie Marshall, our business coordinators Sheryl Mallett and Anni Brunker, and the International Student Services (ISS) at Texas A\&M. 
I have always admired all of their responsibility and enthusiasm at work and, specifically, Anni's ways of keeping herself refreshed and positive. 
The ISS is offering a commendable service in addressing the concerns and needs of all international students here.
Complementing the work of ISS, the student organizations on campus are ensuring a personally comfortable and memorable experience for all students. 
Specifically, I thank the Indian Graduate Students Association (IGSA) and India Association (IA) for having helped in making the transition from India to the United States very smooth.
I also appreciate the Big Event and Aggie Replant communities for doing laudable work for the society and I thank them for giving me a chance to be part of that Aggie spirit.

I deeply thank the university and department committees for having admitted me into the Master of Science (M.S.) program of the Department of Electrical and Computer Engineering and providing a great international exposure early in my career.

Ultimately, I thank my parents, my family and the Almighty for their blessings and continuous support.

\pagebreak{}
%
%
%


\chapter*{NOMENCLATURE}
\addcontentsline{toc}{chapter}{NOMENCLATURE}  

\begin{tabular}{ll}
APP  & A Posteriori Probability\tabularnewline
AWGN & Additive White Gaussian Noise\tabularnewline
BEC  & Binary Erasure Channel\tabularnewline 
BMS & Binary Memoryless Symmetric\tabularnewline
BP & Belief Propagation\tabularnewline
B-DMC & Binary Input DMC\tabularnewline
CN & Check Node\tabularnewline
DMC & Discrete Memoryless Channel\tabularnewline
FFT & Fast Fourier Transform\tabularnewline 
GF & Galois Field\tabularnewline 
LDPC & Low-Density Parity-Check\tabularnewline
MAP & Maximum A Posteriori\tabularnewline 
QEC & $q$-ary Erasure Channel\tabularnewline 
QSC & $q$-ary Symmetric Channel\tabularnewline 
QSCE & $q$-ary Symmetric Channel with Erasures\tabularnewline 
RBC & Random Burst Channel\tabularnewline
RS & Reed-Solomon\tabularnewline 
SC-LDPC & Spatially-Coupled LDPC\tabularnewline
SP & Spatial Position\tabularnewline
SPBC & Single Position Burst Channel\tabularnewline
VN & Variable Node\tabularnewline
w.l.o.g & without loss of generality\tabularnewline
\end{tabular}

\pagebreak{}

%
%
%

\phantomsection
\addcontentsline{toc}{chapter}{TABLE OF CONTENTS}  

\begin{singlespace}
\renewcommand\contentsname{\normalfont} {\centerline{TABLE OF CONTENTS}}


\setlength{\cftaftertoctitleskip}{1em}
\renewcommand{\cftaftertoctitle}{%
\hfill{\normalfont {Page}\par}}

\tableofcontents

\end{singlespace}

\pagebreak{}


\phantomsection
\addcontentsline{toc}{chapter}{LIST OF FIGURES}  

\renewcommand{\cftloftitlefont}{\center\normalfont\MakeUppercase}

\setlength{\cftbeforeloftitleskip}{-12pt} 
\renewcommand{\cftafterloftitleskip}{12pt}

\renewcommand{\cftafterloftitle}{%
\\[4em]\mbox{}\hspace{2pt}FIGURE\hfill{\normalfont Page}\vskip\baselineskip}

\begingroup

\begin{center}
\begin{singlespace}
\setlength{\cftbeforechapskip}{0.4cm}
\setlength{\cftbeforesecskip}{0.30cm}
\setlength{\cftbeforesubsecskip}{0.30cm}
\setlength{\cftbeforefigskip}{0.4cm}
\setlength{\cftbeforetabskip}{0.4cm} 

\listoffigures

\end{singlespace}
\end{center}

\pagebreak{}

%
\phantomsection
\addcontentsline{toc}{chapter}{LIST OF TABLES}  

\renewcommand{\cftlottitlefont}{\center\normalfont\MakeUppercase}

\setlength{\cftbeforelottitleskip}{-12pt} 

\renewcommand{\cftafterlottitleskip}{12pt}

\renewcommand{\cftafterlottitle}{%
\\[4em]\mbox{}\hspace{4pt}TABLE\hfill{\normalfont Page}\vskip\baselineskip}

\begin{center}
\begin{singlespace}

\setlength{\cftbeforechapskip}{0.4cm}
\setlength{\cftbeforesecskip}{0.30cm}
\setlength{\cftbeforesubsecskip}{0.30cm}
\setlength{\cftbeforefigskip}{0.4cm}
\setlength{\cftbeforetabskip}{0.4cm}

\listoftables 

\end{singlespace}
\end{center}
\endgroup
\pagebreak{}  

%
%
%


\pagestyle{plain} 
\pagenumbering{arabic} 
\setcounter{page}{1}

\chapter{\uppercase{Introduction}}
\label{sec:thesis_intro}

In this thesis, we consider two capacity achieving codes that form the state-of-the-art techniques in coding theory today: \emph{Polar codes} and \emph{Spatially-Coupled Low-Density Parity-Check codes}.
Both of these codes are very recent inventions given the six-and-a-half decade history of information and coding theory.
To appreciate the significance of these codes, we need to take a brief look into the motivation behind modern communication systems and get an idea of their current maturity.
Specifically, it is imperative to glimpse through the fundamentals of coding theory in order to understand the notion of capacity and hence capacity achieving codes.
This section attempts to provide an overview that just meets this purpose.
The last subsection outlines the rest of this thesis.

\section{From Telephone Systems to Modern Communication}
\label{sec:shannon}

Telephone systems became prevalent in the United States since the turn of the twentieth century.
The Bell Systems company found it increasingly difficult to maintain good service to all customers given their rapid expansion.
By the 1930s, manual operation of the telephone systems were starting to get replaced by switching circuits for the need of speed.
The introduction of these switching circuits made systems more complex and it became imperative to understand their behavior in theory.
This was the primary motivation for Claude E. Shannon, a young electrical engineer working as a research assistant in the Massachusetts Institute of Technology, Cambridge.
In his own words, ``Examples of these circuits occur in automatic telephone exchanges, industrial motor-control equipment, and in almost any circuits designed to perform complex operations automatically''~\cite{Shannon-aiee1938}.
Shannon is arguably the first person to introduce boolean logic in the representation of switching circuits, which he did in 1937 in his remarkable Master's thesis titled \emph{A Symbolic Analysis of Relay and Switching Circuits}. 
Ironically, the thesis was unpublished though a paper abstracted from it was published in the Transactions of the American Institute for Electrical Engineers in 1938~\cite{Shannon-aiee1938}.
This is still regarded as one of the most important theses ever written.

He introduced the notion of $1$ and $0$ to denote the open and closed state of a circuit switch, respectively.
In his work, he focused mainly on the problem of network synthesis -- how do we synthesize a network that incorporates certain desired characteristics?
He showed that several well-known theorems in impedance networks have roughly analogous versions in relay circuits.
The simple, yet powerful, approach was to represent relays and switches as mathematical variables and construct systems of equations that describe their interactions and behavior.
In this way, it was possible to first write down the desired characteristics in precise mathematical language, solve the set of equations to get the optimal solution using the necessary calculus, and then implement the solution with actual relays and circuits.

Later, Shannon desired to characterize the transport of information through the telephonic systems.
Precisely, he wanted to know how to design telephone systems that could carry maximum amount of information and also account for distortion in the lines.
To answer that question, he needed to quantify the abstract notion of \emph{information} conveyed through any form -- text, sound or image.
This inspired him to invent and formulate the profound mathematics called \emph{Information Theory}.

\section{The Idea of Information Theory}

Shannon's view of information in a message was as follows: 
a message that brought more surprise to the receiver contained more information and 
a message that was more predictable contained very less information.
Concisely, he viewed information as the amount of uncertainty that it resolves upon reception at the receiver.
This notion is indeed very intuitive and he defines \emph{entropy} as the amount of information or randomness that a message contained by itself.
The more profound idea was to represent any kind of data just by a sequence of $1$s and $0$s.
Probably, it made sense to him to represent data just by states of a set of switches given his background and work on relays and switches.
As a New York Times article pointed out~\cite{NYtimes-2001}, he proposed that the information contained in a message had nothing to do with the content but only with the number of $1$s and $0$s necessary to represent it.
All character sets developed for computers rely on this underlying notion of message representation.
Combining the idea of the digital notation of messages, his notion of information and his desire to find the maximum amount of information that can be transmitted over a distortion-prone telephone line, he laid down a seminal mathematical treatise titled \emph{A Mathematical Theory of Communication}~\cite{Shannon-bst1948} which was later published as a book, co-authored by Warren Weaver, with the title \emph{The Mathematical Theory of Communication}.
The small but significant change in the title was to emphasize the generality of the work.

\section{The Communication System Model}

Shannon modeled a communication system as follows: an information source generates data, then it is transmitted appropriately as a signal over a communication channel that corrupts the signal with noise, and eventually the corrupted signal is received and converted back to the original data format and delivered to the desired destination.
The generality of this model is very evident.
Also, such a formulation allows us to work on and refine the individual blocks while keeping their interface to the rest of the system unperturbed, i.e. the model has very good modularity.
His model of communication systems is simple, elegant and has proven to be very successful having stood the test of time.
Even today, research and development in the broad area of communications depends strongly on his model.

\begin{sidewaysfigure}[p]
\begin{center}
\scalebox{1}{\usetikzlibrary{decorations.markings}

\makeatletter
\@ifundefined{vecnot}{%
\newcommand{\vecnot}[1]{\underline{#1}}
}{}
\makeatother

\begin{tikzpicture}

  \draw[thick] (-3,0) rectangle (-1,2);
  \node[draw=none, align=center] at (-2,1) {Digital\\Source\\(Input)};
  
  \draw[thick,->] (-1,1) -- (1,1); 
  
  \draw[thick] (1,0) rectangle (3,2);
  \node[draw=none, align=center] at (2,1) {Source\\Encoder};
  
  \node[draw=none,align=center] at (4,1.5) {Message};
  \draw[thick,->] (3,1) -- (5,1); 
  \node[draw=none,align=center] at (4,0.5) {\color{blue}$\vecnot{u} \in \mathcal{X}^{K}$};

  \draw[fill=yellow,thick] (5,0) rectangle (7,2);
  \node[draw=none, align=center] at (6,1) {Channel\\Encoder};
  
  
%

  \node[draw=none,align=center] at (8.5,1.5) {Codeword};
  \draw[thick] (7,1) -- (10,1); 
  \node[draw=none,align=center] at (8.5,0.5) {\color{blue}$\vecnot{v} \in \mathcal{X}^{N}$};
  \draw[thick,->] (10,1) -- (10,0);  
  \node[draw=none,align=center] at (11,0.5) {$X \in \mathcal{X}$};
  
  \draw[fill=yellow,thick] (9,-2) rectangle (11,0);
  \node[draw=none, align=center] at (10,-1) {Noisy\\Channel\\$p_{Y|X}$};
  \node[draw=none,align=center] at (12,-1) {$W(y|x)$};
  

  \draw[thick] (10,-2) -- (10,-3); 
  \node[draw=none,align=center] at (11,-2.5) {$Y \in \mathcal{Y}$};
  \draw[thick,->] (10,-3) -- (7,-3);  
  \node[draw=none,align=center] at (9,-4) {Codeword with\\some symbols corrupted};
  \node[draw=none,align=center] at (8.5,-2.5) {\color{blue}$\vecnot{y} \in \mathcal{Y}^{N}$};
  

  \draw[fill=yellow,thick] (5,-4) rectangle (7,-2);
  \node[draw=none, align=center] at (6,-3) {Channel\\Decoder};
  
 
  \node[draw=none,align=center] at (4,-2.5) {\color{blue}{$P_B$}};
  \draw[thick,->] (5,-3) -- (3,-3); 
  \node[draw=none,align=center] at (4,-4) {Prob. of\\Block Error};
  
  \draw[thick] (1,-4) rectangle (3,-2);
  \node[draw=none, align=center] at (2,-3) {Source\\Decoder};
  
  \draw[thick,->] (1,-3) -- (-1,-3); 

  \draw[thick] (-3,-4) rectangle (-1,-2);
  \node[draw=none, align=center] at (-2,-3) {Digital\\Sink\\(Output)};

\end{tikzpicture}}
\caption{\label{fig:simplecomm}A communication system highlighting the main blocks from a coding theoretic perspective. It is slightly more detailed than the model considered by Shannon in~\cite{Shannon-bst1948} in the sense that the source and channel coding blocks are shown for emphasis.}
\end{center}
\end{sidewaysfigure}
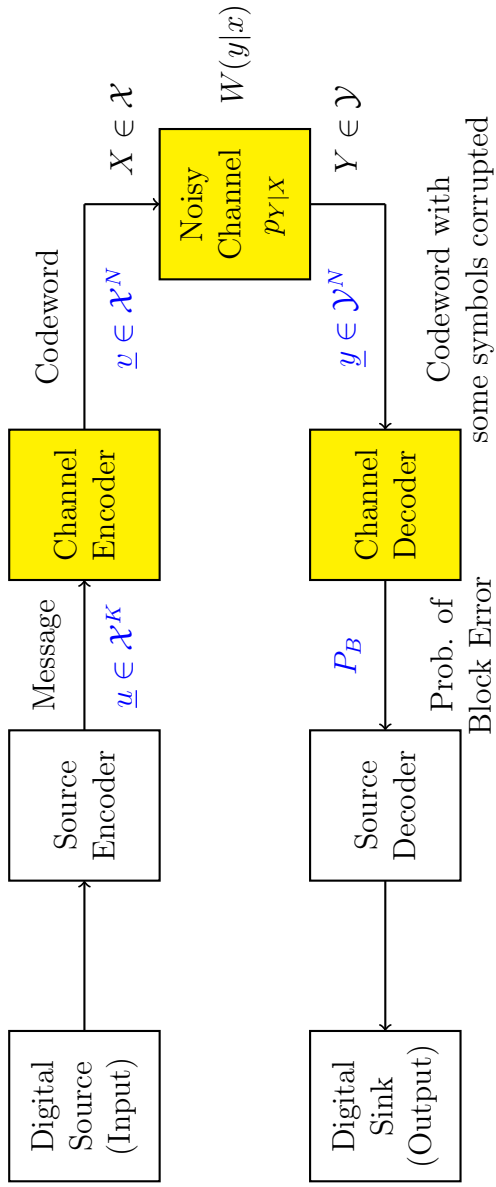

A slightly more detailed model than the one in Shannon's paper~\cite{Shannon-bst1948} is given in Fig.~\ref{fig:simplecomm}.
The source coding block desires to represent the input data in the most concise form but being stringent on the allowance on information loss.
This thesis fits into this big picture in the channel encoder-decoder pair of blocks.
The function of the channel encoding block is to add calculated redundancy to the source-coded data so that the distortion introduced by the channel can be corrected at the channel decoding block, given the mathematical structure of the added redundancy.
The block error probability, $P_B$, at the receiver is a measure of the chance that the channel decoder will fail to correct the errors introduced by the channel in the codeword.
Hence, the goal for coding theory is to construct efficient codes over a particular channel or a class of channels so that, the code introduces minimal redundancy while retaining $P_B$ at an acceptable low level.

\section{Fundamentals of Coding Theory}

\subsection{What is a Code?}

A code is an encoding-decoding pair that explicitly states the details of the two complementary blocks.
A channel encoder would typically encode $K$ information bits, that it receives as input from the source encoder, into $N$ code bits by adding $(N-K)$ bits of calculated redundancy.
Every codeword is a block of encoded bits of data and its length, called \emph{blocklength}, is $N$.
Hence, for every $N$ bits of coded data, there are $K$ bits of information conveyed and the \emph{rate}, $R$, of the code is given as
\begin{equation}
\label{eq:rate}
R = \frac{K}{N} .
\end{equation}
This is the rate of coded transmission into the channel.
Different codes have different ways of calculating the redundancy and different ways of decoding the received (corrupted) word.
The choice for the encoder and decoder ultimately decides the performance of the channel coding block in the communication system, measured by the \emph{probability of decoding error} $P_B$.
This quantity is also termed the \emph{block error probability} since at least one bit in the codeword block is in error after decoding.

\subsection{Capacity Achieving Codes}

To set up the goal of coding theory, mathematically, we need to revisit the fundamentals of information theory.
\emph{Entropy} and \emph{Mutual Information} are two fundamental quantities defined by Shannon to address the information inherently present in a message and the amount of shared, or mutual, information between two correlated messages, respectively.
For this reason entropy is also called \emph{self-information}.

Consider a source emitting messages from a binary alphabet $\mathcal{X} = \{0,1\}$.
The distribution on the alphabet is arbitrary and source-dependent.
Let $X$ be a random variable that denotes a message from this source.
Then, the entropy of the source is given by
\begin{equation}
\label{eq:entropy}
H(X) \triangleq \mathbb{E}_{X} \biggr[ \text{log } \frac{1}{p(x)} \biggr] = - \sum_{x \in \mathcal{X}} p(x) \text{ log } p(x) \hspace{4mm} \text{bits}
\end{equation}
where, $p(x)$ is the probability that $X$ takes the value $x$, and the logarithm is over base-$2$.
Unless specified otherwise, all logarithms in this thesis will be base $2$.
However, the notion of entropy is general and can be extended to non-binary alphabets too.

Now, let the message $X$ be sent over a distortion-prone channel and received as $Y$ which might also belong to the same binary alphabet or to a different one depending on the channel model.
For simplicity, let us assume that the output alphabet $\mathcal{Y}$ is also binary.
Then, the mutual information between $X$ and $Y$ is defined as
\begin{equation}
I(X;Y) \triangleq \mathbb{E}_{(X,Y)} \biggr[ \text{log } \frac{p(x,y)}{p(x) p(y)} \biggr] = - \sum_{y \in \mathcal{Y}} \sum_{x \in \mathcal{X}} p(x,y) \text{ log } \frac{p(x) p(y)}{p(x,y)} \hspace{4mm} \text{bits/channel use}. 
\end{equation}

Ideally, if $1$ bit of information is sent over one use of the channel, then $1$ bit of information must be received.
But, since the channel is distortion-prone, the mutual information is less than $1$ bit/channel use.
Hence, the \emph{capacity} of a channel, $C$, is defined as the maximum amount of mutual information over all input distributions on $\mathcal{X}$.
\begin{equation}
C \triangleq \underset{p(X)}{\max} \hspace{2.5mm} I(X;Y) \hspace{4mm} \text{bits/channel use}.
\end{equation}
This is the maximum rate, $R$, at which information can be transmitted and recovered reliably through an appropriate coding scheme.
The \emph{symmetric capacity} of a channel is the maximum amount of information that can be transmitted reliably over one use of the channel, subject to using the input values $1$ and $0$ with equal frequency.

\emph{In other words, given a noisy channel with capacity $C$, for every transmission rate $R < C$, there exists a coding scheme which guarantees that information can be reliably transmitted over that channel at the rate $R$ and that the maximum probability of decoding error at the receiver can be made arbitrarily small.}

Shannon proved this in his famous noisy channel coding theorem using random coding arguments.
But, to implement a practical communication system we need to design specific codes that can achieve this limit.
Codes that achieve this limit for a given class of channels are called \emph{capacity-achieving codes}.
The following statement formalizes the notion of capacity-achieving codes:

Given a channel with capacity $C$, there exists a sequence of codes, indexed by $n$, with rates $R_n$ such that
\begin{equation}
\underset{n \rightarrow \infty}{\lim} R_n = C \hspace{3.5mm} \text{ with } \hspace{3.5mm} P_{B_n}^{\max} \rightarrow 0.
\end{equation}
Designing such codes with deterministic constructions has been the pursuit of coding theorists over the past six decades.

\section{Outline of the Thesis}

Polar codes are the first codes to have been explicitly shown to achieve the capacity of arbitrary symmetric binary-input discrete memoryless channels (B-DMCs) under low-complexity successive cancellation decoding.
This breakthrough was made by Arikan in 2009~\cite{Arikan-it09}.
Spatial coupling of multiple low-density parity-check (LDPC) codes was shown to be another way to achieve the capacity of binary erasure channels by Kudekar et al.~\cite{Kudekar-it11}, in 2011.
This structure was originally introduced as convolutional LDPC codes by \felstrom~and Zigangirov~\cite{Felstrom-it99} in 1999 but the proof happened to come much later.
These two codes are the state-of-the-art in modern coding theory and are strong competitors for practical applications.
However, these codes achieve capacity only asymptotically, i.e. as the blocklength approaches infinity.
Hence, the finite length performance of these codes is of primary concern among coding theorists and code designers.

Section~\ref{sec:cyclic_polar} proposes a new construction of polar codes which allows us to construct polar codes of arbitrary blocklength that also achieve significantly higher rates than the original polar codes on memoryless erasure channels, at comparable blocklengths.
In addition, these codes are cyclic and, specifically, are subcodes of RS codes so that a suitable cyclic encoder-RS decoder pair can be used if suboptimal performance suffices for the application as a trade-off for lower complexity and higher speed.
The section details the construction of the transform, proves polarization for the construction and discusses a algebraic successive cancellation decoder.
Simulation results are produced for the $q$-ary erasure channel and $q$-ary symmetric channel.

Spatially-coupled LDPC codes can be made more robust towards bursts of erasures than block LDPC codes.
Since there are multiple applications that exhibit a burst erasure phenomenon, not necessarily in the traditional communication transmission sense, it is of interest to understand their performance in such scenarios.
Hence, Section~\ref{sec:scldpc_codes} analyzes the average performance of random regular spatially-coupled LDPC ensembles on burst erasure channels.
A few practical applications are also provided as a motivation.
The stopping sets in the Tanner graphs of the codes are used to characterize the performance in both the unexpurgated and expurgated scenarios.
Since the two sections discuss different coding schemes in detail, conclusions are given at the end of each section for coherence and to avoid breaking the continuity.

Appendix~\ref{sec:cooley_tukey} gives a discussion of the Cooley-Tukey fast Fourier transform and derives its Kronecker product formulation.
Appendix~\ref{sec:channel_polarize} details the channel polarization for cyclic polar codes.
Appendix~\ref{sec:Forney} discusses the modified Forney's decoder for the small blocks in the FFT structure of cyclic polar codes and finally,
Appendix~\ref{sec:qsce_capacity} derives the Shannon capacity for the $q$-ary symmetric channel with erasures.

%
%
%


\chapter[\uppercase {Cyclic Polar Codes}]{\uppercase {Cyclic Polar Codes}$^{\star}$}
\label{sec:cyclic_polar}

\section{Introduction to Polar Codes}
Polar codes, invented by \arikan ~\cite{Arikan-it09}, are binary linear codes that can achieve the symmetric capacity of an arbitrary binary-input discrete memoryless channel (B-DMC) under successive cancellation (SC) decoding. 
The transform used to construct the polar codes is based on the Kronecker product of the $2\times 2$ kernel matrix,
\begin{equation*}
G_{2} = \begin{bmatrix}
1&0\\ 1&1
\end{bmatrix}.
\end{equation*}
The transform equation for blocklength $N = 2^{n}$ is given by
\begin{equation*}
G_{N} =  B_{N}G_{2}^{\otimes n} ,
\end{equation*}
where $B_{N}$ is the bit-reversal permutation matrix of size $N$, $A \otimes B$ is the Kronecker product of matrix $A$ with matrix $B$ and $G_{2}^{\otimes n} = \underset{n \text{ times}}{\underbrace{G_2 \otimes G_2 \otimes \cdots \otimes G_2}}$.
Definition~\ref{def:Kronecker} in Section~\ref{sec:ctfft} gives a precise definition of the Kronecker product.
\blfootnote{\newline $^{\star}$ Reprinted, with permission from N. Rengaswamy and H.D. Pfister, Cyclic Polar Codes, In \emph{Proc. IEEE Int. Symp.
Inform. Theory}, pages 1287-1291, June 2015\nocite{Rengaswamy-isit15}. \copyright~2015~IEEE.}

\begin{figure}[p]
\begin{center}
\large
\scalebox{0.8}{\includegraphics[width=\textwidth,height=0.4\textheight,keepaspectratio]{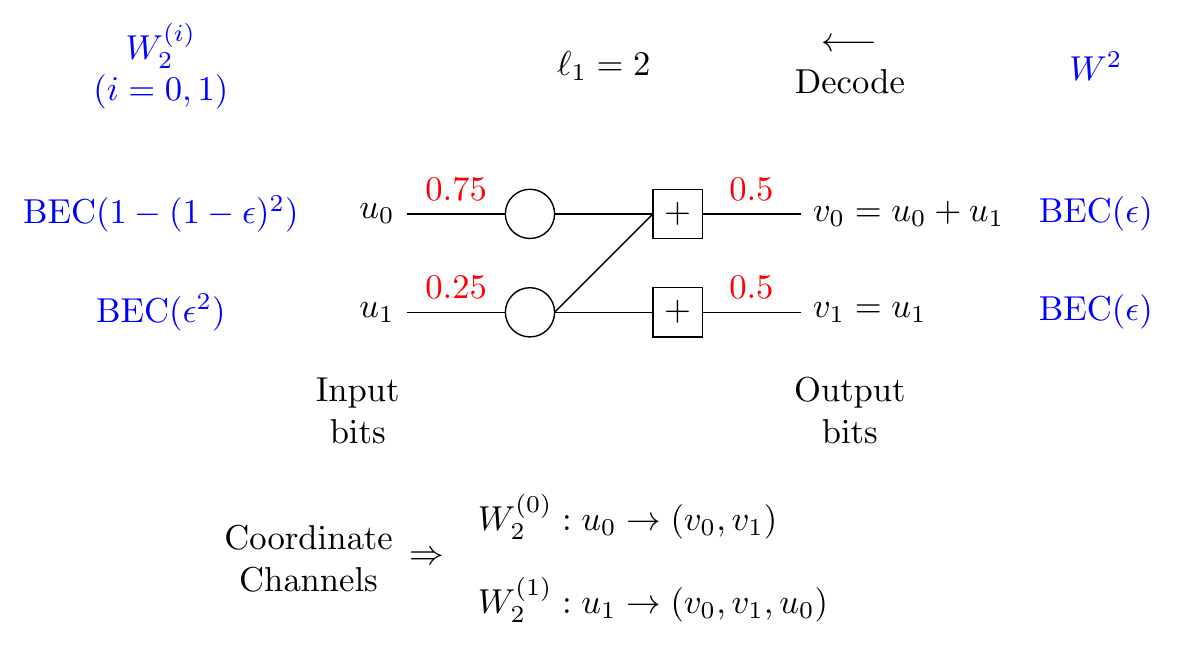}}
\normalsize
\caption{\label{fig:polar_2}Coordinate channels for binary polar code of blocklength $N=2$ over an underlying binary erasure channel $W \triangleq \text{BEC}(\epsilon)$. A numerical example for $W \triangleq \text{BEC}(0.5)$ is also shown. The numbers represent average erasure rates of the corresponding bits under successive cancellation decoding. The input erasure rates are obtained through one stage of density evolution.
}
\end{center}
\end{figure}

\begin{figure}[p]
\begin{center}
\large
\scalebox{0.85}{\includegraphics[width=\textwidth,height=0.4\textheight,keepaspectratio]{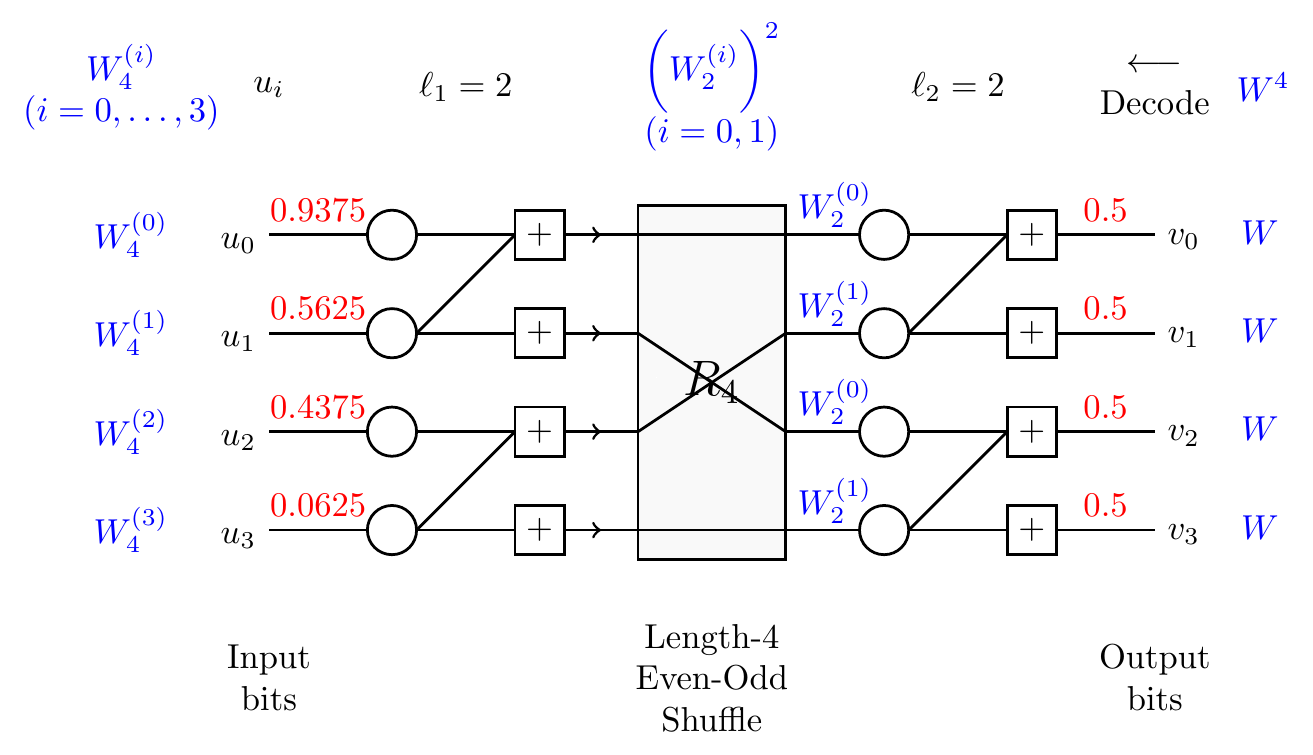}}
\normalsize
\caption{\label{fig:polar_4}Coordinate channels for binary polar code of blocklength $N=4$ over a general channel $W$. A numerical density evolution example over BEC($0.5$) is also shown.
}
\end{center}
\end{figure}

\begin{sidewaysfigure}[p]
\begin{center}
\large
\scalebox{1}{\includegraphics[width=\textwidth,height=\textheight,keepaspectratio]{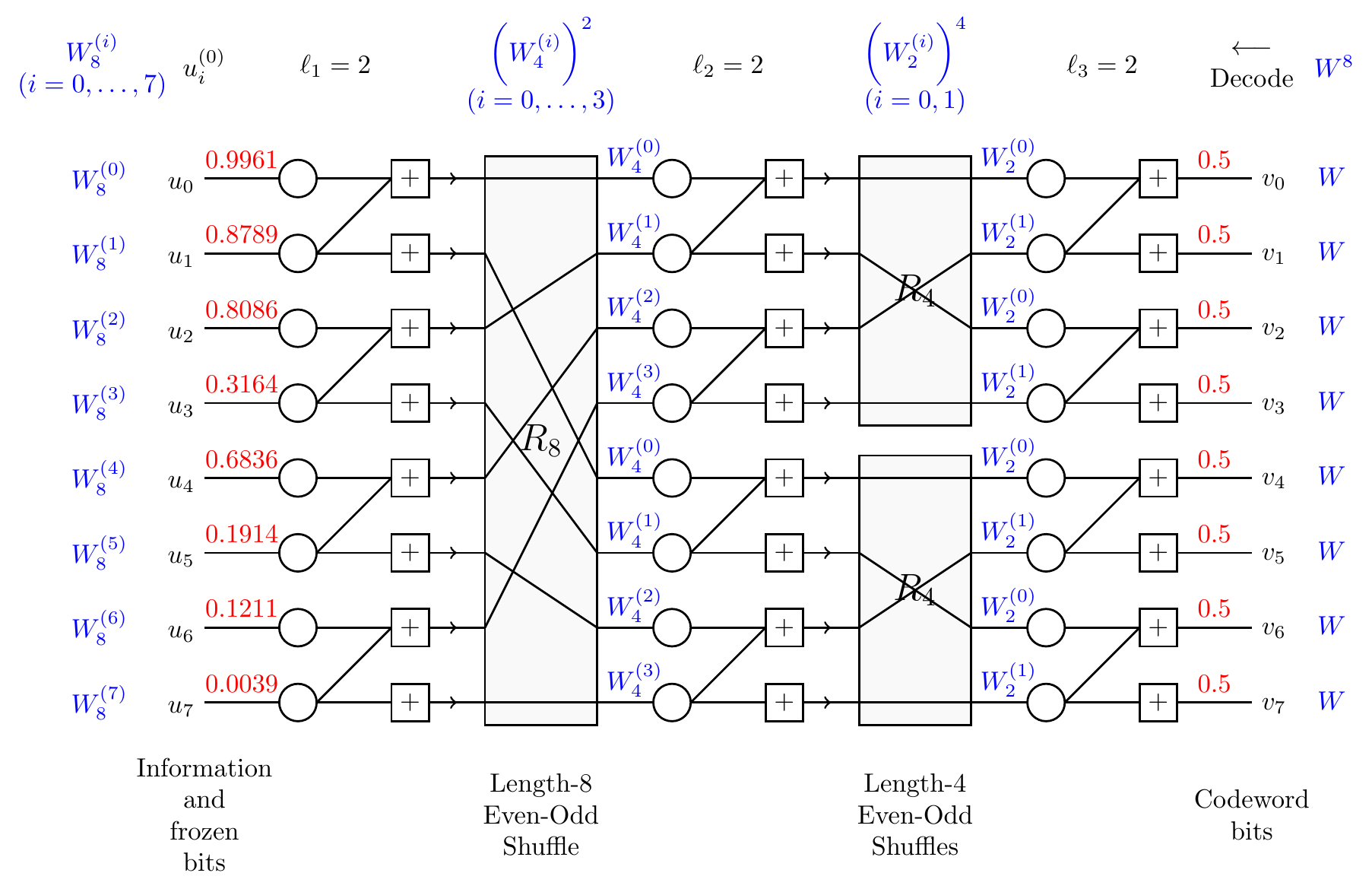}}
\normalsize
\caption{\label{fig:polar_8}Coordinate channels for binary polar code of blocklength $N=8$ over a general channel $W$. A numerical density evolution example for BEC($0.5$) is also shown.
}
\end{center}
\end{sidewaysfigure}

The binary input sequence $\vecnot{u}=(u_1,\ldots,u_N)$ consists of $K$ information bits and $(N-K)$ known frozen bits.
The codeword is encoded using $\vecnot{v} = \vecnot{u} G_{N}$ and then transmitted via $N$ independent uses of the underlying B-DMC $W$.
The successive cancellation decoder attempts to decode the $i^{th}$ bit $u_i$ given the knowledge of the received vector $\vecnot{y}$, which is a noisy observation of $\vecnot{v}$, and all the previously decoded inputs $u_{1}^{i-1}$.
This allows one to view the $i$-th input bit as being transmitted over the coordinate channel $W_{N}^{(i)}:\{0,1\}\longrightarrow \mathcal{Y}^{N} \times \{0,1\}^{i-1}$ with transition probabilities
\begin{equation*}
W_{N}^{(i)}(y_{1}^{N},u_{1}^{i-1}|u_{i}) = \frac{1}{2^{N-1}}\sum_{u_{i+1}^{N}}W^{N}(y_{1}^{N}|u_{1}^{N}),
\end{equation*} 
where $1\leq i\leq N$ and $W^{N}(y_{1}^{N}|u_{1}^{N}) = \prod_{j=1}^{N} W(y_{j}|v_{j})$ for the B-DMC $W:\{0,1\}\longrightarrow \mathcal{Y}$.
Therefore, for coordinate channel $W_{N}^{(i)}$, bit $u_i$ is the input and the output vector $\vecnot{y}$ combined with the $(i-1)$ previously decoded inputs, $u_{1}^{i-1}$, are the outputs.
These are the channels that the successive cancellation decoder ``sees'' even though the actual transmission of $\vecnot{v}$ is over $N$ independent uses of the ``actual'', physical, channel $W$.
Figs.~\ref{fig:polar_2},~\ref{fig:polar_4}~and~\ref{fig:polar_8} show the evolution of coordinate channels for blocklengths $N=2,4 \text{ and } 8$, respectively, for an underlying channel $W$.
A numerical example for $W \triangleq \text{BEC}(\epsilon = 0.5)$ is also shown. 
The numbers in the figures represent average erasure rates of the corresponding bits under successive cancellation decoding.
The process of initializing the channel density for $W$, which is $\epsilon$ for BEC($\epsilon$), and allowing it to evolve until the input stage to determine the average erasure rates of the input coordinate channels is called \emph{density evolution}.
For an arbitrary B-DMC $W$, the coordinate channels are obtained through channel combining operations~\cite{Arikan-it09}.

The idea of \emph{polarization} is that, for large values of $n$, the capacities of these coordinate channels either approach $1$, for ``good'' channels, or approach $0$, for ``bad'' channels, and \emph{no} value in between.
This means that each coordinate channel is polarized to either full capacity or zero capacity.
Hence, information can be transmitted at rate $1$ over the ``good'' channels and the bits in $\vecnot{u}$ corresponding to the ``bad'' channels can be \emph{frozen} to a fixed value, thereby implying a transmission at rate $0$.
It is important to note that the previous statement is valid because the notion of ``channel'' in coding theory is from the decoder's perspective, and not necessarily the physical channel.

For a given blocklength, the design phase for the polar codes computes the capacities of the coordinate channels and decides the information and frozen bits according to a target rate or a target block error rate for the code.
The indices corresponding to information bits are collected in a set $\mathcal{A}$ so that $|\mathcal{A}| = K$.
For example, in Fig.~\ref{fig:polar_8}, if we fix the target block error rate as $\delta=0.1$ so that $P_B \leq \delta$, then we have $\mathcal{A} = \{ 7 \}$ and hence, the rate of the code is $R = 1/8$. 
The design procedure is described in Section~\ref{sec:design}.
The rest of the bits are frozen to values known both to the encoder and decoder, i.e. they carry no information from the source but help the successive cancellation decoder immensely.
Typically, these bits are frozen to zeros.
Hence, for our example, $u_0 = \cdots = u_6 = 0$.

Given the notion of polar coding, let us see how these codes achieve capacity.
$I(W)$ denotes the symmetric capacity of the underlying channel $W$, i.e. the maximum amount of information that can be transmitted reliably over one use of the channel $W$, subject to using the input values $1$ and $0$ with equal frequency.

\emph{Polar codes achieve capacity because the fraction of channels that are ``good'' is equal to the symmetric capacity $I(W)$ of the underlying channel $W$, at sufficiently large blocklengths.
Mathematically, this means that for any $\theta \in (0,1)$,
\begin{equation}
\underset{n \rightarrow \infty}{\lim} R_n = \underset{n \rightarrow \infty}{\lim} \frac{1}{N} \biggr| \biggr\{ i: I(W_{N}^{(i)}) \in (1-\theta,1] \biggr\} \biggr| = I(W) \text{ for } N=2^n.
\end{equation}
}
Refer the notes of Pfister~\cite{Pfister-polar2014} for a very good introduction to polar codes.
In this thesis, the following terminologies refer to \arikan's $G_2$ polar codes: original, standard or binary polar codes.

\section{Background Work}
\label{sec:polar_bg}

Further work has shown that polar codes can be constructed using larger kernel sizes by using an $\ell\times \ell$ binary matrix $G_{\ell}$ as the base matrix in the Kronecker product.
Korada, \sasoglu, and Urbanke established that, as long as the transformation matrix $G$ is not upper triangular it will polarize the coordinate channels~\cite{Korada-it10*2}.
They also showed that the rate of polarization can be strictly better than the original construction when $\ell=16$.

In~\cite{Sasoglu-itw09}, \sasoglu, Telatar, and \arikan~show that the original polar code construction achieves the symmetric capacity of $q$-ary channels when $q$ is prime.
Mori and Tanaka consider polar codes over non-binary alphabets in~\cite{Mori-itw10,Mori-it14} and use Reed-Solomon (RS) and algebraic geometry codes to construct good polarizing kernels.

In this work, we construct polar codes, of arbitrary blocklength, that are also cyclic.
Unlike previously proposed constructions, we use mixed kernel sizes to exploit connection with non-binary Galois field Fourier transforms (GFFTs) and fast Fourier transforms (FFTs).
Using this construction, the information sequence $\vecnot{u}$ and the code sequence $\vecnot{v}$ become Fourier-transform pairs.

The design phase of these codes defines the positions of the frozen symbols.
These symbols are set to zero in the Fourier transform of all codewords.
Using the polynomial representation of messages and codewords, it follows that the frozen symbols define a set of common roots for all the code polynomials.
Thus, the code is cyclic and we refer to these codes as \emph{cyclic polar codes}.
In this work, we show that these codes achieve the symmetric capacity of $q$-ary erasure channels and 
that these codes achieve higher rates than the original polar codes on memoryless erasure channels, at comparable finite blocklengths.

One benefit of these codes is that they can be made backwards compatible for a system that currently uses RS codes.
This is because cyclic polar codes can be designed to be subcodes of higher rate RS codes.
The polarizing matrices used at each stage of the transform act essentially identical to RS codes during the successive cancellation decoding process.
Overall, while a standard RS decoder for the whole code only has one chance to correct all errors and erasures, the decoder of a cyclic polar code can exploit multistage decoding that converts some errors into erasures at each stage.
Thus, the existing RS decoder in the system could be used with some performance penalty and a SC decoder could be used to improve performance.

\iffull
\arikan~discusses systematic polar codes in~\cite{Arikan-comlett11}. Since the codes proposed here are cyclic, a systematic encoder can be realized by implementing suitable message mapping prior to the (non-systematic) encoder, so that the system is backwards compatible. 
\fi

The remainder of this discussion is organized as follows.
Section~\ref{sec:preliminaries} discusses the GFFT, the Cooley-Tukey FFT algorithm and the channel definitions.
Section~\ref{sec:construction} describes the cyclic polar code construction and decoding. 
Section~\ref{sec:design} considers code design and
Section~\ref{sec:results} discusses results.
Finally, Section~\ref{sec:conclusion} concludes this work. 

\section{Preliminaries}
\label{sec:preliminaries}

\subsection{Galois-Field Fourier-Transform}
\label{sec:GFFT}
Let $\mathbb{F}\triangleq \mathbb{F}_q$ denote the Galois field with $q$ elements, $\alpha \in \mathbb{F}$ be a distinguished primitive element, and $\omega_{\ell} =\alpha^{(q-1)/\ell}$ be a primitive $\ell$-th root of unity (i.e., $\ell\mid q-1$).
Then, the length-$\ell$ Galois-field Fourier-transform (GFFT) of the vector $\vecnot{v} = (v_0,\ldots,v_{\ell-1})$ is given by
\begin{equation*}
u_{i} = [ F_{\ell} \vecnot{v}]_i = \sum_{j=0}^{\ell-1}\omega_{\ell}^{ij}v_{j},
\end{equation*}
where the matrix $F_{\ell}$ is defined by $[F_{\ell}]_{i,j} \triangleq \omega_{\ell}^{ij}$.
The inverse Fourier transform is given by
\begin{equation*}
v_{j} = \ell^{-1} [ F_{\ell} ' \vecnot{u}]_j = \ell^{-1}\sum_{i=0}^{\ell-1}\omega_{\ell}^{-ij}u_{i},
\end{equation*}
where the matrix $F_{\ell}'$ is defined by $[F_{\ell}']_{i,j} \triangleq \omega_{\ell}^{-ij}$ and $\ell^{-1}$ is multiplicative inverse of $\ell$ in $\mathbb{F}_q$.
Since the \naive~complexity of this transform is $O(\ell^{2})$, we use the reduced complexity FFT version popularized by Cooley and Tukey~\cite{Cooley-moc65}.


\subsection{Cooley-Tukey Fast Fourier Transform}
\label{sec:ctfft}

Let $a$ and $b$ be arbitrary positive integers and define $\ell= ab$, $\gamma = \omega^{b}_{\ell}$, and $\beta = \omega^{a}_{\ell}$.
It is easy to verify that the elements $\gamma$ and $\beta$ have multiplicative orders $a$ and $b$ in the field $\mathbb{F}_q$.
The Cooley-Tukey formula~\cite{Blahut-1985,Cooley-moc65} for $\vecnot{u} = F_{\ell} \vecnot{v}$ is given by
\begin{equation}
\label{eq:fft_sum}
u_{a i' + i''} = \sum_{j'=0}^{b-1}\beta^{j'i'}\left[\omega_{\ell}^{j'i''}\left(\sum_{j''=0}^{a-1}\gamma^{j''i''}v_{j'+b j''}\right)\right].
\end{equation}

From this equation, one can see there are four steps in the Cooley-Tukey FFT.
First, $b$ Fourier transforms of length $a$ are computed on $b$ interleaved blocks.
Next, the $i$-th element of the resulting vector, which is indexed by $i=b i'' + j'$, is multiplied by the twiddle factor $\omega_{\ell}^{j' i''}=\omega_{\ell}^{\lfloor i/b \rfloor (i \bmod b)}$.
Then, $a$ Fourier transforms of length-$b$ are computed on $a$ adjacent blocks in the resulting vector.
Finally, the output vector is formed by deinterleaving the result of the previous step by $a$.
\iffull
The complexity is now reduced to $O(\ell(a+b))$.
This process is described in more detail in Appendix~\ref{sec:fft}.
\fi


\begin{defn}
\label{def:Kronecker}
Let $A$ and $B$ be $a \times a$ and $b\times b$ square matrices.
The Kronecker product of $A$ and $B$ is defined to be
\[ A\otimes B \triangleq \begin{bmatrix} A_{1,1} B & \cdots & A_{1,a} B \\ \vdots & \ddots & \vdots \\ A_{a,1} B & \cdots & A_{a,a} B \end{bmatrix}. \]
\end{defn}

\begin{defn}
\label{def:shuffle}
For a vector $\vecnot{v}$ of length $ab$, the perfect-shuffle permutation matrix, $S_{a,b}$, is the permutation matrix associated with writing $v$ into an $a \times b$ matrix column-wise and then reading it out row-wise.
Using this definition, one finds that 
\[
S_{a,b}^{T}(A\otimes B)S_{a,b}=(B\otimes A),
\]
where $S_{a,b}^{T}=S_{b,a}$.
\end{defn}

Based on these definitions, we give an expression for the transform using matrix operations.
\begin{lem}
\label{lem:Fab}
The Cooley-Tukey decomposition of the length-$ab$ fast Fourier transform can be expressed in terms of Kronecker products as
\begin{align*}
F_{ab} &= S_{b,a} (I_a \otimes F_b) D_{a,b} (F_a \otimes I_b) \\
&= (F_b \otimes I_a) S_{b,a} D_{a,b} (F_a \otimes I_b),
\end{align*}
where $I_a$ denotes the $a\times a$ identity matrix and the diagonal twiddle-factor matrix is defined by $\left[D_{a,b}\right]_{i,i}=\omega_{ab}^{\left\lfloor i/b\right\rfloor (i\bmod b)}$.
\end{lem}

This can be extended to the general mixed-radix FFT of length $N=\prod_{m=1}^{n}\ell_{m}$ by recursion.
\begin{lem}
\label{lem:generalFFT}
Let $p_{j}=\prod_{m=1}^{j}\ell_{m}$. Then the length-$N$ fast Fourier transform can be decomposed as
\begin{equation}
\label{eq:fft_general}
F_{N} = U_n U_{n-1} \cdots U_1 , 
\end{equation}
where
\begin{equation}
\label{eq:fft_U}
U_{m} =(S_{N/p_{m},\ell_{m}}D_{\ell_{m},N/p_{m}}\otimes I_{p_{m}/\ell_{m}})(F_{\ell_{m}}\otimes I_{N/\ell_{m}}).
\end{equation}
For the inverse transform, $F_N ' = U_n ' U_{n-1} ' \cdots U_1 '$ where $U' _m$ is defined by  replacing $F_{\ell_m}$ by $F_{\ell_m}'$ and $D_{a,b}$ by $D_{a,b}'=D_{a,b}^{-1}$.
\end{lem}
The proofs are given in Appendix~\ref{sec:fft}.
The existence of this Kronecker-product formulation of the GFFT is a key reason that one can construct cyclic polar codes based on the GFFT.
We note that an alternative construction based on the Good-Thomas (or prime-factor) FFT is also possible~\cite{Blahut-1985}.
The main difference is that no twiddle factors are required but the block sizes must be relatively prime.

\subsection{Channels}
\label{sec:channels}

A \emph{$q$-ary symmetric channel with erasures} is determined by the parameters $(q,\beta,\epsilon)$ and is denoted by QSCE($q,\beta,\epsilon$). 
Its transition probabilities are defined, for $x\in \mathbb{F}$ and $y\in \mathbb{F}\cup\{?\}$, by
\begin{equation*}
W(y|x)=
\begin{cases}
1-\beta-\epsilon & \text{if } y=x,\\
\frac{\beta}{q-1} & \text{if } y \in \mathbb{F} \backslash \{x\},\\
\epsilon   & \text{if } y=?
\end{cases}
\end{equation*}
The Shannon capacity of this channel is derived in Appendix~\ref{sec:qsce_capacity}.
Two important special cases of this channel can be obtained by setting either of its parameters to zero.
The \emph{$q$-ary erasure channel} with parameter $\epsilon$ is denoted and defined as QEC($q,\epsilon$) $\triangleq$ QSCE($q,0,\epsilon$).
The \emph{$q$-ary symmetric channel} with parameter $\beta$ is denoted and defined as QSC($q,\beta$) $\triangleq$ QSCE($q,\beta,0$).
For simplicity of notation, we will denote these channels as QSCE($\beta,\epsilon$), QEC($\epsilon$) and QSC($\beta$), respectively.

\section{Cyclic Polar Code Construction}
\label{sec:construction}

\subsection{Overview}

In this section, we describe our construction of cyclic polar codes over the finite field $\mathbb{F}$ with $q$ elements.
The construction depends on the block length $N$, which must divide $q-1$, and the ordered integer factorization $N=\prod_{m=1}^{n} \ell_m$ where each $\ell_m$ is a positive integer.
In contrast to the SC decoder of \arikan's uniform $G_2$ polar codes, some changes are required.
First, the component matrices are not necessarily $2\times 2$ or even the same size.
Next, there are multiplications by twiddle factors after each encoding stage to make the overall transform into a Fourier transform.

The encoder mapping follows from the mixed-radix Cooley-Tukey inverse FFT decomposition for $N=\ell_1  \ell_2 \cdots \ell_n$ based on \eqref{eq:fft_U}.
In particular, let $u^{(0)}_i \in \mathbb{F}$ for $i=0,1,\ldots,N-1$ be the GFFT of a codeword.
Each element of the spectrum is either assigned to carry information or to be frozen to 0.
Let $\mathcal{A} \subseteq \{0,1,\ldots,N-1\}$ be the set of indices that carry information and let its complement $\mathcal{A}^c$ denote the set of indices that are frozen to 0.
The set $\mathcal{A}$ is the output of the code design process discussed in Section~\ref{sec:design}.

Recollect that
\vspace{-2mm}
\begin{equation*}
u_i = \sum_{j=0}^{N-1} \omega_{N}^{ij} v_j,
\end{equation*}
where $\omega_{N}$ has order $N$ in $\mathbb{F}$.
In polynomial notation, with $v(x) = \sum_{j=0}^{N-1} v_j x^{j}$, we have
\begin{equation*}
u(x) = \sum_{i=0}^{N-1} u_i x^{i} = \sum_{i=0}^{N-1} v(\omega_{N}^{i}) x^{i} .
\end{equation*}
So, we see that $u_i$'s are evaluations of $v(x)$.
Given $\mathcal{A}^{c}$, the set of indices frozen to zeros in $u(x)$ such that $u_i = 0 \ \forall \  i \in \mathcal{A}^{c}$, there exists a generator $g(x)$ such that
\begin{equation*}
v(x) = u_{\mathcal{A}}(x) g(x) = u_{\mathcal{A}}(x) \prod_{i \in \mathcal{A}^{c}} (x - \omega_{N}^{i}) ,
\end{equation*}
where $u_{\mathcal{A}}(x)$ represents the information polynomial of degree at most $K-1$.
Hence, we have a cyclic code.
Since we need $N | (q-1)$ for $\omega_N$ to exist in $\mathbb{F}$, the field size must grow with the blocklength.

\begin{sidewaysfigure}[p]
\begin{center}
\large
\scalebox{0.9}{\includegraphics[width=\textwidth,height=\textheight,keepaspectratio]{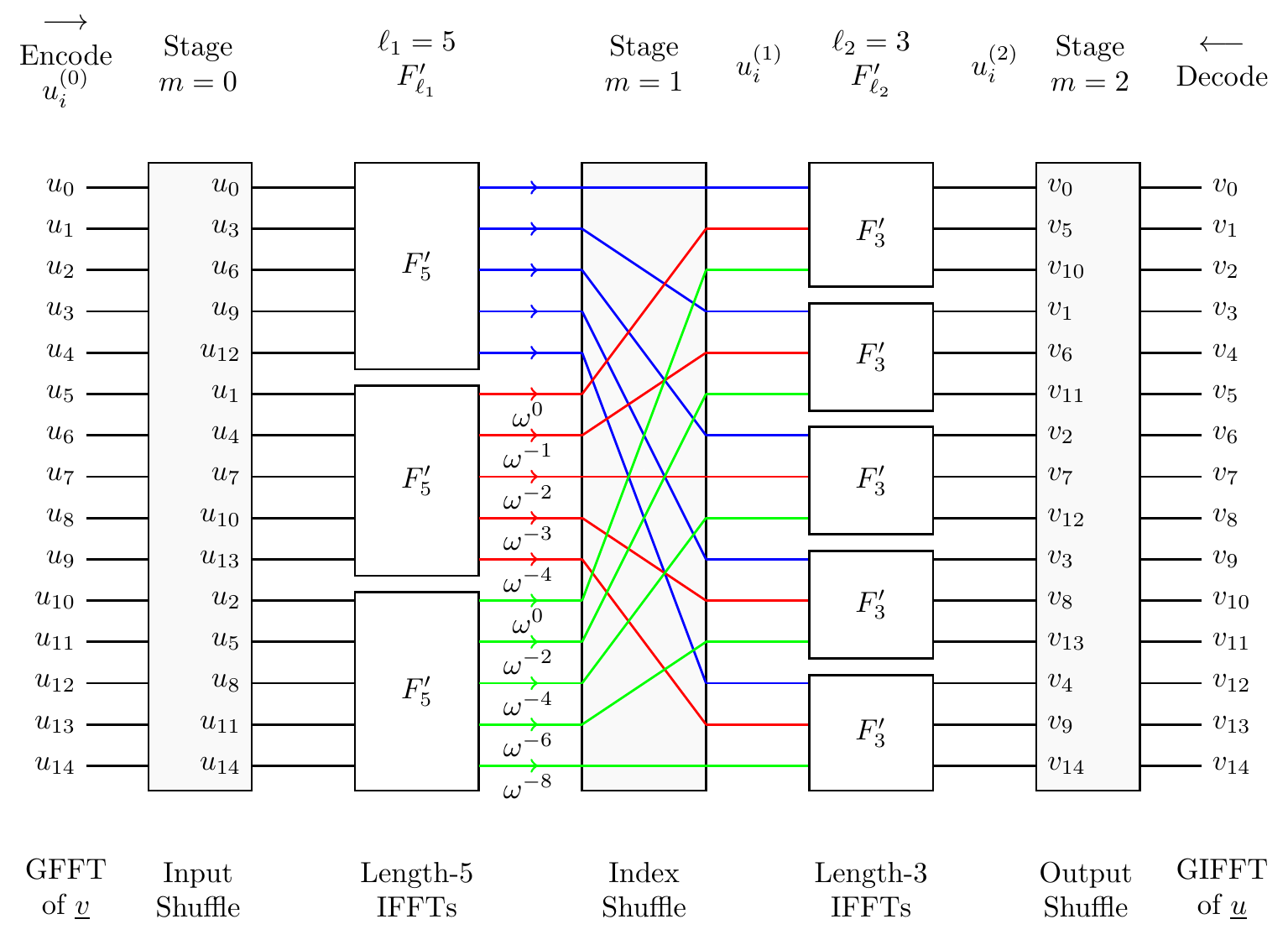}}
\normalsize
\caption{\label{fig:fft}An example for $N = 5 \cdot 3 = 15$ over $\mathbb{F}_{16}$ depicting the Cooley-Tukey fast Fourier transform.
The $F_5'$ and $F_3'$ blocks are a \naive~implementation of the inverse Fourier transform.
$\omega$ is a $N^{th}$ root of unity in $\mathbb{F}_{16}$.
Some lines are colored just for visual clarity as they cross paths during shuffling operations.
}
\end{center}
\end{sidewaysfigure}

\begin{sidewaysfigure}[p]
\begin{center}
\large
\scalebox{1}{\includegraphics[width=\textwidth,height=\textheight,keepaspectratio]{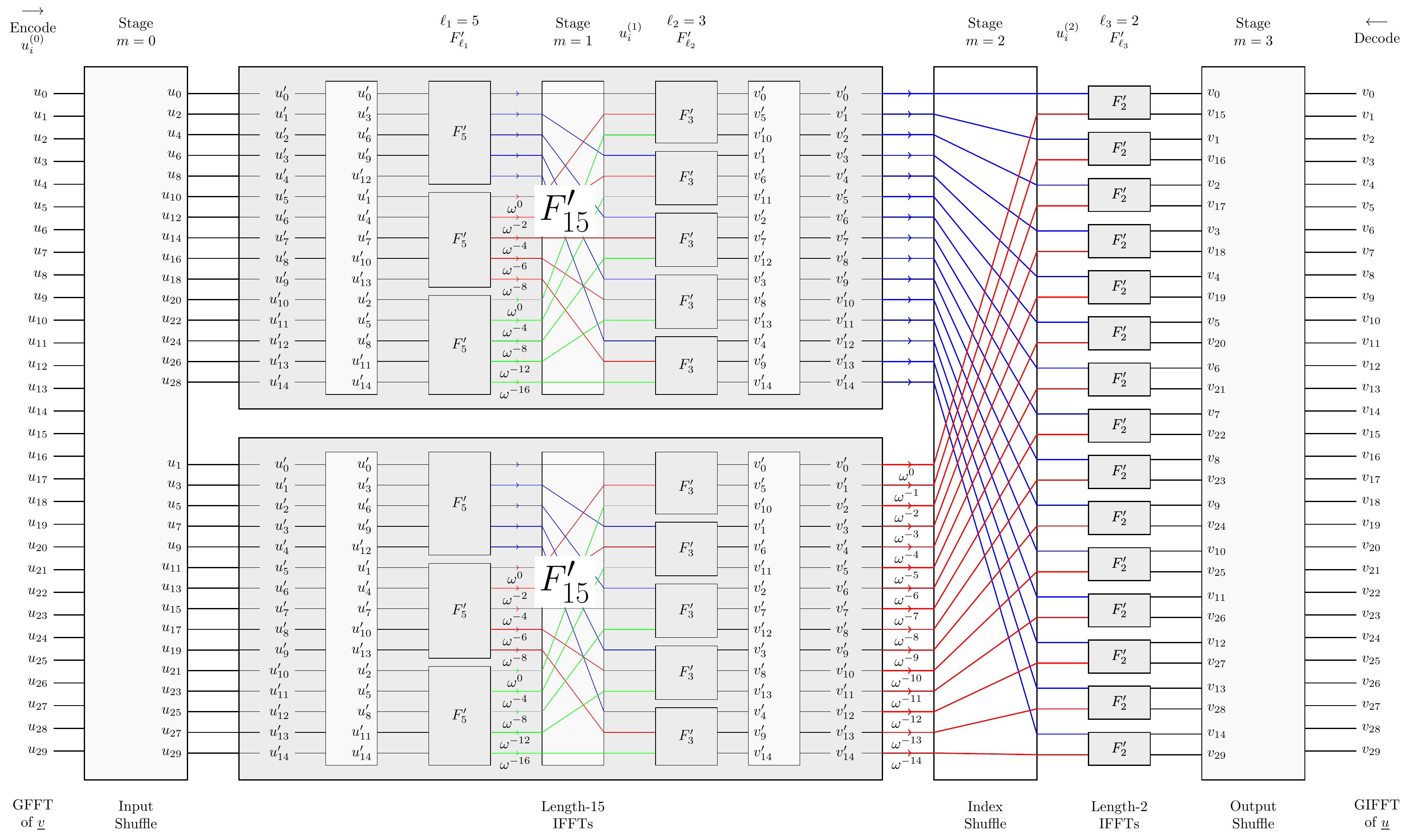}}
\normalsize
\caption{\label{fig:fft_30}An example for $N = 5 \cdot 3 \cdot 2 = 30$ over $\mathbb{F}_{31}$ depicting the Cooley-Tukey fast Fourier transform.
The $F_5'$, $F_3'$ and $F_2'$ blocks are a \naive~implementation of the inverse Fourier transform.
$\omega$ is a $N^{th}$ root of unity in $\mathbb{F}_{31}$.
Some lines are colored just for visual clarity as they cross paths during shuffling operations.
}
\end{center}
\end{sidewaysfigure}

The encoder proceeds by filling the vector $u^{(0)}_i$ and using the mixed-radix Cooley-Tukey algorithm to compute the inverse Fourier transform.
The formula for one stage of the transform is given by,
\begin{equation*}
\vecnot{u}^{(m)} = U_m ' \vecnot{u}^{(m-1)}
\end{equation*}
where $U_m'$ is defined in~\eqref{eq:fft_U} for $m=1,2,\ldots,n$ and $\vecnot{u}^{(n)} = \vecnot{v}$.
An example with $N=5\cdot 3=15$ is shown in Fig.~\ref{fig:fft} and with $N=5\cdot 3 \cdot 2=30$ is shown in Fig.~\ref{fig:fft_30}.

Like other polar code constructions, the set of frozen indices is chosen using a design process that depends on the channel.
In this work, we focus on a number of special cases that allow simplifications.
First, we consider the case where $N=2^n$ is a power of 2 and $q$ is prime.
In this case, polarization is based on the standard radix-2 Cooley-Tukey FFT and the decoder can be implemented efficiently for arbitrary $q$-ary channels.
After that, we consider the $q$-ary erasure channel for arbitrary $\mathbb{F}_q$ because both the decoder and the design process can be implemented efficiently in this case too.
Subsequently, we also discuss a decoding strategy in the presence of errors and erasures.

\subsection{Arbitrary $q$-ary Channel with $q$ prime and $N=2^n$}
\label{sec:soft-decoder}
For $N=2^n$, the design and decoding operations are quite similar to standard polar codes.
Based on the factor-graph perspective on polar codes~\cite{Mori-isit09}, the successive cancellation decoder is equivalent to a particular message-passing schedule on a factor graph with $q$-ary probability messages.
In particular, one needs to keep track of $q$ probabilities for each symbol in the graph.
For the variable denoted by $a$, these will be denoted as $p_a (x) = \Pr (a=x)$ for $x\in \mathbb{F}$.

Consider a $2\times 2$ butterfly operation defined by the input $(a_0,a_1)$, output $(b_0,b_1)$, and the relations
\begin{IEEEeqnarray}{rCl}
b_{0} & = & a_{0}+a_{1}, \\
b_{1} & = & a_{0}+\alpha a_{1}.
\end{IEEEeqnarray}
Now, to estimate $(a_0,a_1)$ from $(b_0,b_1)$ in the polar decoding order, we can write
\begin{IEEEeqnarray}{rCl}
\label{eq:a0}
\hat{a}_{0} & = & (1-\alpha)^{-1}(b_{1}-\alpha b_{0}), \\
\label{eq:a11}
\hat{a}_{1} & = & b_{0}-a_{0}, \\
\label{eq:a12}
\hat{a}_{1} ' & = & \alpha^{-1}(b_{1}-a_{0}).
\end{IEEEeqnarray}
Using these equations, one can use standard techniques from low-density parity-check codes to compute the optimal soft estimates of $(a_0,a_1)$ from soft estimates of $(b_0,b_1)$~\cite[Section 2.4]{Richardson-MCT08}.
Since all arithmetic is modulo the prime $q$, the soft estimate for the addition of two symbols is given by the circular convolution of their probability vectors. For $b_{0} = a_{0}+a_{1}$, we have,
\begin{equation*}
p_{b_{0}} (y) = \textstyle{}\sum_{x\in \mathbb{F}} p_{a_{0}} (x) p_{a_{1}} \left( y-x \right).
\end{equation*}
Similarly, the soft estimate for the multiplication of a symbol by a fixed scalar is given by a permutation of the probability vector. For $b_1' = \beta b_1$, we have
\begin{IEEEeqnarray*}{rCl}
p_{b_1 '} (x) & = & \Pr(b_1 = \beta^{-1} x) = p_{b_1} (\beta^{-1} x).
\end{IEEEeqnarray*}
Also, independent estimates (e.g., $\hat{a}_1$ and $\hat{a}_1 '$) are combined by renormalizing the product of their probability vectors:
\begin{equation*}
p_{a_1 | \hat{a}_1 , \hat{a}_1 '} (x) = \frac{p_{\hat{a}_1} (x) p_{\hat{a}_1 '} (x)}{\sum_{x'\in{\mathbb{F}}} p_{\hat{a}_1} (x') p_{\hat{a}_1 '} (x')}.
\end{equation*}
In this case, the SC decoder can be defined recursively for the whole graph based on these operations.
Hard decisions are made for the information symbols based on the maximum value in their associated probability vectors.


Generalization to the mixed-radix case with arbitrary block sizes is straightforward but computationally expensive.
Soft estimates are stored as vectors of probabilities but a-posteriori-probability (APP) decoding is required for the FFT blocks, which is why we resort to algebraic hard-decision successive-cancellation decoding.



\subsection{Algebraic Erasures Decoding}
\label{sec:erasure}

For erasure channels, cyclic polar codes can be efficiently designed and decoded for an arbitrary $\mathbb{F}_q$ and $N|(q-1)$.
Each factor $\ell$ of $N$ requires the decoding of a $\ell \times \ell$ matrix $F_\ell'$ defined by $[F_\ell']_{i,j} \triangleq \omega_{\ell}^{-ij}$.
Similar to~\cite{Mori-itw10}, polar decoding for $F_\ell'$ essentially requires the decoding of a nested sequence of RS Codes.
Fortunately, erasures only RS decoding can be implemented efficiently using Forney's algorithm~\cite{Forney-it65}.

Let $\vecnot{v} = F_\ell' \vecnot{u} $ and $\vecnot{y}$ be an observation of $\vecnot{v}$ through an erasure channel.
The polar decoding problem for $F_\ell'$ is, for $j=0,1,\ldots,\ell-1$, decode $u_j$ from $\vecnot{y}$ and $u_0,\ldots,u_{j-1}$.
For the $j$-th decode, this can be viewed as decoding a known coset of an $(\ell,\ell-j)$ RS code.
To use Forney's algorithm, we calculate $j$ modified syndromes by removing the contribution of known inputs.
\iffull
The details of this process are discussed in Appendix~\ref{sec:Forney}.
\fi

Due to the nature of Forney's algorithm, decoding either recovers all or none of the unknown inputs.
This leads to the following rule for the SC decoding of each block:
\emph{if $\nu$ inputs of the block are known and at most $\nu$ outputs are erased, then use Forney's decoder to recover all unknown inputs; otherwise, pass an erasure as the decoded input.}
The operations performed during SC decoding are given below:
\begin{itemize}

    \item Begin with the output $\vecnot{v}$ set to the received (hard) values from the channel.
    
	\item While decoding the set of blocks $F_{\ell_{m}}'$ for the $j^{\rm th}$ input, $j=0,\ldots,\ell_m-1$, use the decoding rule above and pass the newly decoded $j^{\rm th}$ inputs at stage $m$ to the $j^{\rm th}$ $F_{N/\ell_{m}}'$ block from the top at the previous stage. Then, recurse and execute the decoder at that stage.
		
	\item While decoding blocks $F_{\ell_1}'$, use the frozen symbols at the input as knowledge to compute syndromes for the Forney decoder. A block $F_{\ell_1}'$ that does not have any frozen symbols must have all outputs already known for successful decoding. Once the outputs for these blocks are determined, pass them forward to the next stage.

	\item When the procedure returns to the set of blocks $F_{\ell_{m}}'$ with an update for the $j^{\rm th}$ input, the updated inputs are used to decode the next input according to the above rule. Based on the SC decoder, the outputs of the block are not updated until all inputs are ``recovered''.
		
\end{itemize}

For the erasure channel, Forney's decoder is run exactly once for each block (when the number of known inputs and outputs equals the block length).
The decoding complexity of Forney's algorithm for a length-$\ell$ block is at most $C\ell^2$ operations for some $C>0$.
Since there are $\prod_{j \neq m} \ell_j = N/\ell_m$ blocks at stage $\ell_m$, the decoding complexity is bounded by
\begin{equation*}
\sum_{m=1}^{n} \prod_{j \neq m} \ell_j \left( C\ell_{m}^{2} \right) = C N \sum_{m=1}^{n} \ell_m \leq C N n \max_m \ell_m .
\end{equation*}


\subsection{Algebraic Errors and Erasures Decoding}
\label{sec:hard-decoder}
The approach in Section~\ref{sec:erasure} can also be extended to handle errors and erasures.
In this case, each small block is decoded using algebraic errors and erasures decoding.
The decoding of each small block results in success, failure, or miscorrection.
In the event of decoder failure, an erasure is passed back to the previous stage.
Otherwise, the value estimated by algebraic decoding is passed back to the previous stage.

Using this decoding strategy, cyclic polar codes can be efficiently designed and decoded for arbitrary $q$ and $N|(q-1)$.
For each of the $\ell_{m}$ decoding iterations of block $F_{\ell_{m}}'$, the Berlekamp-Massey (BM) algorithm is used to obtain the error-erasure locator polynomial which is fed into Forney's decoder to correct errors and erasures~\cite[Section 7.5]{Blahut-ACDT03}\footnote{In Fig. 7.10 of~\cite[Section 7.5]{Blahut-ACDT03}, the update equation in the left-most box above the bottom-most box in the flowchart must be $L \leftarrow r-(L-\rho)$ and not $L \leftarrow r-L-\rho$.}.
Unlike the erasure case, the decoding operation must be executed during each decoding stage.
Thus, the decoding complexity is increased to $C N n \max_m \ell_m^2$.

\iffull
Since each iteration of the SC decoder involves decoding a RS code whose minimum distance depends on the number of inputs already recovered, the decoder efficiency is increased if errors are converted into erasures during the multistage decoding process.
This is because, with $\nu$ known input symbols, the decoder can correct $t$ errors and $e$ erasures iff $\nu\geq 2t+e$.
Hence, this decoding strategy is sub-optimal.
All of the intermediate channels in this process can be modeled as QSCE.

We need to perform density evolution of the error and erasure probabilities to design the code.
Since density evolution is complex to be performed on arbitrary $q$-ary channels, a Monte Carlo design methodology is employed to compute the capacities of input coordinate channels for an arbitrary blocklength $N$.
An example of the design results is discussed in Section~\ref{sec:example}.
\fi

%

\section{Code Design}
\label{sec:design}

\subsection{Erasure Channels}
\label{sec:erasure_code_design}


The input parameters to the design procedure for the QEC($\epsilon$) are $\left(N,q,\epsilon,\delta\right)$ where $\delta$ is the target block erasure rate.
Consider the upper bound on $P_B$ given by
\begin{equation}
P_{B} \leq \sum_{i\in \mathcal{A}} \epsilon_{i}^{(0)},
\label{eq:pB}
\end{equation}
where $\epsilon_{i}^{(0)}$ is the erasure probability of the coordinate channel $W_{N}^{(i)}$ and $\mathcal{A}$ is the set of information symbols~\cite{Arikan-it09}. 
The design procedure chooses $\mathcal{A}$ to be the largest subset $S\subseteq \{0,1,\ldots,N-1\}$ such that $\sum_{i\in \mathcal{S}} \epsilon_{i}^{(0)} \leq \delta$.
This design strategy is applicable to both binary polar codes and cyclic polar codes.

The erasure probabilities of the output symbols are initialized to $\epsilon_{0}^{(n)}=\epsilon$.
The design process commences by performing density evolution -- recursively computing the erasure rates of the coordinate channels from stage $m=n$ down to $m=0$.
Due to the structure of polar codes, there will be at most $\ell_{m+1}\ell_{m+2} \cdots \ell_n$ distinct erasure probabilities in the $m$-th stage.
The $i$-th distinct erasure probability at stage $m$ is denoted by $\epsilon_{i}^{(m)}$ for $i=0,1,\ldots,\ell_{m+1}\ell_{m+2} \cdots \ell_n-1$ and $m=0,1,\ldots,n$.

Consider the erasure decoding of a single block of length $\ell$ as described in Section~\ref{sec:erasure}.
Given the knowledge of $j$ previously decoded inputs, the next input can be computed if and only if at least $\ell-j$ of the outputs are not erased.
This is because the $j$ known symbols imply that the output sequence lies in a known coset of an $(\ell,\ell-j)$ RS code that can correct $j$ erasures.
Thus, if the outputs are erased i.i.d.\ with probability $\epsilon'$ and $j$ previous inputs are known, then next input is erased with probability $\psi(\ell, j, \epsilon')$ given by
\begin{equation}
\label{eq:epsilon0}
\psi(\ell, j, \epsilon') \triangleq  \sum_{i=0}^{(\ell-1)-j} \binom{\ell}{i}(1-\epsilon')^{i} (\epsilon')^{\ell-i}.
\end{equation}
We note that this formula is due to Mori and Tanaka~\cite{Mori-itw10}.

For an ordered factorization $N=\ell_1 \ell_2 \cdots \ell_n$, this implies that the distinct erasure probabilities of the coordinate channels satisfy the recursion
\begin{equation}
\label{eq:epsilon}
\epsilon^{(m-1)}_{\ell_{m}k+j} = \psi \left( \ell_m,j, \epsilon^{(m)}_{k} \right)
\end{equation}
for $j=0,1,\ldots,\ell_{m}-1$ and $k=0,1,\ldots,\ell_{m+1}\ell_{m+2} \cdots \ell_n-1$.

Having established the stage-by-stage evolution of the erasure rates in~\eqref{eq:epsilon0} and \eqref{eq:epsilon}, we re-state the polarization theorem in~\cite{Arikan-it09} for the case of a $q$-ary erasure channel (QEC).

\begin{thm}
\label{thm:polarization}
For a QEC $W$ with erasure rate $\epsilon$, the input coordinate channels $W_{N}^{(i)}$ polarize in the sense that, for any fixed $\theta \in (0,1)$, as $N$ goes to infinity through multiples of positive integers, the fraction of indices $i \in \{0,1,\ldots,N-1\}$ for which $\epsilon_{i}^{(0)} \in (1-\theta,1]$ goes to $\epsilon$ and the fraction for which $\epsilon_{i}^{(0)} \in [0,\theta)$ goes to $(1-\epsilon)$.

\begin{proof}
We will need the following properties of the mapping defined in~\eqref{eq:epsilon0} to motivate this proof.
\begin{lem}
\label{lem:preserve_eps}
Eqn.~\eqref{eq:epsilon0} defines a mapping from $\mathbb{R}$ to $\mathbb{R}^{\ell}$ with the following two properties:
\begin{itemize}

\item[$(i)$] The mapping preserves the mean erasure rate through each stage of density evolution so that 
\begin{equation}
\frac{1}{\ell} \sum_{j=0}^{\ell - 1} \psi(\ell, j, \epsilon') = \epsilon'.
\end{equation}

\item[$(ii)$] If $\epsilon' \in (0,1)$, then the erasure rates of the new coordinate channels polarize away from the mean such that
\begin{equation}
\psi(\ell, \ell-1, \epsilon') < \epsilon ' < \psi(\ell, 0, \epsilon').
\end{equation}

\end{itemize}
\end{lem}
Based on these two properties, the martingale convergence analysis in~\cite{Arikan-it09} can be used to show that the erasure rates must polarize to 0 and 1 as $n\to \infty$ and that the fraction of coordinate channels with erasure rate 0 must be $\epsilon$. A detailed proof is provided in Appendix~\ref{sec:channel_polarize}.   \qedhere
\end{proof}
\end{thm}

To motivate the limit $n\to \infty$, let $p$ be a prime that satisfies $\gcd (N,p)=1$ for some $N=\ell_1 \ell_2 \cdots \ell_n$.
Then, there exists an extension finite field $\mathbb{F}_{p^m}=\mathbb{F}_{q}$ with $m\leq N-1$ such that $N|(q-1)$.
Of course, the field size may be exceedingly large for a given $N$ and $p$.






\subsection{An Example}
\label{sec:example}
\begin{sidewaysfigure}[p]
\begin{center}
\large
\scalebox{0.9}{\includegraphics[width=\textwidth,height=\textheight,keepaspectratio]{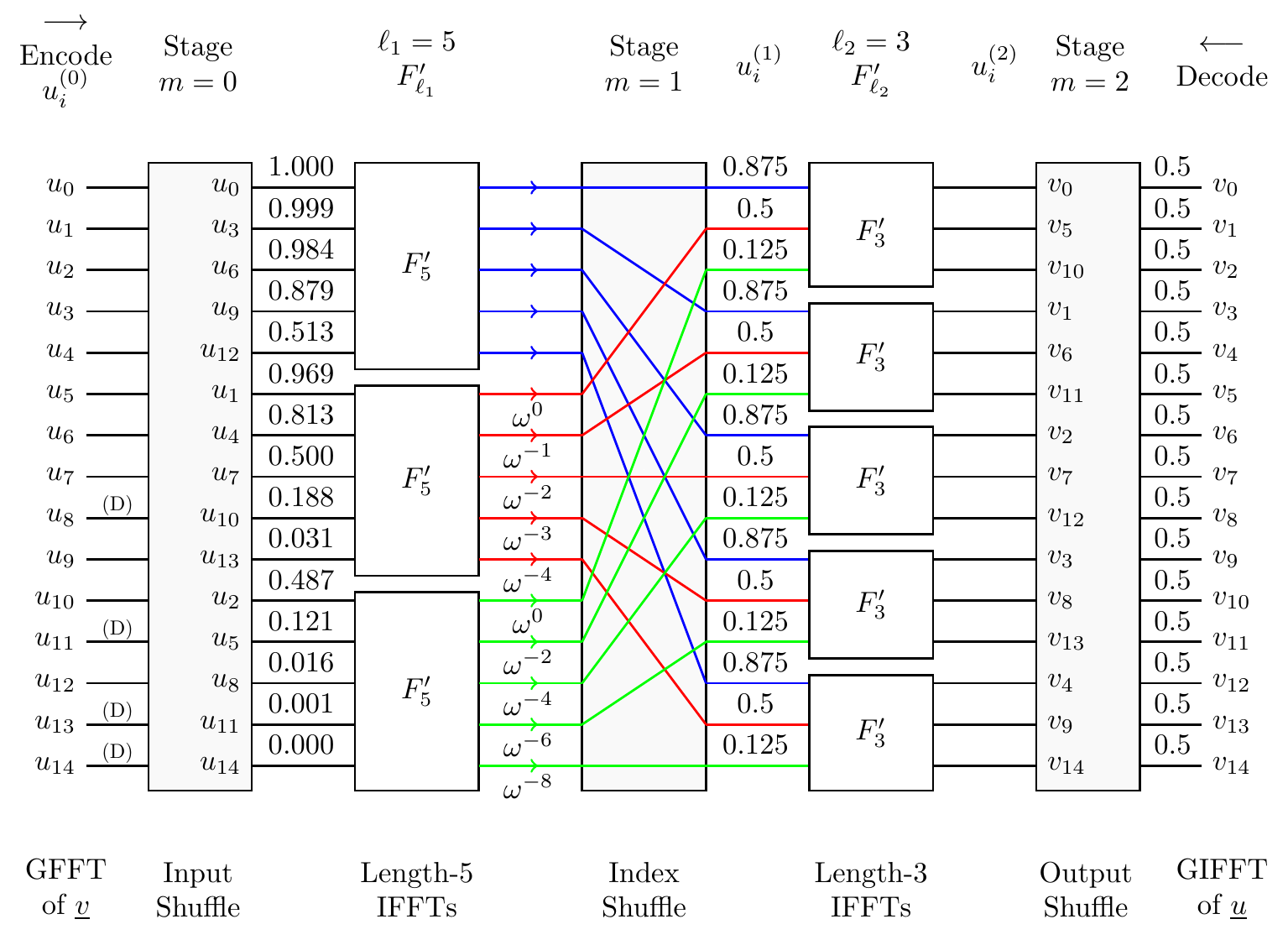}}
\normalsize
\caption{\label{fig:fft_de}An example for $N = 15$ over $\mathbb{F}_{16}$ depicting the transform, density evolution process and code construction. Design parameters: $\epsilon = 0.5$, $\delta = 0.1$. Information symbols are marked with (D).
The $F_5'$ and $F_3'$ blocks are a \naive~implementation of the inverse Fourier transform.
$\omega$ is a $N^{th}$ root of unity in $\mathbb{F}_{16}$.
Some lines are colored just for visual clarity as they cross paths during shuffling operations.
}
\end{center}
\end{sidewaysfigure}

Fig. \ref{fig:fft_de} shows an example for $N = 15$ over $\mathbb{F}_{16}$ depicting the transform, density evolution process and code construction over QEC($\epsilon$).
The design parameters chosen for this example are channel erasure rate $\epsilon = 0.5$ and maximum block erasure rate $\delta = 0.1$.
The design methodology is described in Section~\ref{sec:erasure_code_design}.
The various erasure rates at intermediate stages are shown in the graph for easy comprehension.
For example, for the $F_3'$ block, the output erasure probabilities are $\epsilon_{0}^{(2)}=0.5$ and the input erasure probabilities are given by $\epsilon_0^{(1)} = 0.875$, $\epsilon_1^{(1)} = 0.5$, and $\epsilon_2^{(1)} = 0.125$.
These values are repeated for the other two $F_3'$ blocks as well, because we only track the distinct erasure probabilities at each stage and not all $N$ indices.

According to (\ref{eq:pB}), the information indices are chosen as $\mathcal{A}=\{8,11,13,14\}$ and are represented with the prefix (D), for data, in the input side of the graph. 
Thus, the rate of this code is $\frac{4}{15}=0.2667$. 
The generator polynomial is given by
\begin{equation*}
g(x) = \prod_{i \in \mathcal{A}^c} (x-\omega_{15}^{i}).
\end{equation*}
Now, the encoder fills the indices in $\mathcal{A}^c$ with zeros and the other four indices with information, performs the transform to get the output vector $\vecnot{v}$ and transmits it via $N$ independent uses of the underlying channel.

\iffull
The results obtained for $N=15$ over $\mathbb{F}_{256}$ on QSCE($0.5$,$0$) using Monte Carlo design over $M=10^4$ iterations are below. 
Each column from the far left of the matrix corresponds to the inputs of that $\ell=5$ block from the top in Fig.~\ref{fig:fft}.
The ($p_{\text{error}}, p_{\text{erasure}}$) pair for the input channels are:

\[
\begin{bmatrix}
    (0.9354,0)&        (0,0.9999)&        (0.0176,0.9475)\\
    (0.0613,0.9354)&   (0.0238,0.9751)&   (0.3257,0.4885)\\
    (0.2745,0.7054)&   (0.0140,0.9718)&   (0.0836,0.4620)\\
    (0.0167,0.9799)&   (0.0555,0.8369)&   (0.0516,0.2145)\\
    (0.0471,0.9192)&   (0.0858,0.4822)&   (0.0150,0.0730)\\
\end{bmatrix}
\]

It can be noted that the resultant probabilities of error are significantly lower than the corresponding probabilities of erasure. 
The channels with high error probabilities are those decoded with no input information.
While the capacity of QSCE($0.5,0$) is $0.3753$, the average capacity of the input channels are significantly low at $0.1807$ due to the use of a sub-optimal decoding strategy, as noted in Section~\ref{sec:hard-decoder}. 
The design can be significantly improved if an APP decoder is employed in place of the hard errors and erasures decoder, at the cost of additional computational complexity.
\fi

\section{Results and Discussion}
\label{sec:results}

Binary and cyclic polar codes were designed for various blocklengths on BEC($0.5$) and QEC($0.5$), respectively, for a target block erasure rate $\delta = 0.1$.
The resulting rates are tabulated in Table~\ref{tab:rates}.
We see that cyclic polar codes achieve higher rates at much smaller blocklengths than equivalent rate binary polar codes.
As a more fair comparison, let us consider the cyclic polar code of blocklength $N=1023$ over GF($1024$) and the binary polar code of blocklength $N=2^{16}=65536$ over GF($2$) (or GF($65537$), if we ignore complexity comparisons).
The equivalent binary blocklength for the length-$1023$ code would be $N = 10 \cdot 1023 = 10230$ bits.
So, the cyclic polar code of length $N=10230$ bits can achieve a rate almost equal to that of a binary polar code with length $N=65536$ bits which is more than $6$ times higher.
This shows that this construction allows us to achieve the capacity of the erasure channel, in this case $0.5$, at much smaller blocklengths than binary polar codes, with the only constraint being the complexity introduced by higher field size.
Also, it is to be noted that the channel for cyclic polar codes is assumed to introduce symbol erasures rather than bit erasures.

Experiments also show that the order of the factorization of $N$ affects the code rate.
For $q=1024$, $N=1023$, and $\delta = 0.1$, the order [31 3 11] results in a rate of $0.4340$ while the order [3 11 31] gives a rate of $0.4291$.
This is the reason for multiple rates for some blocklengths in the table.
We note that [3 11 31] implies that the length-$3$ blocks are close to the channel.

\begin{table}[h!]
\centering
\begin{tabular}{@{}cc@{}} 
\toprule
\textbf{Blocklength} $\boldsymbol{N}$ & \textbf{Rate} $\boldsymbol{R}$ \\
\midrule
$2^3 = 8$      & $0.125$ \\
\rowcolor{blue!25}
$12$     & $0.25$ \\
\rowcolor{blue!25}
$13$     & $0.3077$ \\
\rowcolor{blue!25}
$14$     & $0.2857$ \\
$2^4=16$     & $0.25$ \\
\rowcolor{blue!25}
$30$     & $0.2667,0.3$ \\
\rowcolor{blue!25}
$60$     & $0.2833,0.3,0.3167$ \\
$2^6 = 64$     & $0.2812$ \\
\rowcolor{blue!25}
$255$     & $0.3843,0.3882,0.3922,0.3961$ \\
$2^8 = 256$    & $0.3281$ \\
\rowcolor{blue!25}
$1023$    & $0.4291,0.4340$ \\
$2^{16} = 65536$ & $0.4397$ \\
\bottomrule
\end{tabular}
\caption{\label{tab:rates}Rates achieved by Binary and Cyclic Polar Codes when designed over BEC($0.5$) and QEC($0.5$), respectively, for $P_B \leq \delta = 0.1$. The entries for cyclic polar codes are highlighted.}
\end{table}

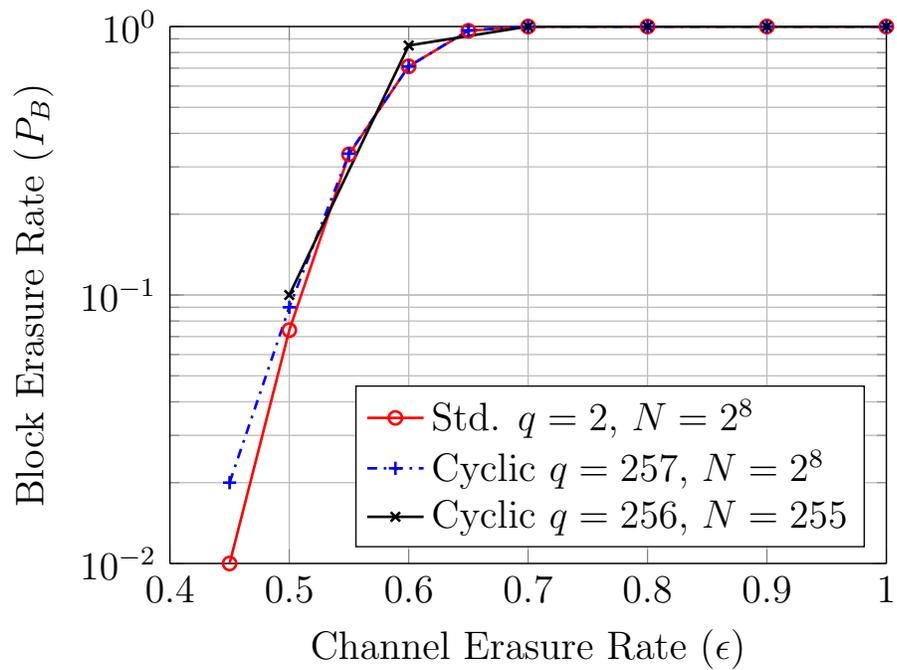
\begin{figure}[p]
\begin{center}
\scalebox{1.25}{
%
%
%
%

\begin{tikzpicture}

\begin{axis}[%
width=3in,
height=2.25in,
scale only axis,
separate axis lines,
every outer x axis line/.append style={black},
every x tick label/.append style={font=\color{black}},
xmin=0.4,
xmax=1,
xlabel={Channel Erasure Rate $(\epsilon)$},
xmajorgrids,
every outer y axis line/.append style={black},
every y tick label/.append style={font=\color{black}},
ymode=log,
ymin=0.01,
ymax=1,
ylabel={Block Erasure Rate $(P_B)$},
ymajorgrids,
yminorgrids,
title style={align=center},
legend style={legend cell align=left,align=left,fill=white},
legend pos = south east
]
\addplot [thick, color=red,solid,mark=o,mark options={solid}]
  table[row sep=crcr]{%
0.4	0\\
0.45 0.01\\
0.5	0.074\\
0.55 0.334\\
0.6	0.711\\
0.65 0.964\\
0.7	1\\
0.8	1\\
0.9	1\\
1	1\\
};
\addlegendentry{Std. $q=2$, $N=2^8$};

\addplot [thick, color=blue,dashdotted,mark=+,mark options={solid}]
  table[row sep=crcr]{%
0.4	0\\
0.45 0.02\\
0.5	0.09\\
0.55 0.336\\
0.6	0.71\\
0.65 0.968\\
0.7	1\\
0.8	1\\
0.9	1\\
1	1\\
};
\addlegendentry{Cyclic $q=257$, $N=2^8$};

\addplot [thick, color=black,solid,mark=x,mark options={solid}]
  table[row sep=crcr]{%
0.4	0\\
0.5	0.1\\
0.6	0.85\\
0.7	1\\
0.8	1\\
0.9	1\\
1	1\\
};
\addlegendentry{Cyclic $q=256$, $N=255$};

\end{axis}
\end{tikzpicture}%
}
\caption{\label{fig:fft_polar}Comparison of performance of standard polar and cyclic polar codes on QEC($\epsilon$). Design parameters were: $\delta = 0.1$; $\epsilon = 0.5$. $R = 0.328$ and $R = 0.384$ for $N = 256$ and $N = 255$, respectively. There were no block erasures observed for $N=256$ and $N=255$ at $\epsilon \leq 0.4$ over 1000 blocks each and 100 blocks each, respectively.}
\end{center}
\end{figure}

A standard polar code with $N=256$ over $\mathbb{F}_{257}$ was designed for the QEC($0.5$) to achieve $P_{B}\leq \delta = 0.1$.
The code was simulated for channels with erasure probabilities $\epsilon=0.1,0.2,0.3,\ldots,1$ and the block erasure rate for each $\epsilon$ (averaged over $M=1000$ blocks) is plotted in Fig.~\ref{fig:fft_polar}.

A cyclic polar code of blocklength $N=2^8$ over $\mathbb{F}_{257}$ (i.e., $\ell_1=\cdots=\ell_8=2$) was designed for the same parameters and the results are plotted in the same figure.
The theory suggests that the performance of these two codes should be identical and simulations support this conclusion. 
Simulation results for an $N=255=3 \cdot 5 \cdot 17$ cyclic polar code over $\mathbb{F}_{256}$ are also presented (averaged over $M=100$ blocks) and the performance validates polarization for our proposed construction.
The code has a rate of $0.384$ compared to $0.328$ for $N=2^8$, for the same design parameters.

\begin{figure}[p]
\begin{center}

\scalebox{1.25}{
%
%



\begin{tikzpicture}

\begin{axis}[%
width=3in,
height=2.25in,
scale only axis,
separate axis lines,
every outer x axis line/.append style={black},
every x tick label/.append style={font=\color{black}},
xmin=0.1,
xmax=0.8,
xlabel={Channel Error Rate $(\beta)$},
xmajorgrids,
every outer y axis line/.append style={black},
every y tick label/.append style={font=\color{black}},
ymode=log,
ymin=0.01,
ymax=1,
yminorticks=true,
ylabel={Block Error Rate $(P_B)$},
ymajorgrids,
yminorgrids,
title style={align=center},
legend pos = {south east},
legend style={legend cell align=left,align=left,draw=black,font=\small}
]
\addplot [thick, color=red,solid,mark=o,mark options={solid}]
  table[row sep=crcr]{%
0.2 0\\
0.25 0.0008\\ 
0.3	0.0940\\
0.35 0.6556\\
0.4	0.9795\\
0.42 0.9964\\
0.44 0.9996\\
0.46 1\\
0.48 1\\
0.5	1\\
0.6	1\\
0.7	1\\
0.8	1\\
0.9	1\\
1	1\\
};
\addlegendentry[align=left]{$(256,84)$ RS code};

\addplot [thick, color=red,solid,mark=+,mark options={solid}]
  table[row sep=crcr]{%
0.3	0.005\\
0.35 0.027\\
0.4	0.146\\
0.42 0.257\\
0.44 0.393\\
0.46 0.574\\
0.48 0.735\\
0.5	0.872\\
0.6	0.999\\
0.7	1\\
0.8	1\\
0.9	1\\
1	1\\
};
\addlegendentry[align=left]{$(256,84)$ CP code};

\addplot [thick, color=black,dashdotted,mark=o,mark options={solid}]
  table[row sep=crcr]{%
0.2 0\\
0.25 0.0181\\ 
0.3	0.3891\\
0.35 0.9220\\
0.4	0.9988\\
0.42 0.9999\\
0.44 1\\
0.46 1\\
0.48 1\\
0.5	1\\
0.6	1\\
0.7	1\\
0.8	1\\
0.9	1\\
1	1\\
};
\addlegendentry[align=left]{$(255,98)$ RS code};

\addplot [thick, color=black,dashdotted,mark=+,mark options={solid}]
  table[row sep=crcr]{%
0.15 0\\
0.18 0\\
0.2	0.07\\
0.22 0.11\\
0.24 0.27\\
0.26 0.5\\
0.28 0.63\\
0.3 0.9\\
0.4 1\\
0.5	1\\
0.6	1\\
0.7	1\\
0.8	1\\
0.9	1\\
1	1\\
};
\addlegendentry[align=left]{$(255,98)$ CP code};

\addplot [thick, color=blue,dashed,mark=o,mark options={solid}]
  table[row sep=crcr]{%
0.1 0.0005\\
0.15 0.0056\\
0.2 0.0267\\
0.25 0.0796\\ 
0.3	0.1753\\
0.35 0.3119\\
0.4	0.4728\\
0.42 0.5387\\
0.44 0.6029\\
0.46 0.6641\\
0.48 0.7210\\
0.5	0.7728\\
0.6	0.9417\\
0.7	0.9929\\
0.8	0.9998\\
0.9	1\\
1	1\\
};
\addlegendentry[align=left]{$(16,4)$ RS code};

\addplot [thick, color=blue,dashed,mark=+,mark options={solid}]
  table[row sep=crcr]{%
0.1 0\\
0.2	0.002\\
0.3	0.015\\
0.35 0.04\\
0.4	0.086\\
0.42 0.131\\
0.44 0.162\\
0.46 0.193\\
0.48 0.232\\
0.5	0.295\\
0.6	0.558\\
0.7	0.802\\
0.8	0.953\\
0.9	1\\
1	1\\
};
\addlegendentry[align=left]{$(16,4)$ CP code};

\end{axis}
\end{tikzpicture}%

}
\caption{\label{fig:fft_qsc}Performance of QEC-designed cyclic polar (CP) codes on QSC($\beta$). The design parameters $\delta = 0.1$; $\epsilon = 0.5$ resulted in code rates $0.328$, $0.384$ and $0.25$ for $N=256$, $N=255$ and $N=16$, respectively. No block errors were observed for $N=256$ and $N=255$ at $\beta \leq 0.2$ over 1000 blocks each and $\beta \leq 0.18$ over 100 blocks each, respectively. The theoretical performance of RS codes is also plotted for comparison.}

\end{center}
\end{figure}
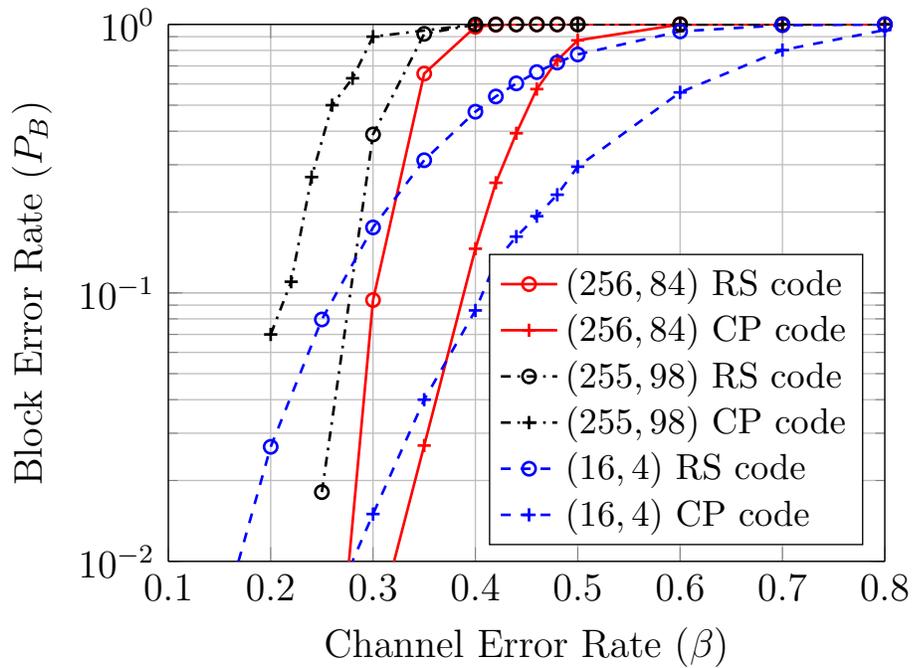

Cyclic polar codes with $N=2^{n}$ over $\mathbb{F}_{q}$, $q$ prime, were designed over QEC($0.5$) and tested on QSC($\beta$) using the soft-decision decoder discussed in Section~\ref{sec:soft-decoder}.
The design parameters were chosen to be the same as that of the simulation discussed previously.
The results obtained for $N=256$ and $N=16$ are averaged over $M=1000$ blocks and shown in Fig. \ref{fig:fft_qsc}.
Similarly, a cyclic polar code with $N=255$ over $\mathbb{F}_{256}$ was constructed for QEC($0.5$).
It was tested with the hard-decision decoder discussed in~\ref{sec:hard-decoder} on QSC($\beta$), and the results (averaged over $M=100$ blocks) are shown in the same figure.

For comparison, a RS code of rate $R\approx 0.328$ can correct a fraction $(1-R)/2 \approx 0.336$ errors and the Shannon limit (i.e., maximum error rate) of the QSC for rate $0.328$ is roughly $0.548$. Similarly, the limit for rate $0.384$ is roughly $0.491$.
The theoretical curves for RS codes of the same rates are also plotted for comparison. 
It is evident that, on the QSC, the cyclic polar code with a soft decoder clearly outperforms an RS code of same rate (cases $N=256$ and $N=16$).
The cyclic polar code with hard decision decoding does not outperform the comparable RS code.
However, designing the cyclic polar code for hard decision decoding may change this.

\iffull
While a RS code has only one chance to correct all errors and erasures, the cyclic polar code construction can exploit the depth in the graph to convert errors into erasures and also leverage polarization to provide multiple chances and perform significantly better. These results show that it may not be trivial to exhibit this theoretical advantage. 

Also, the pattern of errors and erasures at the outputs can have significant effect on code performance. The number of blocks at the output stage that get affected by errors and/or erasures should be minimum for better performance. But, there is an underlying shuffling of indices at the output. Hence, burst errors/erasures will weaken the code as consecutive erroneous indices affect multiple output blocks and the decoder may not have enough information to perform hard-decision decoding at all stages of the decoding process. It might be preferable to transmit the codeword in the shuffled format if the underlying channel is bursty.
\fi

It is interesting to note that, while the decoding of all blocks in the graph is identical to that of RS codes, the cyclic polar code itself is not a RS code because, in general, the design process does not choose a consecutive set of indices for the frozen symbols.
Our cyclic polar codes are always subcodes of a (possibly trivial) RS code though.
For example, the code in Fig.~\ref{fig:fft} has 8 consecutive zeros in its spectrum and, thus, is a subcode of a $(15,7,9)$ generalized RS code.

\section{Conclusion}
\label{sec:conclusion}

This work introduces a method to construct cyclic polar codes over $\mathbb{F}_q$ for any blocklength $N$ satisfying $N|(q-1)$.
For the QEC, these codes can be decoded efficiently using Forney's algebraic decoder to decode the intermediate blocks.
In our simulations, they outperform standard polar codes.
For the case of $N=2^n$, a soft-decision SC decoder was also implemented and tested on the $q$-ary symmetric channel.
Under SC decoding, cyclic polar codes clearly outperform RS codes of the same rate and blocklength.

An algebraic errors and erasures decoding strategy was also considered for the intermediate block codes.
Preliminary results show that this approach is suboptimal when compared to hard decision decoding of a RS code with the same rate and blocklength.
In future work, we plan to consider APP decoding of the intermediate blocks for small lengths while retaining a hard-decision decoder at larger blocks typically placed close to inputs in the graph.
We will also consider the rate of polarization for these codes based on similar work for standard polar codes~\cite{Korada-it10*2,Mori-it14}.
The programs developed for this work can be accessed at \url{https://github.com/nrenga/cyclic_polar}.


\emph{
The interesting part about working on polar codes is that one has to learn its literature with the same strategy as its own successive-cancellation decoder -- going back and forth in building knowledge slowly, but steadily, towards capacity.
Perhaps, this is truly an optimal strategy for learning, in general.
}

\chapter[\uppercase{Spatially-Coupled LDPC Codes}]{\uppercase{Spatially-Coupled LDPC Codes}}
\label{sec:scldpc_codes}

\section{Introduction}
\label{sec:intro}
Low-density parity-check (LDPC) codes are widely used due to
their outstanding performance under low-complexity belief propagation (BP) decoding.  
However, an error probability exceeding that of
maximum-a-posteriori (MAP) decoding has to be tolerated with (sub-optimal) BP decoding. 
Lately, it has been empirically observed 
for spatially coupled LDPC (SC-LDPC) codes---first introduced by 
Felstr\"om and Zigangirov as convolutional LDPC codes \cite{Felstrom-it99}---that the BP performance of 
these codes can improve dramatically towards the MAP performance of the 
underlying LDPC code under many different settings and conditions, e.g.~\cite{Lentmaier-ita09}.
This phenomenon, termed \emph{threshold saturation}, has been proven rigorously in \cite{Kudekar-it11,Kudekar-it13}. 
In particular, the BP threshold of a coupled LDPC ensemble tends to its MAP threshold on any binary memoryless symmetric
(BMS) channel.

Besides their excellent performance on the BEC and AWGN channels, much less is known about the burst error correctability of SC-LDPC codes. 
In~\cite{Jule-isit13}, the authors consider SC-LDPC ensembles over a block erasure channel (BLEC) where the channel erases complete spatial positions instead of individual bits. 
This block erasure model mimics block-fading channels frequently occurring in wireless communications. 
The authors give asymptotic lower and upper bounds for the bit and block erasure probabilities obtained from density evolution. 
In \cite{Iyengar-icc10}, the authors construct protograph-based codes that maximize the correctable burst lengths, while the authors in~\cite{Mori-corr15} apply interleaving (therein denoted band splitting) to a protograph-based SC-LDPC code to increase the correctable burst length. 
If windowed decoding is used, this approach results in an increased required window length and thus complexity. 
Recently, it has been shown that protograph-based LDPC codes can increase the diversity order of block fading channels and are thus good candidates for block erasure channels~\cite{ulHassan-isit14},\cite{ulHassan-itw15}; however, they require large syndrome former memories if the burst length becomes large.
 
In this work, we consider the $(d_v,d_c,w,L,M)$ code ensemble introduced in~\cite{Kudekar-it11} and derive tight lower bounds on the correctability of a long burst of erasures. 
First, we consider the case when a complete spatial position is erased and then generalize the expression to the case where the burst can occur at any position within a codeword. 
We show that estimating the capability of correcting long burst erasures reduces to the problem of finding small stopping sets in the code structure.
Also, we demonstrate that if we properly expurgate the ensemble, then a random code from the ensemble has very good average burst erasure capabilities.
We focus on the general $(d_v,d_c,w,L,M)$ code ensemble as the common protograph-based approach contains unavoidable small stopping sets in each spatial position, which are not recoverable if erased~\cite{Olmos-isit11}. 

The discussion is organized as follows: 
Section~\ref{sec:prelims} reviews essential technical background, 
Sections~\ref{sec:spbcAnalysis}~and~\ref{sec:RBC} provide finite-length analysis of the random ensemble on burst erasure channels, 
Section~\ref{sec:floorBEC} gives the error floor for the ensemble on the BEC, 
Section~\ref{sec:expurgate} details the effects of expurgating the ensemble, 
Section~\ref{sec:observe} compares ensembles and highlights important observations from this work, and 
Section~\ref{sec:conclude} concludes the work mentioning potential problems for future research.

\section{Preliminaries}
\label{sec:prelims}

\subsection{The Regular $(d_v,d_c,w,L,M)$ SC-LDPC Ensemble}
\label{sec:randomSCLDPC}

\begin{figure}[p]
\begin{center}
\large
\scalebox{0.7}{\includegraphics[width=\textwidth,height=0.4\textheight,keepaspectratio]{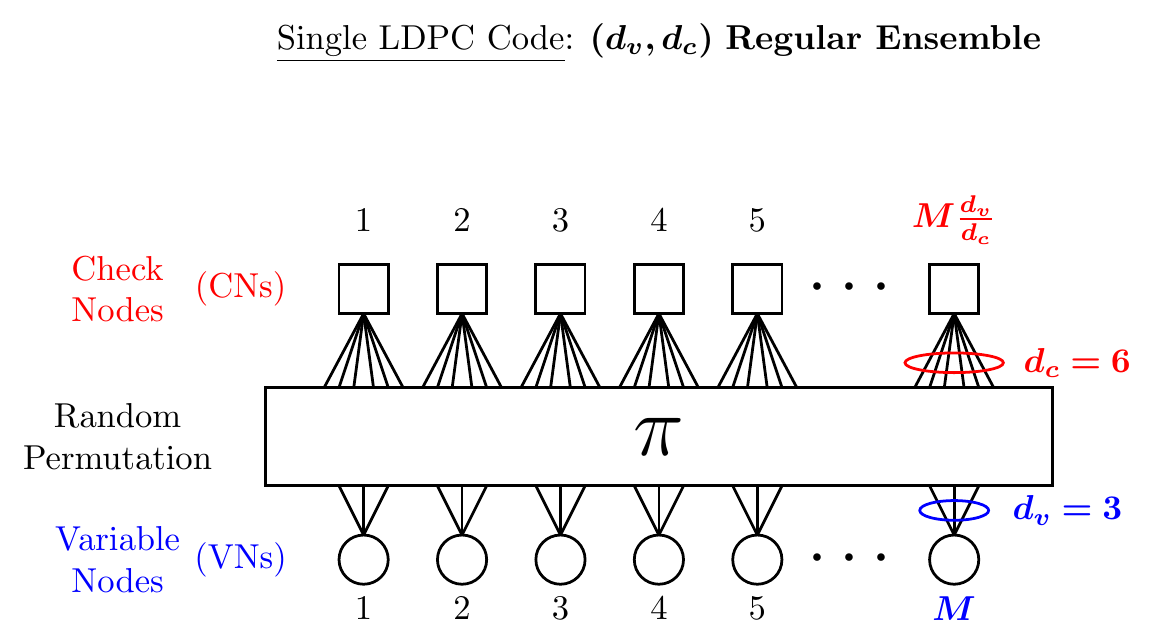}}

\vspace{2cm}

\hspace{-0.5cm}
\scalebox{1}{\includegraphics[width=\textwidth,height=0.4\textheight,keepaspectratio]{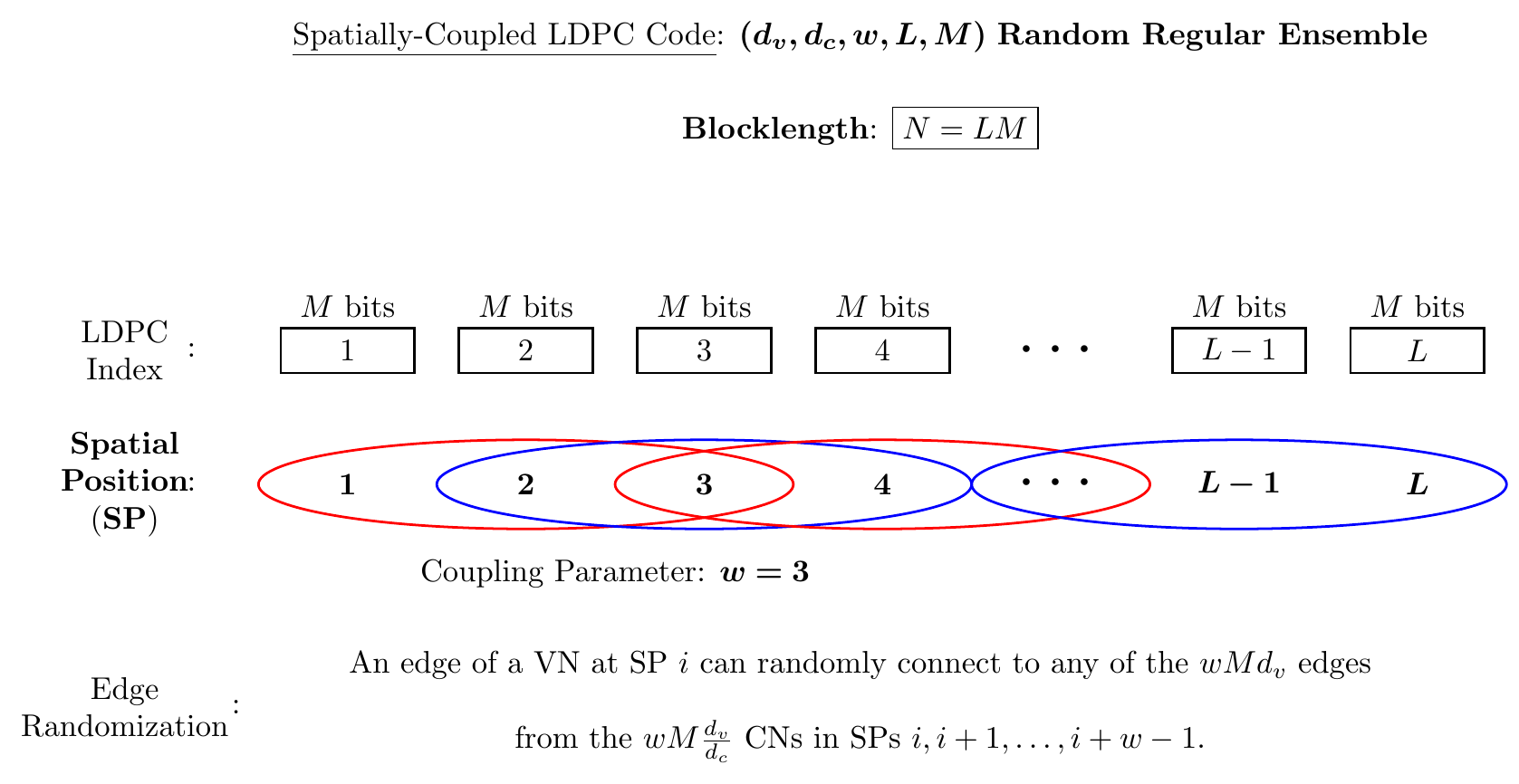}}
\normalsize
\caption{\label{fig:scldpc_fig}A depiction of a random regular $(d_v,d_c,w,L,M)$ SC-LDPC ensemble constructed from a regular $(d_v,d_c)$ LDPC ensemble.
}
\end{center}
\end{figure}

We now briefly review how to sample a code from a random regular ($d_v,d_c,w,L,M$) SC-LDPC ensemble~\cite{Kudekar-it11}. 
We first lay out a set of positions indexed from $z=1$ to $L$ on a \emph{spatial dimension}. 
At each spatial position (SP), $z$, there are $M$ variable nodes (VNs) and $M\frac{d_v}{d_c}$ check nodes (CNs), 
where $M\frac{d_v}{d_c} \in \mathbb{N}$ and $d_v$ and $d_c$ denote the variable and check node degrees, respectively.
Let $w>1$ denote the smoothing (coupling) parameter. 
Then, we additionally consider $w-1$ sets of $M\frac{d_v}{d_c}$ CNs in SPs $L+1,\dots,L+w-1$. 
Every CN is assigned with $d_c$ ``sockets'' and made to impose an even parity constraint on its $d_c$ neighboring VNs. 
Each VN in SP $z$ is connected to $d_v$ CNs in SPs $z,\dots,z+w-1$ as follows: 
each of the $d_v$ edges of this VN is allowed to randomly and uniformly connect to any of the $wMd_v$ sockets arising from the CNs in SPs $z,\dots,z+w-1$, such that multiple edges are avoided in the resultant bipartite graph.
This graph represents the code so that we have $N=LM$ code bits, over $L$ SPs. 
Because of additional check nodes in SPs $z>L$, the code rate $r = 1-\frac{d_v}{d_c}-\delta$, where $\delta=O(\frac{w}{L})$.
Fig.~\ref{fig:scldpc_fig} gives a pictorial depiction of this ensemble.
Throughout this work, we assume that $d_v \geq 3$ and $wM> 2(d_v+1)d_c$.

Let us define \emph{constellation} and \emph{type} for each VN as introduced in~\cite{Kudekar-it11}. 
Again, consider a VN in SP $i$. 
Assume that the $d_v$ edges are indexed by $k \in \{1,2,\ldots,d_v\}$. 
We define an associated $d_v$-tuple vector, called its \emph{constellation}, as $c=(c_1,c_2,\ldots,c_{d_v})$ where $c_k \in \{0,1,\ldots,w-1\}$ and the $k^{\text{th}}$ edge connects to a CN at position $i+c_k$. 
Clearly, there are $w^{d_v}$ constellations. 
We define an associated $type$ vector $t=(t_0,t_1,\ldots,t_{w-1})$ where $t_j$ indicates the number of edges of this VN that connect to a CN in position $i+j$. 
Hence $\sum_{j=0}^{w-1} t_j = d_v$ and there are $\binom{d_v+w-1}{w-1}$ types. 

Note that there exists a many-to-one mapping between constellations and types. 
In our random ensemble, all constellations are possible while more structured ensembles might have only few constellations that are allowed. 
We impose a uniform distribution on the set of all constellations and, owing to the many-to-one mapping, this introduces a distribution on the set of all types. 
Let $\tau(c)$ denote the type of a constellation. 
Then, the distribution on the types can be expressed using the probability of a type,
\begin{equation*}
p(t) = \frac{|{c: \tau(c)=t|}}{w^{d_v}}.
\end{equation*}

\subsection{Stopping Sets}
\label{sec:ss}

A subset $\mathcal{A}$ of the set of VNs in a code is a \emph{stopping set} if all the neighboring CNs of (the VNs in) $\mathcal{A}$ connect to $\mathcal{A}$ at least twice~\cite[Def. 3.137]{Richardson-MCT08}. 
In such a case, if all VNs in $\mathcal{A}$ have been erased by the channel, then the peeling decoder will fail as all the neighboring CNs are connected to at least two erased VNs. 
Therefore, such a set will stop the decoding process and hence is called a \emph{stopping set}. 
The cardinality of the set $\mathcal{A}$ is also its size.
A \emph{minimal stopping set} is one which does not contain a smaller size non-empty stopping set within itself.

\subsection{Binary Erasure Channel}
\label{sec:bec}

The symmetric binary erasure channel with parameter $\epsilon$ is denoted by BEC($\epsilon$) and its transition probabilities are defined, for $x\in \{0,1\}$ and $y\in \{0,1,?\}$, by
\begin{equation*}
W(y|x)=
\begin{cases}
1-\epsilon & \text{if } y=x\\
\epsilon   & \text{if } y=?
\end{cases}
\end{equation*}
Hence, approximately, a fraction $\epsilon$ of the transmission (in bits) is erased randomly.

\subsection{Single-Burst-Erasure Channel Models}
\label{sec:SPBC}

We introduce two channel models for computing the burst erasure recoverability. 
First, the \emph{Single Position Burst Channel} (SPBC) erases all $M$ VNs of exactly one SP in the transmitted codeword and leaves all other bits undisturbed. %

The second model is the more general \emph{Random Burst Channel} (RBC) whose burst pattern is denoted by RBC($\ell$,$s$,$b$) where $s \in \{1,\ldots,M\}$ is the starting bit index of the burst in SP $\ell\in\{1,\ldots,L\}$, indicating the offset from the first VN of the SP $\ell$, and $b$ is the length of the burst.
Note that in general $0 < b \leq (L-\ell)M-s$. As for the SPBC, all VNs in the random burst are erased while all other VNs are received correctly.
We sometimes omit the SP $\ell$ when referring to the RBC for the following reason: neglecting boundary effects in the limit of large enough $L$, all SPs are structured identically.
With some abuse of terminology, we will use the same notation to refer to the channel itself, rather than the burst introduced by it.

While multiple models exist for a correlated erasure channel, like the Gilbert-Elliott model used in~\cite{Iyengar-icc10}, we use this model because it is sufficient to describe the scenarios that we consider: for instance, the SPBC can be used to model a slotted-ALOHA multiple access scheme where each user transmits an SC-LDPC codeword over $L$ time slots, but one SP might be erased in the case of a collision. 
Additionally, long burst erasures might occur in block fading scenarios, or in optical communications which are subject to polarization dependent loss.


\section{Error Analysis on the SPBC}
\label{sec:spbcAnalysis}



Let $P_\B^\SPBC (d_v,d_c,w,L,M)$ denote the average block erasure (decoding error) probability of the $(d_v,d_c,w,L,M)$ ensemble on the SPBC under BP decoding i.e. the probability that the iterative decoder fails to recover the codeword. 
For large enough $M$, size-$2$ stopping sets (each of which also form a codeword) are the dominant structures in the graph that cause the BP decoder to fail~\cite{Olmos-isit11}. 
Hence, the number of size-$2$ stopping sets per SP, denoted $\mathbb{N}_2^{\rm\scriptscriptstyle SP}$, is a good starting point for analyzing the performance of the ensemble. 
We have
\begin{IEEEeqnarray}{rCl}
P_\B^\SPBC & = & \text{Prob [At least one stopping set in a SP]} \nonumber \\
			& \geq & \text{Prob [$\mathbb{N}_2^{\rm\scriptscriptstyle SP} \geq 1$]} \nonumber \\
			& \overset{(a)}{\geq} & \frac{\mathbb{E}[\mathbb{N}_2^{\rm\scriptscriptstyle SP}]^2}{\mathbb{E}[\mathbb{N}_2^{{\rm\scriptscriptstyle SP}^2}]}		
			\overset{(b)}{\geq} \mathbb{E}[\mathbb{N}_2^{\rm\scriptscriptstyle SP}] \biggr( 1-\frac{M^2}{(\frac{w}{d_c}M-3)^{d_v}} \biggr) \nonumber\\
			& = & \mathbb{E}[\mathbb{N}_2^{\rm\scriptscriptstyle SP}] \biggr( 1-O\biggr(\frac{1}{M^{d_v-2}}\biggr) \biggr)\approx \mathbb{E}[\mathbb{N}_2^{\rm\scriptscriptstyle SP}]\doteq \lambda_{\rm SP},
\label{eq:lower_bound}
\end{IEEEeqnarray}
where $(a)$ is the application of the second moment method and $(b)$ can be shown as follows:
Define $U_{ij}=1$ if VNs $i$ and $j$ form a stopping set, otherwise $U_{ij}=0$. 
Then $\mathbb{N}_2^{\rm\scriptscriptstyle SP}=\sum_{1\leq i<j\leq M} U_{ij}$
where the summation is over all $\binom{M}{2}$ pairs of VNs from a SP. 
We can see that $\lambda_{\rm SP}=\mathbb{E}[\mathbb{N}_2^{\rm\scriptscriptstyle SP}]=\binom{M}{2}p$, where $p=\mathbb{E}[U_{ij}]$ is the probability of forming a size-$2$ stopping set. 
\begin{IEEEeqnarray}{rCl}
\mathbb{E}\left[\mathbb{N}_2^{{\rm\scriptscriptstyle SP}^2}\right] &=
\mathbb{E}\left[ \left(\sum_{1\leq i<j\leq M} U_{ij}\right)^2\right]\nonumber\\
	&= \sum_{1\leq i<j\leq M} \mathbb{E}[U_{ij}^2]
	&+ \sum_{\underset{(i,j)\neq(k,l)}{i<j,k<l}} \mathbb{E}[U_{ij}U_{kl}],\nonumber
\end{IEEEeqnarray}
where in the last step, $\sum_{1\leq i<j\leq M} \mathbb{E}[U_{ij}^2]=\binom{M}{2} p$ as
$U_{ij}\in\{0,1\}$ and the second term is over the remaining $\binom{M}{2} \biggr( \binom{M}{2}-1 \biggr)$
combinations.
Using some combinatorial arguments, we can show that
$\mathbb{E}[U_{ij}U_{kl}]=\mathbb{P}(U_{ij}=1)\mathbb{P}(U_{kl}=1|U_{ij}=1)\leq 2 p/\binom{wM\frac{d_v}{d_c}-2d_v}{d_v}$. 
As a result, we have
\begin{IEEEeqnarray}{rCl}
\mathbb{E}\left[\mathbb{N}_2^{{\rm\scriptscriptstyle SP}^2}\right] & < &  
\mathbb{E}[\mathbb{N}_2^{\rm\scriptscriptstyle SP}] \biggr( 1+\frac{2\binom{M}{2}}{\binom{wM\frac{d_v}{d_c}-2d_v}{d_v}} \biggr) \nonumber\\
&<& \mathbb{E}[\mathbb{N}_2^{\rm\scriptscriptstyle SP}] \biggr( 1+\frac{M^2}{(\frac{w}{d_c}M-3)^{d_v}} \biggr),\nonumber
\end{IEEEeqnarray}
which eventually implies \eqref{eq:lower_bound}. 
Note that following standard arguments~\cite{Olmos-isit11}, \cite[Appendix C]{Richardson-MCT08}, we can also approximate the bound on $P_\B^\SPBC$ by a Poisson distribution with mean $\lambda_{\rm SP}$, for a large $M$, so that
\begin{equation}
\label{eq:SPBCLowerBound}
P_\B^\SPBC \approx 1-e^{-\lambda_{\rm SP}} \approx \lambda_{\rm SP}.
\end{equation}
Both~\eqref{eq:lower_bound}~and~\eqref{eq:SPBCLowerBound} are very tight when $w \geq d_v$ (which is a prerequisite for constructing capacity-achieving codes~\cite{Kudekar-it11}) as otherwise, we have observed that the contribution of larger stopping sets becomes non-negligible.
We use this observation later in Section~\ref{sec:floorBEC} to characterize the number of size-$2$ stopping sets in the code, $\mathbb{N}_2^{H}$ (instead of one SP). 

\begin{figure}
\begin{center}
\large
\scalebox{1}{\hspace*{-4.5mm} 

\makeatletter
\@ifundefined{vecnot}{%
\newcommand{\vecnot}[1]{\underline{#1}}
}{}
\makeatother

\scalebox{1}{%
\begin{tikzpicture}

  \draw[color=black] (0,0) circle (.25);
  \draw[color=black] (2.5,0) circle (.25);  
  
  \node[draw=none,align=center] at (0,-0.5) {$v_1$};
  \node[draw=none,align=center] at (2.5,-0.5) {$v_2$};

  \fill[black] (-2,2) rectangle (-1.75,2.505);
  \draw[color=black] (-1.75,2) rectangle (-1.5,2.5);
  \draw[color=black] (-1.5,2) rectangle (-1.25,2.5);
  \draw[color=black] (-1.25,2) rectangle (-1,2.5);
  \draw[color=black] (-1,2) rectangle (-0.75,2.5);  
  \draw[color=black] (-0.75,2) rectangle (-0.5,2.5);  
  
  \node[draw=none,align=center] at (-1.25,3) {$c_1$};

  \fill[black] (0.5,2) rectangle (0.75,2.505);
  \draw[color=black] (0.75,2) rectangle (1,2.5);
  \draw[color=black] (1,2) rectangle (1.25,2.5);
  \draw[color=black] (1.25,2) rectangle (1.5,2.5);
  \draw[color=black] (1.5,2) rectangle (1.75,2.5);
  \draw[color=black] (1.75,2) rectangle (2,2.5);  
  
  \node[draw=none,align=center] at (1.25,3) {$c_2$};

  \fill[black] (3,2) rectangle (3.25,2.505);
  \draw[color=black] (3.25,2) rectangle (3.5,2.5);
  \draw[color=black] (3.5,2) rectangle (3.75,2.5);
  \draw[color=black] (3.75,2) rectangle (4,2.5);
  \draw[color=black] (4,2) rectangle (4.25,2.5);
  \draw[color=black] (4.25,2) rectangle (4.5,2.5);  
  
  \node[draw=none,align=center] at (3.75,3) {$c_3$};
  
  \draw[color=black] (0,0.25) -- (-1.875,2);
  \draw[color=black] (0,0.25) -- (0.625,2);
  \draw[color=black] (0,0.25) -- (3.125,2);

  \draw[dashed,color=black] (2.5,0.25) -- (-1.625,2);
  \draw[dashed,color=black] (2.5,0.25) -- (0.875,2);
  \draw[dashed,color=black] (2.5,0.25) -- (3.375,2);

\node at (-0.75,2.75) [above right] {\footnotesize Socket};
\draw [densely dotted,thin,color=black] (-0.625,2.5) -- (-0.5,2.75);
\end{tikzpicture}}}
\normalsize
\caption{\label{fig:size2ss}A size-$2$ stopping set from a $(3,6)$ random ensemble. CNs $\{c_1,c_2,c_3\}$ and VNs $\{v_1,v_2\}$ have been labeled for convenience. CNs have been expanded to show all their $d_c=6$ sockets. The solid edges indicate definite connections and the dashed edges complete one configuration to form a stopping set. Multiple edges are not allowed in the ensemble.
}
\end{center}
\vspace{-4mm}
\end{figure}
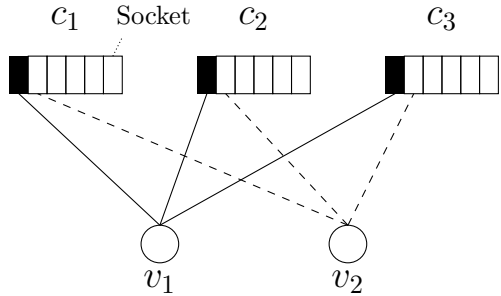

\subsection{Calculation of $p$}
\label{sec:wt2s}

We now calculate the probability $p$ of finding a size-$2$ stopping set within an SP of a code uniformly sampled from an ensemble. 
As example, we randomly choose two VNs $v_1$ and $v_2$ from an SP of the $(d_v=3,d_v=6,w,L,M)$ ensemble. 
First, we connect the $d_v=3$ edges of $v_1$ to randomly chosen empty sockets of $d_v$ distinct CNs 
as described in Section~\ref{sec:randomSCLDPC}. Let $c_1,c_2,c_3$ denote the CNs adjacent to $v_1$. 
A stopping set (and in this case, also a low-weight codeword) is formed if and only if the edges of $v_2$ are connected 
to the same CNs, i.e. $c_1,c_2,c_3$. This situation is shown in Fig.~\ref{fig:size2ss}: once we have assigned $d_v$ CNs to $v_1$, we have $d_c-1=5$ free distinct sockets each for CNs $c_1,c_2,c_3$. 
Thus, the first edge of $v_2$ has $d_v(d_c-1)=15$ ways to attach to these sockets, the second edge has $(d_v-1)(d_c-1)=10$ ways and the last edge has $(d_v-2)(d_c-1)=5$ ways. 
In general, the edges of $v_2$ can be connected to any of the $(wMd_v-d_v)$ possible sockets.

By a counting argument, we can compute $p = \frac{T_{ss}}{T}$ where $T_{ss}$ is the total number of combinations 
by which the edges of $v_2$ can form a stopping set with $v_1$ and $T$ is the total number of combinations 
by which the edges of $v_2$ can be fit to the possible CN sockets without forming multiple edges. 
We have 
\begin{IEEEeqnarray*}{rl}
T_{ss} = & \hspace{3mm} 15 \times 10 \times 5 ,\\
T = & \hspace{3mm} 15 \times 10 \times 5 \\
    & + \left[15 \times 10 \times (3wM-18)\right] \times 3 \\
    & + \left[15 \times (3wM-18) \times (3wM-24)\right] \times 3 \\
    & + (3wM-18) \times (3wM-24) \times (3wM-30)
\end{IEEEeqnarray*}
that give
\begin{equation*}
p \approx \frac{15 \times 10 \times 5}{(3wM-18)(3wM-24)(3wM-30)}.
\end{equation*}
Hence, for a general $(d_v,d_c,w,M)$ ensemble we can calculate $p=\frac{T_{ss}}{T}$ with
\begin{IEEEeqnarray*}{rCl}
\label{eq:p2}
T_{ss} & = & \prod_{i=0}^{d_v-1} (d_v-i) (d_c-1) = d_v! (d_c-1)^{d_v} , \\
T & = & \sum_{i=0}^{d_v} \frac{ (d_c-1)^i d_v!}{(d_v-i)!}\binom{d_v}{i} \left[ \prod_{k = 0}^{d_v-1-i} (wMd_v - (d_v+k) d_c) \right]. 
\end{IEEEeqnarray*}
For large $M$, $T$ can be well approximated by the dominating summand ($i=0$) leading to
 \begin{equation}
\label{eq:p2approx}
p \approx \prod_{i=0}^{d_v-1} \frac{(d_v-i) (d_c-1)}{(wMd_v - (d_v+i)d_c)} \approx \frac{d_v!(d_c-1)^{d_v}}{((wM-d_c)d_v)^{d_v}}.
\end{equation}
We observe that $\lambda_{SP} = \binom{M}{2}p \sim O(M^{2-d_v})$.

\subsubsection{Poisson Ensemble}
\label{sec:poisson}

We make note of a significant change to~(\ref{eq:p2}) when this random ensemble is slightly relaxed. 
Retaining the construction of the random ensemble, if there is no limit placed on the check degree then we get the so-called Poisson ensemble $\mathcal{C}_P$. 
For this ensemble, sockets are not distinct and therefore the calculation of $p$ is much simpler. 
Let $v_1$ and $v_2$ belong to SP $i$. 
Assume that the edges of $v_2$ are assigned to CNs sequentially. 
The first edge can connect to any of the $(wM\frac{d_v}{d_c})$ CNs from SPs $i,i+1,\ldots,i+w-1$. 
The second edge has one CN less to choose from, the third edge has two CNs less to choose from and so on. 
But, there is exactly one way in which the edges can connect exactly to the same CNs as $v_1$. 
Hence the probability of $v_2$ forming a stopping set with $v_1$ is
\begin{equation}
\label{eq:p2poisson}
p' = \left[ \binom{wM\frac{d_v}{d_c}}{d_v} \right]^{-1}.
\end{equation}
Let us compare this with~(\ref{eq:p2approx}). First, we rewrite $p$ as
\begin{equation*}
p \cong \left( \frac{d_c-1}{d_c} \right)^{d_v} \times \frac{d_v!}{\prod_{i=0}^{d_v-1} \left( wM\frac{d_v}{d_c} - (d_v+i) \right)}.
\end{equation*}
Then, we can compare the two ensembles as below.
\begin{IEEEeqnarray*}{ll}
& \hspace{4mm} p' - p \\
& = \frac{d_v!}{\prod_{i=0}^{d_v-1} \left( wM\frac{d_v}{d_c} - i \right)} - \left( \frac{d_c-1}{d_c} \right)^{d_v}  \frac{d_v!}{\prod_{i=0}^{d_v-1} \left( wM\frac{d_v}{d_c} - (d_v+i) \right)} \\
& = \frac{d_v!}{\prod_{i=0}^{2d_v-1} \left( wM\frac{d_v}{d_c} - i \right)} \times  \prod_{i=0}^{d_v-1} \left[ \left( wM\frac{d_v}{d_c} - (d_v + i) \right) - \left( \frac{d_c-1}{d_c} \right) \left( wM\frac{d_v}{d_c} - i \right) \right].
\end{IEEEeqnarray*}
Now, analyze the individual product terms as
\begin{equation*}
a' = \left( wM\frac{d_v}{d_c} - i \right) - d_v \hspace{2mm} \text{and} \hspace{2mm} a = \left( \frac{d_c-1}{d_c} \right)  \left( wM\frac{d_v}{d_c} - i \right),
\end{equation*}
where $i=0,1,\ldots,d_v-1$. 
We immediately see that
\begin{equation*}
wM\frac{d_v}{d_c} - i \geq d_v d_c \Rightarrow a' \geq a.
\end{equation*}
Evaluating this condition for the worst case, $i=d_v-1$, we can conclude that
\begin{equation*}
M \geq \frac{(d_v d_c + d_v - 1)d_c}{w d_v} \Rightarrow p'-p \geq 0.
\end{equation*}
Hence, the Poisson ensemble performs worse than the random ensemble under this condition.

\subsection{Simulations}
\label{sec:SPBCSim}

\begin{figure}[p]
\begin{center}
\large
\scalebox{0.9}{
%
%
\begin{tikzpicture}

\begin{axis}[%
width=4.34in,
height=3.354in,
at={(0.728in,0.518in)},
scale only axis,
xmin=0,
xmax=1000,
xlabel={\# VNs per Spatial Position $(M)$},
xmajorgrids,
ymode=log,
ymin=0.0001,
ymax=0.01,
yminorticks=true,
ylabel={Output Block Erasure Probability $(P_{B})$},
ymajorgrids,
yminorgrids,
axis background/.style={fill=white},
title style={font=\bfseries},
legend style={legend cell align=left,align=left,draw=white!15!black}
]
\addplot [color=blue,mark=x,mark options={solid}]
  table[row sep=crcr]{%
80	0.00836505081768372\\
100	0.00615646028159649\\
120	0.00493201221166224\\
140	0.00427145981410607\\
160	0.00361059058430187\\
180	0.00300954628080271\\
200	0.00288426317170884\\
220	0.00247873861949131\\
240	0.00228241717108086\\
260	0.00196469444089708\\
280	0.00204769071679414\\
300	0.00189782490287881\\
320	0.00171404868583887\\
340	0.00158365164731444\\
360	0.00143705415394874\\
380	0.00139981914336668\\
400	0.00139019992465116\\
420	0.00141781828957237\\
440	0.00120793662681281\\
460	0.00114362696718134\\
480	0.00114350403942802\\
500	0.00106487880614307\\
520	0.00103376268943701\\
540	0.000965748754666981\\
560	0.000908295737458934\\
580	0.000899757155543719\\
600	0.000848531954864888\\
620	0.000826675153864913\\
640	0.000849685531384834\\
660	0.00077874766473044\\
680	0.000742627013354662\\
700	0.000747285857764599\\
720	0.000778932675291009\\
740	0.000682861407814393\\
760	0.000646324257001792\\
780	0.000670168969702331\\
800	0.000641344257563854\\
820	0.000616826661222143\\
840	0.000604991787236488\\
860	0.000606130035501036\\
880	0.000607551869740879\\
900	0.000573297850018403\\
920	0.000570324748615109\\
940	0.000535952784703479\\
960	0.000529523868023589\\
980	0.000540780530170416\\
1000	0.000538920574424662\\
};
\addlegendentry{$w = 3$ Simulation, $1000$ failures};

\addplot [color=blue,dashdotted,thick]
  table[row sep=crcr]{%
80	0.00700530843085789\\
100	0.00550776963784583\\
120	0.00453726993839598\\
140	0.00385735812556298\\
160	0.00335457075347456\\
180	0.00296769357169491\\
200	0.00266079849144518\\
220	0.0024114121242732\\
240	0.00220475783501195\\
260	0.0020307211422359\\
280	0.00188214572343171\\
300	0.00175382578065186\\
320	0.00164188396642928\\
340	0.00154337308613295\\
360	0.00145601298236253\\
380	0.00137801194386034\\
400	0.00130794257255629\\
420	0.00124465367331927\\
440	0.00118720653394799\\
460	0.00113482806637288\\
480	0.00108687582387768\\
500	0.0010428115252884\\
520	0.00100218076677372\\
540	0.000964597297468006\\
560	0.000929730704511522\\
580	0.000897296675168135\\
600	0.000867049228077676\\
620	0.000838774464250625\\
640	0.000812285501924248\\
660	0.000787418341634649\\
680	0.000764028468123623\\
700	0.000741988040316\\
720	0.000721183553966775\\
740	0.00070151388675066\\
760	0.00068288865472943\\
780	0.000665226823831189\\
800	0.000648455531344849\\
820	0.000632509081281762\\
840	0.000617328084396629\\
860	0.000602858719138455\\
880	0.00058905209414406\\
900	0.000575863696364198\\
920	0.000563252911695455\\
940	0.000551182607243184\\
960	0.000539618766165151\\
980	0.000528530167532959\\
1000	0.000517888104867525\\
};
\addlegendentry{$w = 3$ Lower Bound};

\addplot [color=red,mark=o,mark options={solid}]
  table[row sep=crcr]{%
80	0.00311523568315561\\
100	0.00237245109785175\\
120	0.00199762282883369\\
140	0.00159312534550906\\
160	0.00147793225454132\\
180	0.00137683583848614\\
200	0.00110015116076949\\
220	0.00105401177963565\\
240	0.000952797460985326\\
260	0.000883717765289864\\
280	0.000834856250276546\\
300	0.000762065594033941\\
320	0.000708281510736131\\
340	0.000667831364566471\\
360	0.000633750911809124\\
380	0.000592632628219032\\
400	0.000538802683452885\\
420	0.000512744784360053\\
440	0.00048843961128022\\
460	0.000492352291850683\\
480	0.000464089884928913\\
500	0.00043193241812613\\
520	0.000428487567433231\\
540	0.00040168999012646\\
560	0.000384712450525979\\
580	0.00037542314881661\\
600	0.000347196061963387\\
620	0.000336515083447328\\
640	0.00033169773181774\\
660	0.00031478593611986\\
680	0.000333484068132129\\
700	0.000300571988494104\\
720	0.000308896528929704\\
740	0.000322154569762572\\
760	0.000289062872620109\\
780	0.000282009963976047\\
800	0.000282123975995764\\
820	0.000267267341106893\\
840	0.000256543330040172\\
860	0.000237791431233334\\
880	0.000244813503521153\\
900	0.000260872650320039\\
920	0.000233408397473961\\
940	0.000212023019763302\\
960	0.000226141812626131\\
980	0.000213214206291994\\
1000	0.000216418081644154\\
};
\addlegendentry{$w = 4$ Simulation, $1000$ failures};

\addplot [color=red,dashed,thick]
  table[row sep=crcr]{%
80	0.00288609872263434\\
100	0.00228013313927089\\
120	0.00188437790639862\\
140	0.00160564640911054\\
160	0.00139872976718547\\
180	0.0012390456039254\\
200	0.00111208044257172\\
220	0.00100871388159918\\
240	0.000922926741573571\\
260	0.000850586306754497\\
280	0.000788761034113628\\
300	0.000735313742538102\\
320	0.000688649637771155\\
340	0.00064755452385723\\
360	0.000611087663467513\\
380	0.000578508899918395\\
400	0.000549227905697114\\
420	0.000522768097769188\\
440	0.000498740502035022\\
460	0.000476824507292917\\
480	0.000456753479126104\\
500	0.000438303859831901\\
520	0.000421286807142263\\
540	0.000405541707656387\\
560	0.000390931092264846\\
580	0.00037733661234074\\
600	0.000364655827188254\\
620	0.000352799618137944\\
640	0.000341690091170577\\
660	0.00033125886368468\\
680	0.000321445655756092\\
700	0.000312197124570091\\
720	0.000303465894423982\\
740	0.000295209745059144\\
760	0.000287390928969344\\
780	0.000279975594393189\\
800	0.000272933295382582\\
820	0.000266236573994494\\
840	0.000259860602513262\\
860	0.000253782875879294\\
880	0.000247982946289582\\
900	0.000242442193377101\\
920	0.000237143624524094\\
940	0.000232071700800529\\
960	0.000227212184769288\\
980	0.000222552007020593\\
1000	0.000218079148800121\\
};
\addlegendentry{$w = 4$ Lower Bound};

\end{axis}
\end{tikzpicture}%
}
\normalsize
\caption{\label{fig:SPBC_w_3_w_4}Monte Carlo simulations on the SPBC with a $(3,6)$ random ensemble for $w=3$ and $w=4$, along with their respective theoretical lower bound~(\ref{eq:lower_bound}). The bound becomes tight very quickly with $M$.
}
\end{center}
\end{figure}
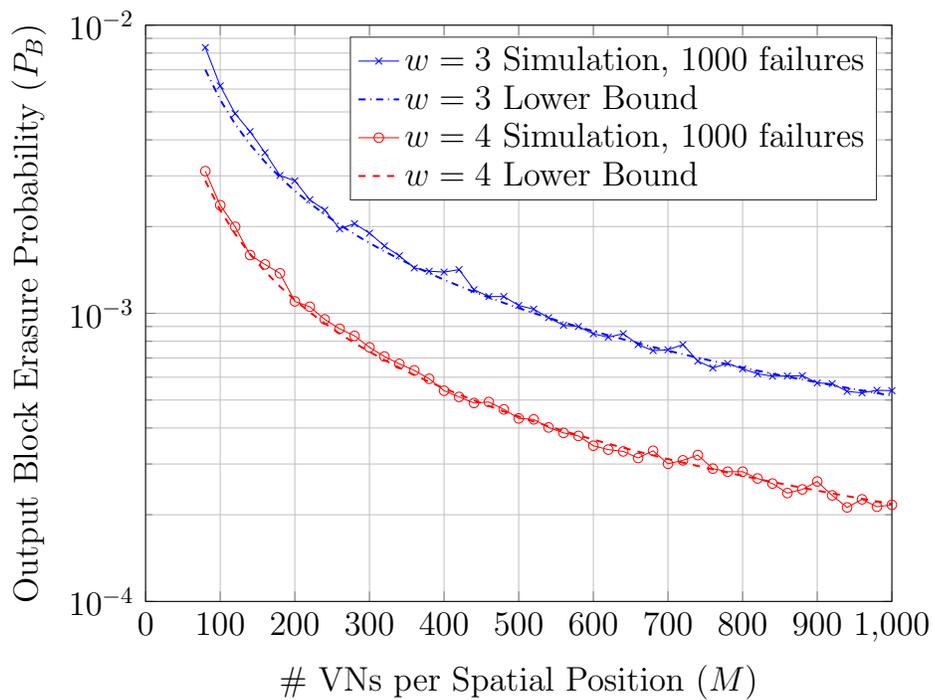

We performed Monte-Carlo simulations where we randomly selected a spatial position from the middle of the graph (to avoid boundary effects) to be erased, for each transmitted codeword. 
At the receiver we performed BP decoding and averaged over the ensemble. 
We counted $1000$ decoding failures for each $M$ to assess the average block erasure probability $P_\B^\SPBC$. 
The simulation results for a $(3,6)$ random ensemble with $w=3$ and $w=4$ are shown in Fig.~\ref{fig:SPBC_w_3_w_4} along with their respective lower bounds calculated using~\eqref{eq:lower_bound} and \eqref{eq:p2approx}. 
We observe that the bound indeed becomes very good for large $M$, since large-size stopping sets (larger than $2$) vanish. 
The simulation curve is slightly unstable because counting $1000$ failures is not enough to keep the sample variance small as $P_\B^\SPBC$ decreases by $O(M^{2-d_v})$.

\begin{figure}[p]
\begin{center}
\large
\scalebox{0.9}{
%
%
\begin{tikzpicture}

\begin{axis}[%
width=4.34in,
height=3.354in,
at={(0.728in,0.518in)},
scale only axis,
xmin=0,
xmax=1000,
xlabel={\# VNs per Spatial Position $(M)$},
xmajorgrids,
ymode=log,
ymin=0.0001,
ymax=0.1,
yminorticks=true,
ylabel={Output Block Erasure Probability $(P_B)$},
ymajorgrids,
yminorgrids,
axis background/.style={fill=white},
title style={font=\bfseries},
legend style={legend cell align=left,align=left,draw=white!15!black}
]
\addplot [color=blue,solid,mark=x,mark options={solid}]
  table[row sep=crcr]{%
80	0.0344839477223353\\
100	0.0184328399476507\\
120	0.0119195194049776\\
140	0.00904903673004009\\
160	0.00692727058611636\\
180	0.00613515752016933\\
200	0.00529714325064493\\
220	0.00455394143631313\\
240	0.00430835778326375\\
260	0.0037654004879959\\
280	0.00343154412622592\\
300	0.00323467572375869\\
320	0.00302796324052626\\
340	0.00277297129420116\\
360	0.00269387117369273\\
380	0.00258325841883919\\
400	0.002409435348826\\
420	0.00234636032586252\\
440	0.00203683825671087\\
460	0.00208107883126613\\
480	0.00201288244766506\\
500	0.00194144221976738\\
520	0.00182539378307385\\
540	0.00167543196824721\\
560	0.00169660035221423\\
580	0.00165022764890417\\
600	0.00152820142917398\\
620	0.00151661527868564\\
640	0.00135275895188236\\
660	0.00134368327772724\\
680	0.00135012711446783\\
700	0.00122181875051927\\
720	0.00123332391896075\\
740	0.00122977935298849\\
760	0.00119522436154103\\
780	0.00122025476478979\\
800	0.00113293008575148\\
820	0.00111914407860868\\
840	0.00111855071620802\\
860	0.00114501500542165\\
880	0.000975362347111952\\
900	0.00098997158781543\\
920	0.000975019037246702\\
940	0.000996902623548634\\
960	0.000954041893887644\\
980	0.000862261487694235\\
1000	0.000901337765511572\\
};
\addlegendentry{Simulation for $\mathcal{C}_{P}$};

\addplot [color=red,dashdotted,thick]
  table[row sep=crcr]{%
80	0.0109122469464132\\
100	0.00876139332923276\\
120	0.00731876545204768\\
140	0.00628402726747068\\
160	0.00550562153978618\\
180	0.00489879831817419\\
200	0.00441245859514272\\
220	0.00401396100342732\\
240	0.00368147828565757\\
260	0.00339986151250626\\
280	0.00315826745365733\\
300	0.00294873034905796\\
320	0.00276526680892053\\
340	0.00260329526911096\\
360	0.00245924817418963\\
380	0.00233030615743435\\
400	0.00221421167541191\\
420	0.00210913570461424\\
440	0.00201358067293567\\
460	0.00192630863305565\\
480	0.00184628733792358\\
500	0.00177264922039777\\
520	0.00170465981259937\\
540	0.00164169316425944\\
560	0.00158321251486837\\
580	0.00152875495465632\\
600	0.00147791914593809\\
620	0.00143035541539649\\
640	0.00138575769984095\\
660	0.00134385695314843\\
680	0.0013044157141977\\
700	0.0012672236040816\\
720	0.00123209357227005\\
740	0.00119885875030801\\
760	0.00116736980134802\\
780	0.00113749267668872\\
800	0.00110910670822117\\
820	0.00108210297953526\\
840	0.0010563829293232\\
860	0.00103185714932774\\
880	0.00100844434593672\\
900	0.000986070440007869\\
920	0.000964667783923234\\
940	0.000944174478438975\\
960	0.000924533774800951\\
980	0.000905693549961284\\
1000	0.000887605844678618\\
};
\addlegendentry{Bound for $\mathcal{C}_{P}$};

\addplot [color=red,dashed,thick,mark=o,mark options={solid}]
  table[row sep=crcr]{%
80	0.00700530843085789\\
100	0.00550776963784583\\
120	0.00453726993839598\\
140	0.00385735812556298\\
160	0.00335457075347456\\
180	0.00296769357169491\\
200	0.00266079849144518\\
220	0.0024114121242732\\
240	0.00220475783501195\\
260	0.0020307211422359\\
280	0.00188214572343171\\
300	0.00175382578065186\\
320	0.00164188396642928\\
340	0.00154337308613295\\
360	0.00145601298236253\\
380	0.00137801194386034\\
400	0.00130794257255629\\
420	0.00124465367331927\\
440	0.00118720653394799\\
460	0.00113482806637288\\
480	0.00108687582387768\\
500	0.0010428115252884\\
520	0.00100218076677372\\
540	0.000964597297468006\\
560	0.000929730704511522\\
580	0.000897296675168135\\
600	0.000867049228077676\\
620	0.000838774464250625\\
640	0.000812285501924248\\
660	0.000787418341634649\\
680	0.000764028468123623\\
700	0.000741988040316\\
720	0.000721183553966775\\
740	0.00070151388675066\\
760	0.00068288865472943\\
780	0.000665226823831189\\
800	0.000648455531344849\\
820	0.000632509081281762\\
840	0.000617328084396629\\
860	0.000602858719138455\\
880	0.00058905209414406\\
900	0.000575863696364198\\
920	0.000563252911695455\\
940	0.000551182607243184\\
960	0.000539618766165151\\
980	0.000528530167532959\\
1000	0.000517888104867525\\
};
\addlegendentry{Bound for $\mathcal{C}_{R}$};
\addplot [color=blue,solid,mark=triangle,mark options={solid}]
  table[row sep=crcr]{%
80	0.0238731856378915\\
100	0.0170366458251699\\
120	0.0141232963773745\\
140	0.0122012225625008\\
160	0.0103924176920519\\
180	0.00994154372290929\\
200	0.00852769368524283\\
220	0.00748687924412467\\
240	0.00704190638489652\\
260	0.00645694509013895\\
280	0.0055890276209745\\
300	0.00557640511467877\\
320	0.00518392568324141\\
340	0.00482769940812405\\
360	0.00432249252208794\\
380	0.00409357922099187\\
400	0.00395849909548296\\
420	0.00381855742537584\\
440	0.00367035782318418\\
460	0.00335217254302513\\
480	0.00312865661742161\\
500	0.00317616604996109\\
520	0.00294628914881706\\
540	0.00291146242757737\\
560	0.00282735865327253\\
580	0.00261321082600981\\
600	0.00263126734991909\\
620	0.00264282486263918\\
640	0.00248443501463332\\
660	0.00236848400444328\\
680	0.00223472176596653\\
700	0.0022816568479367\\
720	0.00209740341457276\\
740	0.00208167233228486\\
760	0.00200922636747947\\
780	0.00199502838925398\\
800	0.00201014318249889\\
820	0.00194341551389736\\
840	0.00189370268110426\\
860	0.00188661447033299\\
880	0.00170493100144237\\
900	0.00173931495341245\\
920	0.00168803151892452\\
940	0.00160738367766171\\
960	0.00167535056710617\\
980	0.00157398021821662\\
1000	0.00157442627906391\\
};
\addlegendentry{Simulation for $C_{R}$ with Multiple Edges};

\end{axis}
\end{tikzpicture}%
}
\normalsize
\caption{\label{fig:SPBC_poisson_w_3}Monte Carlo simulations on the SPBC with the Poisson ensemble $\mathcal{C}_P$ for $d_v=3,w=3$, along with the theoretical lower bound calculated using~(\ref{eq:p2poisson})~as $p$ in~(\ref{eq:SPBCLowerBound}). The bound for a $(3,6,w=3)$ random ensemble $\mathcal{C}_R$ and the simulation results for $\mathcal{C}_R$ with multiple edges, for the same scenario, is also plotted for comparison.
}
\end{center}
\end{figure}
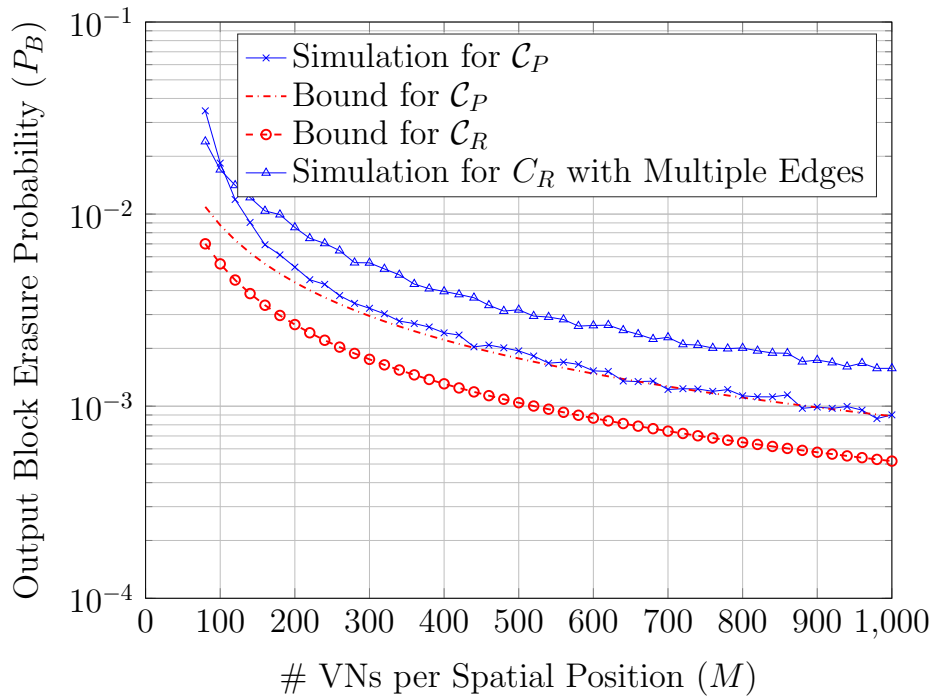

We performed the above experiment again for a $d_v=3,w=3$ Poisson ensemble and plotted the results in Fig.~\ref{fig:SPBC_poisson_w_3} along with the lower bound calculated using~(\ref{eq:p2poisson})~as $p$ in~(\ref{eq:SPBCLowerBound}). 
For comparison, we have also plotted the lower bound from the previous simulation for a $(3,6,w=3)$ random ensemble. 
As noted in Section~\ref{sec:poisson}, the Poisson ensemble performs worse than the equivalent random ensemble since our condition for this case is $M \geq 14$. 
Moreover, we performed simulations for the same scenario with the random ensemble by allowing multiple edges in the graph and plotted the results in the same figure. 
We see that multiple edges degrade the performance even more. 
This gives a complete picture of the relative performances of these ensembles on the SPBC.

\section{Error Analysis on the RBC}
\label{sec:RBC}

We now generalize our results to the RBC, where a burst can span multiple spatial positions and can be of arbitrary length. Besides the stopping sets within a single spatial position, we first have to derive an expression for stopping sets that span multiple SPs.

\subsection{Size-$2$ Stopping Sets across Coupled SPs}
\label{sec:wt2sCoupled}

The results from Section~\ref{sec:spbcAnalysis} can be extended when the channel is a RBC, i.e., the burst occurs at arbitrary location and is of arbitrary length. This means that size-$2$ stopping sets formed across coupled SPs will also contribute to decoding failures. Hence, we will now calculate the probability that two VNs chosen each from two coupled spatial positions form a stopping set.

Let us first consider two VNs chosen from two adjacent SPs: w.l.o.g, call them $v_1$ and $v_2$ chosen from SPs $1$ and $2$, respectively. 
We immediately notice that the check positions adjacent to $v_1$ are $1,2,\ldots,w$ and to $v_2$ are $2,3,\ldots,w+1$. 
Hence, to form a stopping set, $v_1$ should not have any edge connected to check position $1$. 
This restricts the number of favorable constellations~\cite{Kudekar-it11} for $v_1$ to be $(w-1)^{d_v}$. 
Using the same ideas as in Section~\ref{sec:wt2s} and restricting the constellations for $v_1$, we have
\begin{equation*}
p_{(1,2)} = \frac{(w-1)^{d_v}}{w^{d_v}}p ,
\end{equation*}
where $p$ can be approximated by~(\ref{eq:p2}). 
This idea can now be extended to VNs chosen from positions $(1,3), (1,4), \ldots, (1,w)$ by restricting the number of favorable constellations for $v_1$. 
Hereafter, we will refer to these as size-$2$ $(1,i)$-stopping sets. 
Hence, a $(d_v,d_c,w,L,M)$ ensemble can be completely characterized on erasure channels, for large enough $M$, by the vector
\begin{IEEEeqnarray}{rCl}
\label{eq:pvec}
\vecnot{p}(d_v,d_c,w,L,M) & = & (p_{(1,1)}, p_{(1,2)}, \ldots, p_{(1,w)}) \\
          \text{with\ }      p_{(1,i)} & = & \left( \frac{w-(i-1)}{w} \right)^{d_v} p .\nonumber
\end{IEEEeqnarray}
The average number of size-$2$ stopping sets of each type, $\lambda_{(1,i)}$, can be calculated as
\begin{equation}
\label{eq:lambdas}
\lambda_{(1,1)} = \binom{M}{2} p_{(1,1)} = \lambda_{SP} \hspace{2.5mm} ; \hspace{2.5mm} \lambda_{(1,i)} = M^2 p_{(1,i)} ,
\end{equation}
where $i = 2,3,\ldots,w$. Again, we see that $\lambda_{(1,i)} \sim O(M^{2-d_v})$.

\subsection{Performance on the RBC}
\label{sec:RBCBound}

Now let us see the effect of RBC($s,b$) on the ensemble in terms of the average block erasure probability, $P_\B^{\RBC}$. 
For keeping the expressions simple, let us assume in the example that $w=3$ and $0 < b \leq 2M$. 
This means that the burst can span a maximum of $3$ SPs. 
Applying the same argument as in Section~\ref{sec:spbcAnalysis} and assuming all values for $s$ are equally likely,
\begin{IEEEeqnarray}{rCl}
\label{eq:RBCLowerBound}
\hspace{-5mm}P_\B^{\RBC} & \approx & \sum_{s=1}^{M} \frac{1 - P_{(1,1)}P_{(2,2)}P_{(3,3)} P_{(1,2)}P_{(2,3)} P_{(1,3)}}{M} \hspace{2.5mm} ;\\
      P_{(k,k)} &   =  & 1-\binom{m_k}{2} p_{(1,1)} \hspace{5.2mm} \text{for } k = 1,2,3 ,\nonumber \\
      P_{(k,k+1)} &   =  & 1-m_k m_{k+1} p_{(1,2)} \hspace{2mm} \text{for } k = 1,2 ,\nonumber \\
      P_{(k,k+2)} &   =  & 1-m_k m_{k+2} p_{(1,3)} \hspace{2mm} \text{for } k = 1 ,  \nonumber
\end{IEEEeqnarray}
where $m_1 = (M-s), m_2 = \text{min}(b-m_1,M), m_3 = (b-m_1-m_2)$ are the lengths of the burst in each SP that it affects, progressing from left to right. If any of these lengths is zero, all probabilities involving that length are $1$, i.e., the probability of forming no size-$2$ stopping sets involving the SP corresponding to this (zero) length is $1$. 
For general $w$ and longer bursts, this strategy can be extended for finding a very good approximation for the average block erasure probability for the ensemble. 

\subsection{Simulations}
\label{sec:RBCSim}

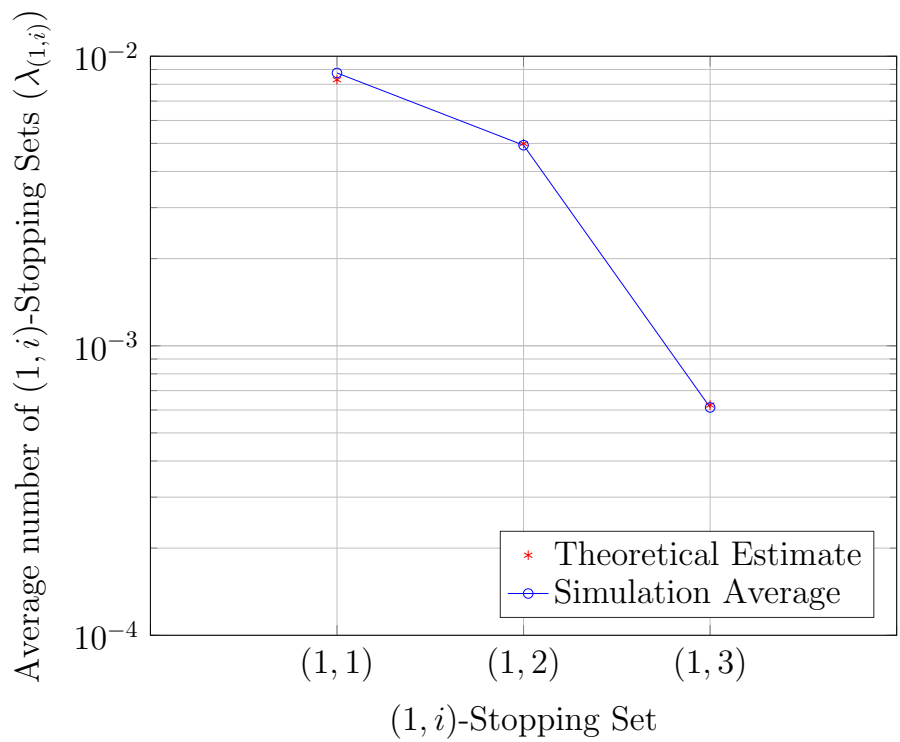
\begin{figure}[p]
\begin{center}
\large
\scalebox{0.9}{
%
%
\begin{tikzpicture}

\begin{axis}[%
width=4.34in,
height=3.37in,
at={(0.728in,0.51in)},
scale only axis,
xmin=0,
xmax=4,
xtick={0,1,2,3,4},
xticklabels={{},{$(1,1)$},{$(1,2)$},{$(1,3)$},{}},
xlabel={$(1,i)$-Stopping Set},
xmajorgrids,
ymode=log,
ymin=0.0001,
ymax=0.01,
yminorticks=true,
ylabel={Average number of $(1,i)$-Stopping Sets $(\lambda_{(1,i)})$},
ymajorgrids,
yminorgrids,
axis background/.style={fill=white},
title style={font=\bfseries},
legend style={at={(0.97,0.03)},anchor=south east,legend cell align=left,align=left,draw=white!15!black}
]
\addplot [color=red,only marks,mark=asterisk,mark options={solid}]
  table[row sep=crcr]{%
1	0.00829450741339703\\
2	0.00499328371035777\\
3	0.000624160463794721\\
};
\addlegendentry{Theoretical Estimate};

\addplot [color=blue,solid,mark=o,mark options={solid}] 
  table[row sep=crcr] {%
1	0.00875510204081633\\
2	0.00492857142857143\\
3	0.000612244897959184\\
};
\addlegendentry{Simulation Average};

\end{axis}
\end{tikzpicture}%
}
\normalsize
\caption{\label{fig:RBC_poisson_dist}Average number of size-$2$ $(1,i)$-stopping sets in a code from the random $(3,6,3,100,64)$ ensemble, along with theoretical estimates calculated using~(\ref{eq:RBCLambdasExample}).
}
\end{center}
\end{figure}

\begin{figure}[p]
\begin{center}
\large
\scalebox{0.9}{
%
%
\begin{tikzpicture}

\begin{axis}[%
width=4.34in,
height=3.354in,
at={(0.728in,0.518in)},
scale only axis,
xmin=0,
xmax=1000,
xlabel={\# VNs per Spatial Position $(M)$},
xmajorgrids,
ymode=log,
ymin=0.0001,
ymax=0.1,
yminorticks=true,
ylabel={Output Block Erasure Probability $(P_{B})$},
ymajorgrids,
yminorgrids,
axis background/.style={fill=white},
title style={font=\bfseries},
legend style={legend cell align=left,align=left,draw=white!15!black}
]
\addplot [color=blue,solid,mark=x,mark options={solid}]
  table[row sep=crcr]{%
80	0.0123661365716123\\
100	0.00783336858349196\\
120	0.00577054023797708\\
140	0.00455304986044902\\
160	0.00376644997024505\\
180	0.0035124816033776\\
200	0.00319560285047774\\
220	0.00305797305314146\\
240	0.00260336667378254\\
260	0.00223241692618513\\
280	0.00213337429411978\\
300	0.00198786606553597\\
320	0.00176499139566695\\
340	0.00169556186681362\\
360	0.00169004277498263\\
380	0.00152307844615537\\
400	0.00142947199592886\\
420	0.00151405267994895\\
440	0.0012978231612117\\
460	0.00130063197707766\\
480	0.00120442312347866\\
500	0.00119927228157954\\
520	0.00117815743246507\\
540	0.00109247709348654\\
560	0.00102987566311119\\
580	0.00107514280584319\\
600	0.00096332431661773\\
620	0.000888109720627325\\
640	0.000864400611649873\\
660	0.000888001676547165\\
680	0.000900954471166754\\
700	0.000810560632366983\\
720	0.000808273487417203\\
740	0.000759048426530564\\
760	0.000796885135382816\\
780	0.000734018760051469\\
800	0.000697001083139683\\
820	0.000706385798254803\\
840	0.000740271901869557\\
860	0.000667799254201793\\
880	0.000654548786793824\\
900	0.000670630540246551\\
920	0.000656515191105007\\
940	0.000599831807161272\\
960	0.000603642378109513\\
980	0.000599512476454147\\
1000	0.000559225316353761\\
};
\addlegendentry{Simulation, $1000$ failures};

\addplot [color=red,dashed]
  table[row sep=crcr]{%
80	0.00773585610425343\\
100	0.00608495828889875\\
120	0.00501425760097355\\
140	0.00426376447269486\\
160	0.00370858000676118\\
180	0.00328126755602252\\
200	0.00294222396718851\\
220	0.00266666596837007\\
240	0.00243829233144523\\
260	0.00224594176937808\\
280	0.00208171539630813\\
300	0.00193986607793466\\
320	0.00181611252500677\\
340	0.00170720002342324\\
360	0.00161061014492266\\
380	0.00152436373086606\\
400	0.00144688389174667\\
420	0.00137689873847291\\
440	0.00131337105510266\\
460	0.0012554465935102\\
480	0.00120241536814327\\
500	0.00115368251237312\\
520	0.00110874578791502\\
540	0.00106717831464064\\
560	0.0010286148435268\\
580	0.000992741117303664\\
600	0.000959285222869727\\
620	0.00092801070750309\\
640	0.000898710931736819\\
660	0.000871204634368372\\
680	0.000845332012640643\\
700	0.000820951738930576\\
720	0.000797938323153228\\
740	0.000776179941152291\\
760	0.000755576627250715\\
780	0.000736038857791449\\
800	0.000717486018704065\\
820	0.000699845423032556\\
840	0.000683051441406264\\
860	0.000667044571789542\\
880	0.000651770739070697\\
900	0.000637180729546081\\
920	0.000623229583998924\\
940	0.000609876277667876\\
960	0.000597083138743877\\
980	0.000584815719961428\\
1000	0.000573042236703935\\
};
\addlegendentry{Theoretical Approximation};

\end{axis}
\end{tikzpicture}%
}
\normalsize
\caption{\label{fig:RBC_Bound}Monte Carlo simulations for a $(3,6,3,20,M)$ random ensemble on the RBC with burst length $b=1.25M$, along with the theoretical approximation~(\ref{eq:RBCLowerBound}).
}
\end{center}
\end{figure}
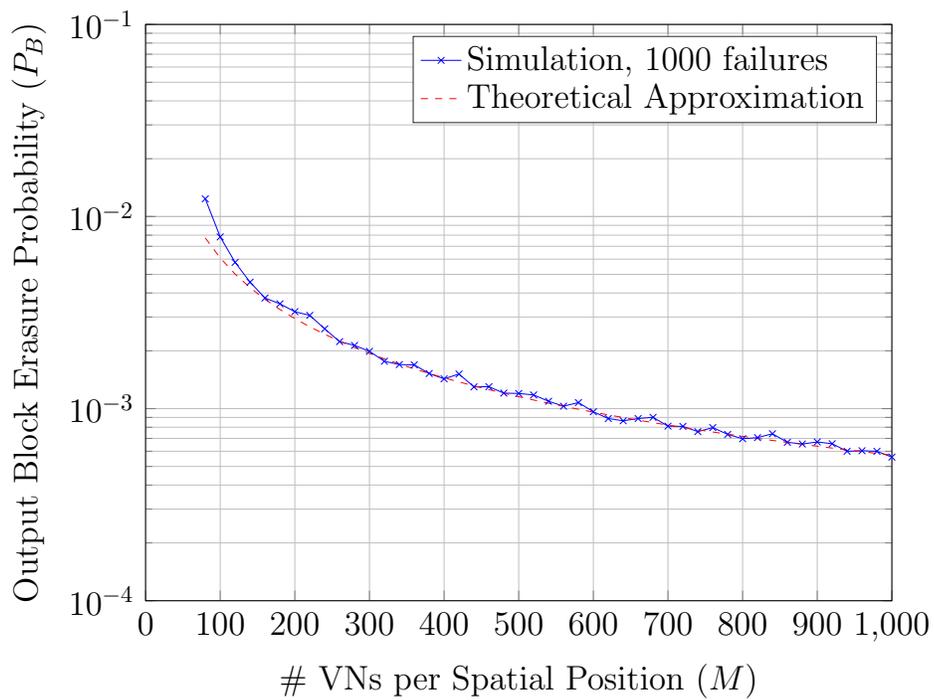

First, we show that the individual components of $\vecnot{p}(d_v,d_c,w,L,M)$ given in~(\ref{eq:pvec}) are accurate for even a small value of $M=64$. 
With $w=3$, $L=100$ for a $(3,6)$ random ensemble, we estimate the average number of size-$2$ $(1,i)$-stopping sets corresponding to each component of $\vecnot{p}$ by averaging over all the SPs of $1000$ codes sampled from the ensemble. 
The experimental histogram and the theoretical averages
\begin{equation}
\label{eq:RBCLambdasExample}
(\lambda_{(1,1)}, \lambda_{(1,2)}, \lambda_{(1,3)}) = \left( \binom{M}{2} p_{(1,1)}, M^2 p_{(1,2)}, M^2 p_{(1,3)} \right)
\end{equation}
are plotted in Fig.~\ref{fig:RBC_poisson_dist}.

To verify the tightness of \eqref{eq:RBCLowerBound}, we again performed Monte-Carlo simulations and counted $1000$ decoding failures for each $M$ to assess the average block erasure probability $P_\B^{\RBC}$. 
For the sake of example, we fixed the burst length to be $b=1.25M$.  
We selected a value for $s$, uniformly from $\{1,\ldots,M\}$, for each codeword. 
The simulation results for the $(3,6,3,20,M)$ ensemble are shown in Fig.~\ref{fig:RBC_Bound} along with \eqref{eq:RBCLowerBound}. 
We see that \eqref{eq:RBCLowerBound} is indeed a tight approximation.

\section{Error Floor on BEC}
\label{sec:floorBEC}

\subsection{Distribution of $\mathbb{N}_2^{H}$}
\label{sec:n2h}

The approach described here is from~\cite{Olmos-isit11}. 
We know that stopping sets of size larger than $2$ vanish for large enough $M$. 
This means that with random erasures on BEC($\epsilon$), size-$2$ stopping sets in the code are, with high probability, the cause of decoder failures. 
As mentioned earlier, the ensemble is completely characterized by the vector $\vecnot{p}(d_v,d_c,w,L,M)$, as given in~(\ref{eq:pvec}). 
Using these we also know the average number of size-$2$ stopping sets of each type in the code, which has been expressed in~(\ref{eq:lambdas}). 
Therefore, the average number of size-$2$ stopping sets in a code is
\begin{equation*}
\lambda = \mathbb{E}[\mathbb{N}_2^{H}] = L \left[ \lambda_{(1,1)} + \sum_{i=2}^{w} \lambda_{(1,i)} \right].
\end{equation*}
and we observe that $\lambda \propto LM^{2-d_v}$. More carefully, if we take into account the boundary effects, we calculate this as
\begin{equation}
\label{eq:n2hPoisson}
\lambda = \mathbb{E}[\mathbb{N}_2^{H}] = L \lambda_{(1,1)} + (L-w+1) \sum_{i=2}^{w} \lambda_{(1,i)} + \sum_{j=1}^{w-2} \sum_{i=j+1}^{w-1} \lambda_{(1,w-i+1)}.
\end{equation}
Since each of the $(1,i)$-stopping sets form a Poisson distribution with mean $\lambda_{(1,i)}$ and the correlation between them is negligible, we conjecture that $\mathbb{N}_2^{H} \sim \text{Poisson}(\lambda)$.

Given that size-$2$ stopping sets are dominantly responsible for decoder failures on BEC($\epsilon$), the expected error floor for a $(d_v,d_c,w,L,M)$ random ensemble is given by
\begin{equation}
\label{eq:errorFloor}
P_{\rm b}(d_v,d_c,w,L,M) = \frac{2 \lambda \epsilon^{2}}{LM},
\end{equation}
where $P_{\rm b}$ is the average bit error rate for the ensemble.

\subsection{Simulations}
\label{sec:becSim}

\begin{figure}[p]
\begin{center}
\large
\scalebox{0.9}{
%
%
\begin{tikzpicture}

\begin{axis}[%
width=4.34in,
height=3.37in,
at={(0.775in,0.518in)},
scale only axis,
xmin=-0.5,
xmax=3.5,
xlabel={Number of Size-$2$ Stopping Sets $(i)$},
xmajorgrids,
ymode=log,
ymin=0.0001,
ymax=1,
yminorticks=true,
ylabel={Prob [$N_{2}^{H} = i$]},
ymajorgrids,
yminorgrids,
axis background/.style={fill=white},
title style={font=\bfseries},
legend style={legend cell align=left,align=left,draw=white!15!black}
]
\addplot [color=red,only marks,mark=asterisk,mark options={solid}]
  table[row sep=crcr]{%
0	0.875571983083397\\
1	0.11634417625024\\
2	0.00772978556239248\\
3	0.000342372586614956\\
};
\addlegendentry{Theoretical Estimate};

\addplot [color=blue,solid,mark=o,mark options={solid}]
  table[row sep=crcr]{%
0	0.877472808234178\\
1	0.114594739275373\\
2	0.00754312966883228\\
3	0.000352823807090542\\
};
\addlegendentry{Simulation Average};

\end{axis}
\end{tikzpicture}%
}
\normalsize
\caption{\label{fig:n2h_poisson_dist}Poisson distribution of $\mathbb{N}_2^{H}$ for a $(3,6,3,10,64)$ ensemble, with theoretical estimates calculated using~(\ref{eq:n2hPoisson}).
}
\end{center}
\end{figure}
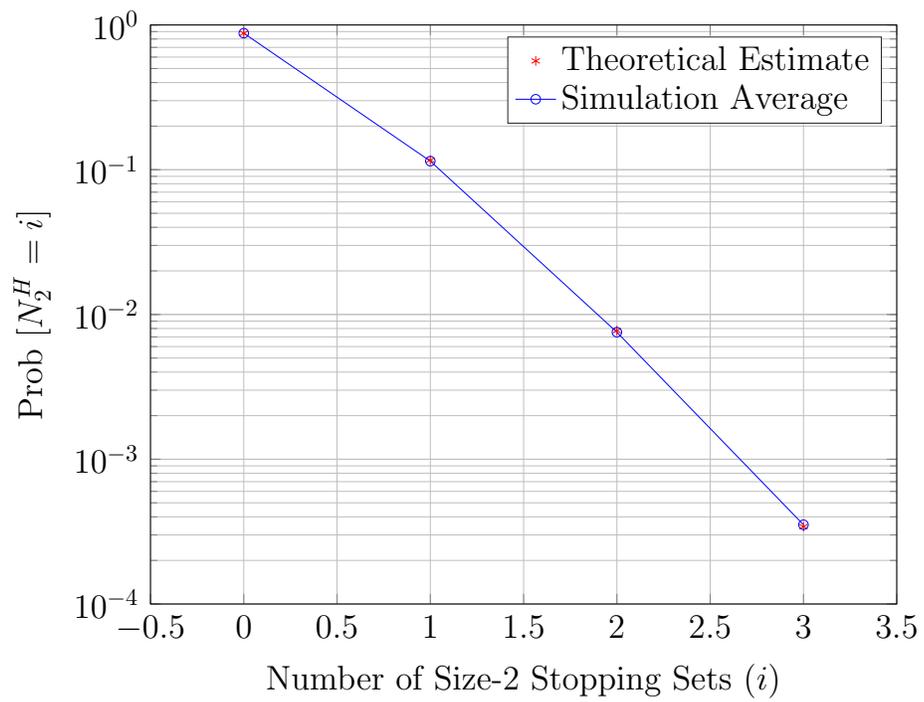

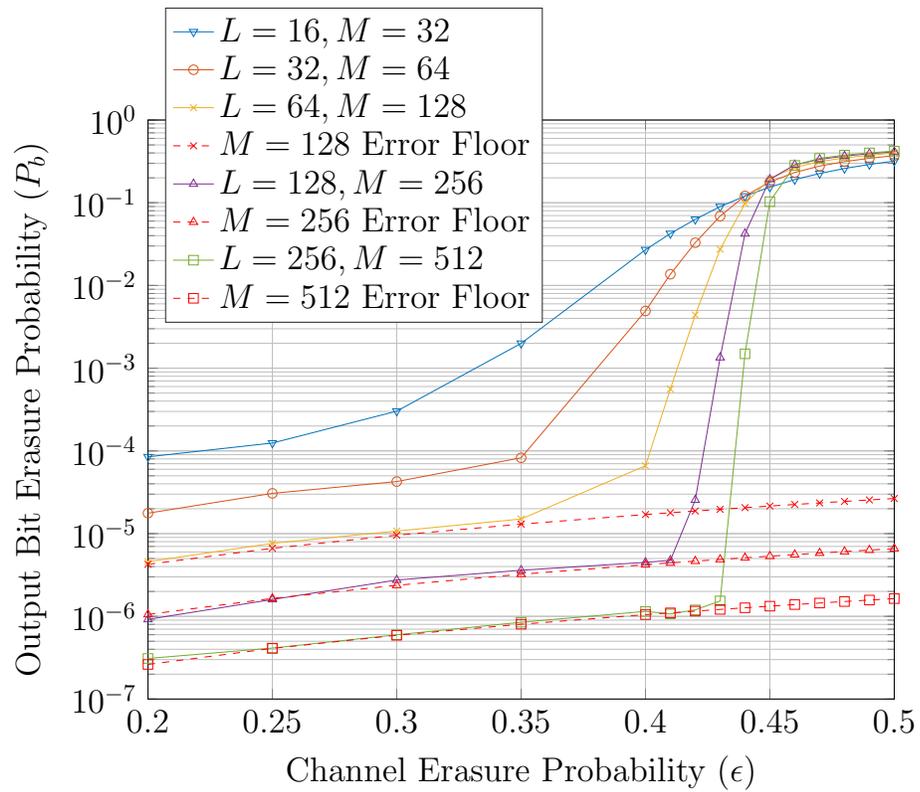
\begin{figure}[p]
\begin{center}
\large
\scalebox{0.9}{
%
%
\definecolor{mycolor1}{rgb}{0.00000,0.44700,0.74100}%
\definecolor{mycolor2}{rgb}{0.85000,0.32500,0.09800}%
\definecolor{mycolor3}{rgb}{0.92900,0.69400,0.12500}%
\definecolor{mycolor4}{rgb}{0.49400,0.18400,0.55600}%
\definecolor{mycolor5}{rgb}{0.46600,0.67400,0.18800}%
\begin{tikzpicture}

\begin{axis}[%
width=4.34in,
height=3.37in,
at={(0.728in,0.51in)},
scale only axis,
xmin=0.2,
xmax=0.5,
xlabel={Channel Erasure Probability $(\epsilon)$},
xmajorgrids,
ymode=log,
ymin=1e-07,
ymax=1,
yminorticks=true,
ylabel={Output Bit Erasure Probability $(P_{b})$},
ymajorgrids,
yminorgrids,
axis background/.style={fill=white},
title style={font=\bfseries},
legend style={at={(0.023,0.65)},anchor=south west,legend cell align=left,align=left,draw=white!15!black}
]
\addplot [color=mycolor1,solid,mark=triangle,mark options={solid,rotate=180}]
  table[row sep=crcr]{%
0.2	8.48046875e-05\\
0.25	0.00012447265625\\
0.3	0.000302294921875\\
0.35	0.00199568359375\\
0.4	0.027009228515625\\
0.41	0.04241599609375\\
0.42	0.062997099609375\\
0.43	0.08969294921875\\
0.44	0.120306259765625\\
0.45	0.15468681640625\\
0.46	0.1906114453125\\
0.47	0.225845849609375\\
0.48	0.2597284375\\
0.49	0.290420673828125\\
0.5	0.318956669921875\\
};
\addlegendentry{$L=16,M=32$};

\addplot [color=mycolor2,solid,mark=o,mark options={solid}]
  table[row sep=crcr]{%
0.2	1.767578125e-05\\
0.25	3.06787109375e-05\\
0.3	4.259521484375e-05\\
0.35	8.2529296875e-05\\
0.4	0.0049197314453125\\
0.41	0.0136787036132813\\
0.42	0.0328832006835938\\
0.43	0.0689346752929687\\
0.44	0.120944426269531\\
0.45	0.178174958496094\\
0.46	0.232404860839844\\
0.47	0.277735173339844\\
0.48	0.314172858886719\\
0.49	0.345026750488281\\
0.5	0.371210490722656\\
};
\addlegendentry{$L=32,M=64$};

\addplot [color=mycolor3,solid,mark=x,mark options={solid}]
  table[row sep=crcr]{%
0.2	4.57763671875e-06\\
0.25	7.630615234375e-06\\
0.3	1.073974609375e-05\\
0.35	1.50262451171875e-05\\
0.4	6.61590576171875e-05\\
0.41	0.000559996337890625\\
0.42	0.00440268737792969\\
0.43	0.0274147290039062\\
0.44	0.0977825543212891\\
0.45	0.194755729370117\\
0.46	0.267834309082031\\
0.47	0.314642916870117\\
0.48	0.348336395874023\\
0.49	0.375849093017578\\
0.5	0.399037845458984\\
};
\addlegendentry{$L=64,M=128$};

\addplot [color=red,dashed,mark=x,mark options={solid}]
  table[row sep=crcr]{%
0.2	4.25495036579848e-06\\
0.25	6.64835994656013e-06\\
0.3	9.57363832304659e-06\\
0.35	1.30307854952579e-05\\
0.4	1.70198014631939e-05\\
0.41	1.78814289122681e-05\\
0.42	1.87643311131713e-05\\
0.43	1.96685080659035e-05\\
0.44	2.05939597704647e-05\\
0.45	2.15406862268548e-05\\
0.46	2.2508687435074e-05\\
0.47	2.34979633951221e-05\\
0.48	2.45085141069993e-05\\
0.49	2.55403395707054e-05\\
0.5	2.65934397862405e-05\\
};
\addlegendentry{$M=128$ Error Floor};

\addplot [color=mycolor4,solid,mark=triangle,mark options={solid}]
  table[row sep=crcr]{%
0.2	9.19036865234375e-07\\
0.25	1.59820556640625e-06\\
0.3	2.74765014648438e-06\\
0.35	3.604736328125e-06\\
0.4	4.49310302734375e-06\\
0.41	4.75540161132812e-06\\
0.42	2.54975891113281e-05\\
0.43	0.00134444458007813\\
0.44	0.0422098208618164\\
0.45	0.192203114471436\\
0.46	0.287621301116943\\
0.47	0.334826372375488\\
0.48	0.366956143341064\\
0.49	0.392271343688965\\
0.5	0.414007606201172\\
};
\addlegendentry{$L=128,M=256$};

\addplot [color=red,dashed,mark=triangle,mark options={solid}]
  table[row sep=crcr]{%
0.2	1.05375928519998e-06\\
0.25	1.64649888312496e-06\\
0.3	2.37095839169995e-06\\
0.35	3.22713781092493e-06\\
0.4	4.21503714079991e-06\\
0.41	4.4284233960529e-06\\
0.42	4.6470784477319e-06\\
0.43	4.87100229583689e-06\\
0.44	5.10019494036789e-06\\
0.45	5.33465638132488e-06\\
0.46	5.57438661870788e-06\\
0.47	5.81938565251687e-06\\
0.48	6.06965348275187e-06\\
0.49	6.32519010941286e-06\\
0.5	6.58599553249986e-06\\
};
\addlegendentry{$M=256$ Error Floor};

\addplot [color=mycolor5,solid,mark=square,mark options={solid}]
  table[row sep=crcr]{%
0.2	3.09600830078125e-07\\
0.25	4.13475036621094e-07\\
0.3	5.98297119140625e-07\\
0.35	8.4808349609375e-07\\
0.4	1.14833831787109e-06\\
0.41	1.06948852539063e-06\\
0.42	1.19644165039062e-06\\
0.43	1.53549194335938e-06\\
0.44	0.00149215751647949\\
0.45	0.103164820251465\\
0.46	0.282932458343506\\
0.47	0.344371552848816\\
0.48	0.376822533111572\\
0.49	0.40098105266571\\
0.5	0.421637405281067\\
};
\addlegendentry{$L=256,M=512$};

\addplot [color=red,dashed,mark=square,mark options={solid}]
  table[row sep=crcr]{%
0.2	2.62204979165678e-07\\
0.25	4.09695279946372e-07\\
0.3	5.89961203122777e-07\\
0.35	8.0300274869489e-07\\
0.4	1.04881991666271e-06\\
0.41	1.10191642494376e-06\\
0.42	1.15632395812064e-06\\
0.43	1.21204251619335e-06\\
0.44	1.26907209916188e-06\\
0.45	1.32741270702625e-06\\
0.46	1.38706433978644e-06\\
0.47	1.44802699744246e-06\\
0.48	1.51030067999431e-06\\
0.49	1.57388538744198e-06\\
0.5	1.63878111978549e-06\\
};
\addlegendentry{$M=512$ Error Floor};

\end{axis}
\end{tikzpicture}%
}
\normalsize
\caption{\label{fig:bec_floors}The expected error floors for a $(3,6,3,L,M)$ random ensemble on the BEC.
}
\end{center}
\end{figure}

First, we show in Fig.~\ref{fig:n2h_poisson_dist} that $\mathbb{N}_2^{H} \sim \text{Poisson}(\lambda)$ through the simulation histogram averaged over $10^5$ code blocks from the $(3,6,3,10,64)$ random ensemble. 
Equation~(\ref{eq:n2hPoisson}) is used to calculate the theoretical Poisson distribution.

We verified the error floor calculation through standard simulations on the BEC for a $(3,6,3,L,M)$ random ensemble with $M = 128,256,512$ and $L=M/2$. 
The results and the predicted error floor~(\ref{eq:errorFloor}) are plotted in Fig.~\ref{fig:bec_floors}. 
In comparison with the observations in~\cite{Olmos-isit11}, the error floor of the random ensemble seems to be slightly worse than the protograph-based ensemble since the latter is more structured. 

Also, it is worth noting again that the performance of a typical code from the random ensemble is concentrated around the ensemble average and hence, this is the expected behavior for a code uniformly sampled from this ensemble.

\section{Effects of Expurgation}
\label{sec:expurgate}

\subsection{Minimal Stopping Set Size}
\label{sec:minimalSS}

As the performance is mainly dominated by size-$2$ stopping sets, we can improve the burst erasure correction capability by expurgating the ensemble and thereby removing all small stopping sets. 
Observing that a size-$2$ stopping set, as shown in Fig.~\ref{fig:size2ss}, is built around 4-cycles, we can reduce the size of the minimal stopping sets by removing small cycles from the graph. 
For example, increasing the girth of the graph to $6$ leads to minimal stopping sets of size $s_{\rm min} = d_v+1$~\cite{Orlitsky-isit02}.

\begin{figure}[p]
\begin{center}

\makeatletter
\@ifundefined{vecnot}{%
\newcommand{\vecnot}[1]{\underline{#1}}
}{}
\makeatother
\scalebox{1.2}{
\begin{tikzpicture}

  \draw[color=black] (-0.25,0) circle (.25);
  \draw[color=black] (1.25,0) circle (.25);  
  \draw[color=black] (2.75,0) circle (.25);  
  \draw[color=black] (4.25,0) circle (.25);  
  
  \node[draw=none,align=center] at (-0.25,-0.5) {$v_1$};
  \node[draw=none,align=center] at (1.25,-0.5) {$v_2$};
  \node[draw=none,align=center] at (2.75,-0.5) {$v_3$};
  \node[draw=none,align=center] at (4.25,-0.5) {$v_4$};

  \draw[color=black] (-2,2) rectangle (-1.5,2.5);
  \node[draw=none,align=center] at (-1.75,3) {$c_1$};

  \draw[color=black] (-0.5,2) rectangle (0,2.5);
  \node[draw=none,align=center] at (-0.25,3) {$c_2$};

  \draw[color=black] (1,2) rectangle (1.5,2.5);
  \node[draw=none,align=center] at (1.25,3) {$c_3$};

  \draw[color=black] (2.5,2) rectangle (3,2.5);
  \node[draw=none,align=center] at (2.75,3) {$c_4$};

  \draw[color=black] (4,2) rectangle (4.5,2.5);
  \node[draw=none,align=center] at (4.25,3) {$c_5$};

  \draw[color=black] (5.5,2) rectangle (6,2.5);
  \node[draw=none,align=center] at (5.75,3) {$c_6$};

  \draw[color=black] (-0.25,0.25) -- (-1.75,2);
  \draw[color=black] (-0.25,0.25) -- (-0.25,2);
  \draw[color=black] (-0.25,0.25) -- (1.25,2);

  \draw[dashed,color=black] (1.25,0.25) -- (-1.75,2);
  \draw[color=black] (1.25,0.25) -- (2.75,2);
  \draw[color=black] (1.25,0.25) -- (4.25,2);
  
  \draw[dashed,color=black] (2.75,0.25) -- (-0.25,2);
  \draw[dashed,color=black] (2.75,0.25) -- (2.75,2);
  \draw[color=black] (2.75,0.25) -- (5.75,2);

  \draw[dashed,color=black] (4.25,0.25) -- (1.25,2);
  \draw[color=black] (4.25,0.25) -- (4.25,2);
  \draw[color=black] (4.25,0.25) -- (5.75,2);
  
  
  \node[draw=none,align=center] at (2,-3) {
     $\begin{bmatrix} 
          1 & 1 & 1 & 0 & 0 & 0 \\  
          1 & 0 & 0 & 1 & 1 & 0 \\ 
          0 & 1 & 0 & 1 & 0 & 1 \\ 
          0 & 0 & 1 & 0 & 1 & 1 \\ 
     \end{bmatrix}$
   };
   
   
   \draw[dashed,color=black] (1.95,-2.05) -- (1.95,-3.9);
   \draw[dashed,color=black] (0.5,-2.45) -- (1.9,-2.45);
   

   \draw[dashed,color=black] (3,-2.05) -- (3,-3.9);
   \draw[dashed,color=black] (2.05,-2.95) -- (2.9,-2.95);
   

   \draw[dashed,color=black] (3.1,-3.45) -- (3.4,-3.45);
   
  \node[draw=none,align=center] at (0,-2.35) {$v_1$};
  \node[draw=none,align=center] at (0,-2.8) {$v_2$};
  \node[draw=none,align=center] at (0,-3.25) {$v_3$};
  \node[draw=none,align=center] at (0,-3.7) {$v_4$};

  \node[draw=none,align=center] at (0.65,-1.75) {$c_1$};
  \node[draw=none,align=center] at (1.2,-1.75) {$c_2$};
  \node[draw=none,align=center] at (1.7,-1.75) {$c_3$};
  \node[draw=none,align=center] at (2.2,-1.75) {$c_4$};
  \node[draw=none,align=center] at (2.75,-1.75) {$c_5$};
  \node[draw=none,align=center] at (3.3,-1.75) {$c_6$};

\end{tikzpicture}}
\normalsize
\caption{\label{fig:size4ss}A size-$4$ stopping set from an expurgated $(3,6,w,L,M)$ random ensemble. CNs $\{c_1,c_2,c_3,c_4,c_5,c_6\}$ and VNs $\{v_1,v_2,v_3,v_4\}$ have been labeled for convenience. The solid edges indicate definite connections and the dashed edges complete one configuration to form a stopping set. Multiple edges are not allowed in the ensemble. The bi-adjacency matrix is also shown with its pattern highlighted.
}
\end{center}
\end{figure}

We give a simple construction which we will use to find the probability of a size $(d_v+1)$ stopping set in a SP of a SC-LDPC code.
Let us consider a $(3,6)$ random ensemble as an example. 
We immediately notice that size-$3$ stopping sets vanish once $girth=6$. 
A size-$4$ stopping set is shown in Fig.~\ref{fig:size4ss} along with its bi-adjacency matrix that describes the neighbors of each VN in the corresponding row. 
We can notice a pattern in this matrix that can be generalized to get a $(d_v+1) \times (\frac{s_{\rm min}d_v}{2})$ matrix for a $(d_v,d_c)$ LDPC (or SC-LDPC) ensemble. 
The pattern has been highlighted using dashed lines in the matrix: row $i \in \{ 1,2,\ldots,d_v \}$ has one subset of $(d_v-(i-1))$ columns with all $1$s and an identity matrix $I_{d_v-i+1}$ spanning these columns starting from row $i+1$. 
Such a construction always corresponds to a minimal stopping set of size $(d_v+1)$ and involves exactly $\frac{s_{min}d_v}{2} = \frac{(d_v+1)d_v}{2}$ neighboring CNs.

\subsection{Performance on the SPBC}
\label{sec:no4SPBCAnalysis}


We can use the same approach as in Section~\ref{sec:wt2s} to calculate the probability of occurrence of the stopping set shown in Fig.~\ref{fig:size4ss} within a spatial position of a code sampled uniformly from the ensemble. 
Once again we have $p = \frac{T_{ss}}{T}$,
where $T_{ss}$ is the total number of combinations of the edges of $v_1,v_2,v_3,v_4$ that form a stopping set and $T$ is the total number of combinations by which these edges can fit to the available CN sockets.
For an expurgated $(3,6,w,L,M)$ random ensemble, we have
\begin{IEEEeqnarray*}{rl}
T_{ss} = & \hspace{3mm} [(1)] \times [(3wM)(3wM-6)(3wM-12)] \times \frac{3!}{0! \times 3!} \\
    & \times [(15)] \times [(3wM-18)(3wM-24)] \times \frac{3!}{1! \times 2!} \\
    & \times [(20)(10)] \times [(3wM-30)] \times \frac{3!}{2! \times 1!} \\
    & \times [(15)(10)(5)] \times [(1)] \times \frac{3!}{3! \times 0!}.
\end{IEEEeqnarray*}
Since $T$ is the total number of combinations in which the edges of $(d_v+1)$ VNs can be assigned to sockets ensuring no $4$-cycles, we can again approximate it by its dominant term as
\begin{equation*}
T \approx \prod_{j=0}^{d_v(d_v+1)-1} (wMd_v - jd_c).
\end{equation*}
For a general $(d_v,d_c,w,L,M)$ random ensemble, the expression for $T_{ss}$ can be calculated as 
\begin{IEEEeqnarray*}{rl}
T_{ss} = & \hspace{3mm} \prod_{i=0}^{d_v} \left[ \prod_{j=1}^{i} j(d_c-1)(d_v-i+1) \right] \\
    & \times \left[ \prod_{k = \sum_{m=0}^{i-1}(d_v-m)}^{\sum_{m=0}^{i-1}(d_v-m) + (d_v-i-1)} (wMd_v - kd_c) \right] \binom{d_v}{i}.
\end{IEEEeqnarray*}
It can be verified that the last value for $k$ in the above expression is $k=\frac{d_v(d_v+1)}{2}-1$. Then, we can simplify and rearrange the expression as
\begin{IEEEeqnarray*}{rCl}
T_{ss} & = & \hspace{3mm} \left[ \prod_{k = 0}^{\frac{d_v(d_v+1)}{2}-1} (wMd_v - kd_c) \right] \\
  &   & \times \prod_{i=0}^{d_v} \left[ \prod_{j=1}^{i} j \right] \left[ \prod_{j=1}^{i} (d_c-1)(d_v-i+1) \right] \\
  &   & \times \frac{d_v!}{i! \times (d_v-i)!} \\
  & = & T_{1/2} \times \prod_{i=1}^{d_v} i! \left[ (d_c-1)(d_v-i+1) \right]^{i} \times \frac{d_v!}{i! \times (d_v-i)!} \\
T_{ss}  & = & T_{1/2} \times \prod_{i=1}^{d_v} \left[ (d_c-1)(d_v-i+1) \right]^{i} \times \frac{d_v!}{(d_v-i)!} ,
\end{IEEEeqnarray*}
where $T_{1/2} = \prod_{k = 0}^{\frac{d_v(d_v+1)}{2}-1} (wMd_v - kd_c)$ is the first half of the products in $T$ which can be canceled while calculating $p$, so that
\begin{equation*}
\frac{T}{T_{1/2}} \cong \prod_{j = \frac{d_v(d_v+1)}{2}}^{d_v(d_v+1)-1} (wMd_v - jd_c).
\end{equation*}
For a general $(d_v,d_c,w,M)$ ensemble, the probability of forming such a minimal stopping set of size $(d_v+1)$ can be shown to be
\begin{equation}
\label{eq:no4pSS}
p = \frac{T_{ss}}{T} \approx \frac{\prod_{i=1}^{d_v} \left[ (d_c-1)(d_v-i+1) \right]^{i} \times \frac{d_v!}{(d_v-i)!}}{\prod_{j = \frac{d_v(d_v+1)}{2}}^{d_v(d_v+1)-1} (wMd_v - jd_c)} ,
\end{equation}
which means the expected number of such stopping sets within a SP of the code is $\lambda_{SP} = \binom{M}{d_v+1} p$. %
Using similar arguments as in Section~\ref{sec:spbcAnalysis}, we have 
\begin{equation*}
\mathbb{N}_{d_v+1}^{SP} \sim \text{Poisson}(\lambda_{SP}) .
\end{equation*}
A tight approximation for the average block erasure probability on the SPBC, $P_\B^\SPBC$, can be calculated as
\begin{equation}
\label{eq:no4SPBCLowerBound}
P_{B,\text{exp}}^\SPBC \approx 1-e^{-\lambda_{\rm SP}}\approx \lambda_{\rm SP}.
\end{equation}

\subsection{Simulations}
\label{sec:no4SPBCSim}

\begin{figure}[p]
\begin{center}
\large
\scalebox{0.9}{
%
%
\begin{tikzpicture}

\begin{axis}[%
width=4.34in,
height=3.354in,
at={(0.728in,0.518in)},
scale only axis,
xmin=40,
xmax=180,
xlabel={\# VNs per Spatial Position $(M)$},
xmajorgrids,
ymode=log,
ymin=1e-05,
ymax=0.1,
yminorticks=true,
ylabel={Output Block Erasure Probability $(P_{B})$},
ymajorgrids,
yminorgrids,
axis background/.style={fill=white},
title style={font=\bfseries},
legend style={legend cell align=left,align=left,draw=white!15!black}
]
\addplot [color=blue,solid,mark=x,mark options={solid}]
  table[row sep=crcr]{%
40	0.0112637981527371\\
60	0.00277223331115547\\
80	0.000674340663416345\\
100	0.000237645409284806\\
120	0.000164829096950827\\
140	0.000102051542151879\\
160	6.50775100260039e-05\\
180	5.92931881391389e-05\\
};
\addlegendentry{Simulation, $100$ failures};

\addplot [color=red,dashed]
  table[row sep=crcr]{%
40	0.00213080574547264\\
60	0.000722906354645447\\
80	0.000357427188013459\\
100	0.000212101824417776\\
120	0.000140157514968653\\
140	9.94192492347334e-05\\
160	7.41530409357205e-05\\
180	5.74166743598781e-05\\
};
\addlegendentry{Theoretical Approximation};

\end{axis}
\end{tikzpicture}%
}
\normalsize
\caption{\label{fig:SPBC_no4cycles}Monte Carlo simulations on the SPBC with an expurgated $(3,6)$ random ensemble for $w=3$ along with the theoretical approximation. The approximation becomes tight very quickly with $M$.
}
\end{center}
\end{figure}
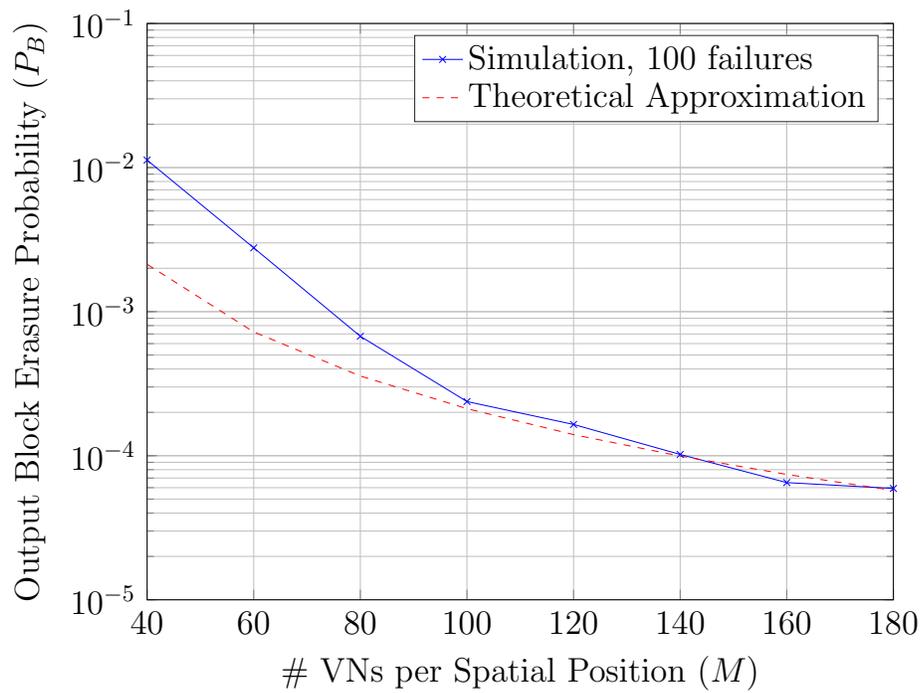

We performed Monte-Carlo simulations for an expurgated $(3,6)$ random ensemble with $w=3$ and counted $100$ decoding failures on the SPBC. 
The simulation averages for varying $M$ and their respective lower bounds calculated  using~(\ref{eq:no4pSS})~and~(\ref{eq:no4SPBCLowerBound}) are plotted in Fig.~\ref{fig:SPBC_no4cycles}. 
It is evident that the bound becomes tight very quickly which reassures that the decoder performance is indeed dominated by minimal stopping sets.

\section{Finite Length Observations}
\label{sec:observe}


\begin{figure}[p]
\begin{center}
\large
\scalebox{0.9}{
%
%
\begin{tikzpicture}

\begin{axis}[%
width=4.284in,
height=3.262in,
at={(0.784in,0.518in)},
scale only axis,
xmin=0,
xmax=1000,
xlabel={\# VNs per Spatial Position $(M)$},
xmajorgrids,
ymode=log,
ymin=1e-16,
ymax=0.01,
yminorticks=true,
ylabel={Output Block Erasure Probability $(P_{B})$},
ymajorgrids,
yminorgrids,
axis background/.style={fill=white},
title style={font=\bfseries},
legend style={at={(0.97,0.97)},anchor=south east,legend cell align=left,align=left,draw=white!15!black}
]
\addplot [color=blue,solid,mark=x,mark options={solid}]
  table[row sep=crcr]{%
80	0.00700530843085789\\
100	0.00550776963784583\\
120	0.00453726993839598\\
140	0.00385735812556298\\
160	0.00335457075347456\\
180	0.00296769357169491\\
200	0.00266079849144518\\
220	0.0024114121242732\\
240	0.00220475783501195\\
260	0.0020307211422359\\
280	0.00188214572343171\\
300	0.00175382578065186\\
320	0.00164188396642928\\
340	0.00154337308613295\\
360	0.00145601298236253\\
380	0.00137801194386034\\
400	0.00130794257255629\\
420	0.00124465367331927\\
440	0.00118720653394799\\
460	0.00113482806637288\\
480	0.00108687582387768\\
500	0.0010428115252884\\
520	0.00100218076677372\\
540	0.000964597297468006\\
560	0.000929730704511522\\
580	0.000897296675168135\\
600	0.000867049228077676\\
620	0.000838774464250625\\
640	0.000812285501924248\\
660	0.000787418341634649\\
680	0.000764028468123623\\
700	0.000741988040316\\
720	0.000721183553966775\\
740	0.00070151388675066\\
760	0.00068288865472943\\
780	0.000665226823831189\\
800	0.000648455531344849\\
820	0.000632509081281762\\
840	0.000617328084396629\\
860	0.000602858719138455\\
880	0.00058905209414406\\
900	0.000575863696364198\\
920	0.000563252911695455\\
940	0.000551182607243184\\
960	0.000539618766165151\\
980	0.000528530167532959\\
1000	0.000517888104867525\\
};
\addlegendentry{$(3,6)$, With $4$-cycles};

\addplot [color=red,solid,mark=x,mark options={solid}]
  table[row sep=crcr]{%
80	0.000357427188013459\\
100	0.000212101824417776\\
120	0.000140157514968653\\
140	9.94192492347334e-05\\
160	7.41530409357205e-05\\
180	5.74166743598781e-05\\
200	4.57640237649937e-05\\
220	3.7327774932927e-05\\
240	3.10250702468817e-05\\
260	2.61930090850182e-05\\
280	2.24074079995251e-05\\
300	1.93866369890827e-05\\
320	1.69377828327377e-05\\
340	1.49251170062925e-05\\
360	1.32509018502258e-05\\
380	1.18433116558236e-05\\
400	1.06486071355105e-05\\
420	9.62593408970758e-06\\
440	8.74378684112553e-06\\
460	7.97755427461855e-06\\
480	7.30778569968393e-06\\
500	6.71894498827097e-06\\
520	6.19850197480787e-06\\
540	5.73626068400834e-06\\
560	5.3238563836322e-06\\
580	4.95437466330184e-06\\
600	4.62205984241848e-06\\
620	4.32208954570523e-06\\
640	4.05039882900304e-06\\
660	3.8035417876392e-06\\
680	3.57858178900994e-06\\
700	3.37300375874428e-06\\
720	3.18464359783111e-06\\
740	3.01163101112945e-06\\
760	2.85234291297343e-06\\
780	2.70536523083731e-06\\
800	2.56946142029868e-06\\
820	2.44354637812805e-06\\
840	2.32666471911003e-06\\
860	2.21797260435697e-06\\
880	2.11672247041328e-06\\
900	2.02225014322988e-06\\
920	1.93396392067502e-06\\
940	1.85133528596282e-06\\
960	1.77389098121594e-06\\
980	1.70120621656444e-06\\
1000	1.6328988335923e-06\\
};
\addlegendentry{$(3,6)$, No $4$-cycles};

\addplot [color=blue,dashdotted,mark=triangle,mark options={solid}]
  table[row sep=crcr]{%
80	7.80270422120477e-05\\
100	4.86618218087376e-05\\
120	3.32187134569617e-05\\
140	2.41099368500919e-05\\
160	1.82917379527581e-05\\
180	1.43509139215459e-05\\
200	1.15588057810756e-05\\
220	9.50878038186165e-06\\
240	7.95940910736093e-06\\
260	6.76001543398197e-06\\
280	5.8126176970541e-06\\
300	5.05126834249303e-06\\
320	4.43026136809355e-06\\
340	3.91711023195462e-06\\
360	3.4882117790902e-06\\
380	3.12608308639106e-06\\
400	2.81755139774909e-06\\
420	2.55253915226739e-06\\
440	2.32323057625106e-06\\
460	2.12348879280455e-06\\
480	1.94844094170143e-06\\
500	1.79417816248417e-06\\
520	1.65753548320158e-06\\
540	1.53592818763482e-06\\
560	1.42722868567979e-06\\
580	1.32967282273633e-06\\
600	1.24178784877138e-06\\
620	1.16233651004105e-06\\
640	1.09027326500311e-06\\
660	1.02470971186097e-06\\
680	9.64887077237186e-07\\
700	9.10154163591415e-07\\
720	8.59949553788653e-07\\
740	8.13787156550561e-07\\
760	7.71244396235993e-07\\
780	7.31952508270517e-07\\
800	6.95588520560619e-07\\
820	6.61868595819293e-07\\
840	6.30542476010021e-07\\
860	6.01388825516302e-07\\
880	5.74211311055173e-07\\
900	5.48835287439431e-07\\
920	5.25104984827607e-07\\
940	5.02881112862674e-07\\
960	4.82038811644436e-07\\
980	4.62465894579545e-07\\
1000	4.44061335036494e-07\\
};
\addlegendentry{$(4,8)$, With $4$-cycles};

\addplot [color=red,dashdotted,mark=triangle,mark options={solid}]
  table[row sep=crcr]{%
80	1.18655553160707e-08\\
100	3.27296467705906e-09\\
120	1.17515852604555e-09\\
140	5.0216431013439e-10\\
160	2.42791675653109e-10\\
180	1.28723143255627e-10\\
200	7.32998106656169e-11\\
220	4.41870984246862e-11\\
240	2.79053447016508e-11\\
260	1.83181247948028e-11\\
280	1.24240617793703e-11\\
300	8.66551275180427e-12\\
320	6.19215789754435e-12\\
340	4.51938486634162e-12\\
360	3.36064509554035e-12\\
380	2.54074539185467e-12\\
400	1.94955163124177e-12\\
420	1.51589851782319e-12\\
440	1.19304566226219e-12\\
460	9.49351708356971e-13\\
480	7.62945262522408e-13\\
500	6.18838313926062e-13\\
520	5.06150676926609e-13\\
540	4.17221812654134e-13\\
560	3.46389583683049e-13\\
580	2.89546164822241e-13\\
600	2.43471909300297e-13\\
620	2.05946371067967e-13\\
640	1.7519319328585e-13\\
660	1.49769086021934e-13\\
680	1.28563826251593e-13\\
700	1.10911280160053e-13\\
720	9.61453139325386e-14\\
740	8.35997937542743e-14\\
760	7.30526750203353e-14\\
780	6.3948846218409e-14\\
800	5.62883073484954e-14\\
820	4.96269692007445e-14\\
840	4.39648317751562e-14\\
860	3.8968828164343e-14\\
880	3.46389583683049e-14\\
900	3.09752223870419e-14\\
920	2.76445533131664e-14\\
940	2.48689957516035e-14\\
960	2.23154827949656e-14\\
980	2.00950367457153e-14\\
1000	1.80966353013901e-14\\
};
\addlegendentry{$(4,8)$, No $4$-cycles};

\addplot [color=blue,dashed,mark=square,mark options={solid}]
  table[row sep=crcr]{%
80	8.36281420002472e-07\\
100	4.14003782922556e-07\\
120	2.34303700108462e-07\\
140	1.4522937197281e-07\\
160	9.6146214878523e-08\\
180	6.69085707949435e-08\\
200	4.84195999961301e-08\\
220	3.61608643029143e-08\\
240	2.77144103222327e-08\\
260	2.17063745866497e-08\\
280	1.73166698758109e-08\\
300	1.40351190669463e-08\\
320	1.15329915617579e-08\\
340	9.59196522209993e-09\\
360	8.06317668189394e-09\\
380	6.84274858997469e-09\\
400	5.85669546193657e-09\\
420	5.05135122619294e-09\\
440	4.3871450916555e-09\\
460	3.83446541185606e-09\\
480	3.37085337420717e-09\\
500	2.97906588286878e-09\\
520	2.64571564656535e-09\\
540	2.36030139877386e-09\\
560	2.11450723597295e-09\\
580	1.90169013869479e-09\\
600	1.7164999421837e-09\\
620	1.5545955633911e-09\\
640	1.4124289515749e-09\\
660	1.28708177449255e-09\\
680	1.17613840888708e-09\\
700	1.0775881298386e-09\\
720	9.89747506174865e-10\\
740	9.11200448427962e-10\\
760	8.40749692088139e-10\\
780	7.7737838388714e-10\\
800	7.20219106575826e-10\\
820	6.6852900992842e-10\\
840	6.21669271616554e-10\\
860	5.790886659085e-10\\
880	5.40309574859066e-10\\
900	5.04917108123948e-10\\
920	4.72549444019421e-10\\
940	4.4289028000577e-10\\
960	4.1566217134914e-10\\
980	3.90620980006418e-10\\
1000	3.67551544755429e-10\\
};
\addlegendentry{$(5,10)$, With $4$-cycles};

\addplot [color=red,dashed,mark=square,mark options={solid}]
  table[row sep=crcr]{%
80	3.25295346215171e-14\\
100	3.10862446895044e-15\\
120	4.44089209850063e-16\\
140	1.11022302462516e-16\\
160	0\\
180	0\\
200	0\\
220	0\\
240	0\\
260	0\\
280	0\\
300	0\\
320	0\\
340	0\\
360	0\\
380	0\\
400	0\\
420	0\\
440	0\\
460	0\\
480	0\\
500	0\\
520	0\\
540	0\\
560	0\\
580	0\\
600	0\\
620	0\\
640	0\\
660	0\\
680	0\\
700	0\\
720	0\\
740	0\\
760	0\\
780	0\\
800	0\\
820	0\\
840	0\\
860	0\\
880	0\\
900	0\\
920	0\\
940	0\\
960	0\\
980	0\\
1000	0\\
};
\addlegendentry{$(5,10)$, No $4$-cycles};

\end{axis}
\end{tikzpicture}%
}
\normalsize
\caption{\label{fig:effect_of_no4cycles}The theoretical approximations on $P_{B}^\SPBC$ for various ensembles in both the unexpurgated and expurgated scenarios.
}
\end{center}
\end{figure}
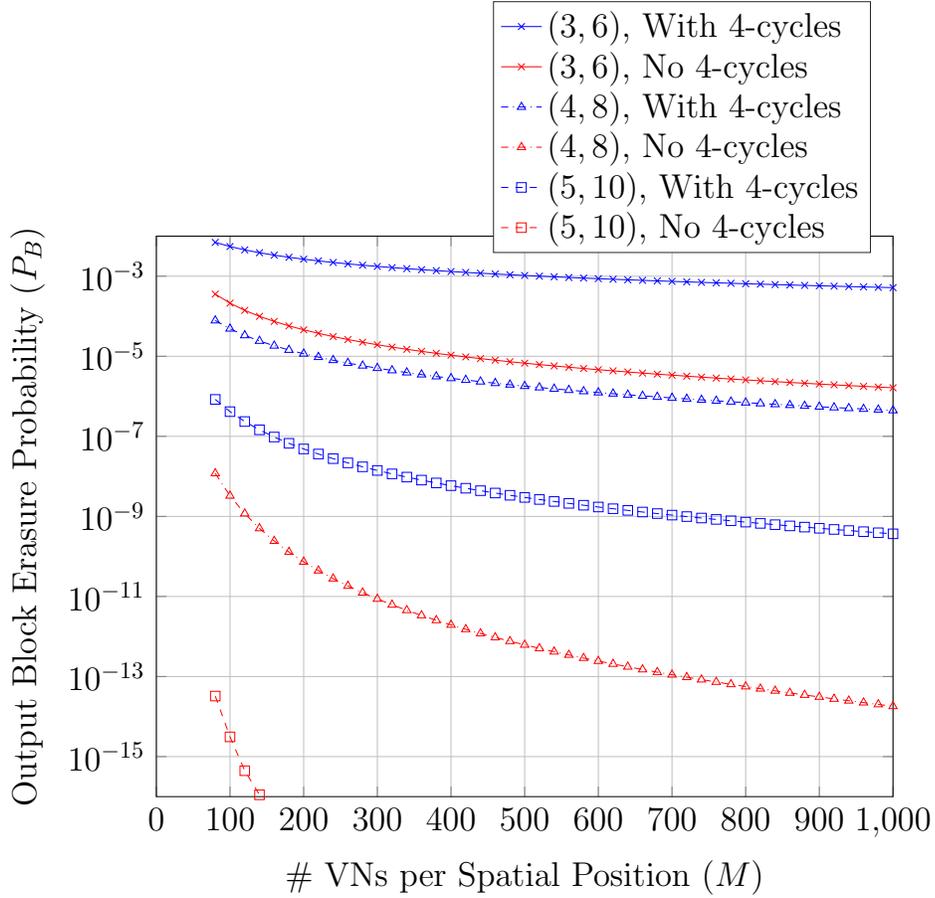

We now compare the average performance of different SC-LDPC ensembles on the SPBC. 
We fix the asymptotic code rate as $\frac{1}{2}$, the smoothing parameter as $w=d_v$ and plot the (tight) approximations on $P_{B}^\SPBC$ of three ensembles, namely $(3,6)$, $(4,8)$ and $(5,10)$, for both the unexpurgated and the expurgated cases in Fig.~\ref{fig:effect_of_no4cycles}.
\begin{itemize}

\item For the unexpurgated case, the average block erasure probability varies as
\begin{equation*}
P_\B^\SPBC \sim O(M^{2-d_v)}).
\end{equation*}
Hence, linearly increasing $d_v$, for a constant rate $\frac{1}{2}$, keeps improving the performance by  multiples of $1/M$. \\

\item When the ensemble is expurgated so that $girth = 6$, the improvement is by an order of $\frac{d_v+1}{2}$ in $M$. Now, we have 
\begin{equation*}
P_\B^\SPBC \sim O(M^{(d_v+1)(2-d_v)/2}).
\end{equation*}
Therefore, for a fixed rate $\frac{1}{2}$, a unit increase in $d_v$ improves the performance by a factor of about $M^{-d_v}$. \\

\item As $d_v$ is increased, it was observed that the performance is worse if $w$ is kept constant. This is because higher size stopping sets dominated when $w < d_v$. All the bounds presented in this work are tight only when $w \geq d_v$. 

\end{itemize}

%
%
%

\section{Conclusion}
\label{sec:conclude}

We have analyzed random SC-LDPC ensembles on the burst erasure channel and provided insights into improving the block erasure probability through increased VN degree and expurgation. 
The expected error floor for the ensemble has been characterized and verified on the BEC. 
We have shown through these results that the vector in~(\ref{eq:pvec}) completely characterizes the ensemble performance on the erasure channel. 

There is more work to be done to arrive at tighter bounds for the block erasure channel. 
We also need to analyze the expurgated ensemble on the random burst channel. 
One method to do that would be to find the vector in~(\ref{eq:pvec}) for the expurgated ensemble. 
Since that is very tedious, the main challenge in this direction is finding a simpler way of characterizing the performance. Also, we have observed that higher size stopping sets dominate when $w < d_v$. 
An explicit proof for this could be insightful.

%
%
%


\chapter{\uppercase {Conclusions}}
\label{sec:total_conclusion}

In Section~\ref{sec:cyclic_polar}, this work introduces a method to construct cyclic polar codes over $\mathbb{F}_q$ for any blocklength $N$ satisfying $N|(q-1)$.
For the QEC, these codes can be decoded efficiently using Forney's algebraic decoder to decode the intermediate blocks.
In our simulations, they outperform standard polar codes.
For the case of $N=2^n$, a soft-decision SC decoder was also implemented and tested on the $q$-ary symmetric channel.
Under SC decoding, cyclic polar codes clearly outperform RS codes of the same rate and blocklength.

An algebraic errors and erasures decoding strategy was also considered for the intermediate block codes.
Preliminary results show that this approach is suboptimal when compared to hard decision decoding of a RS code with the same rate and blocklength.
\iffull
In future work, we plan to consider APP decoding of the intermediate blocks for small lengths while retaining a hard-decision decoder at larger blocks typically placed close to inputs in the graph.
We will also consider the rate of polarization for these codes based on similar work for standard polar codes~\cite{Korada-it10*2,Mori-it14}.
\fi
The programs developed for this work can be accessed at \url{https://github.com/nrenga/cyclic_polar}.

In Section~\ref{sec:scldpc_codes}, we have analyzed random SC-LDPC ensembles on the burst erasure channel and provided insights into improving the block erasure probability through increased VN degree and expurgation. 
The expected error floor for the ensemble has been characterized and verified on the BEC. 
We have shown through these results that the vector in~(\ref{eq:pvec}) completely characterizes the ensemble performance on the erasure channel. 

There is more work to be done to arrive at tighter bounds for the block erasure channel. 
We also need to analyze the expurgated ensemble on the random burst channel. 
One method to do that would be to find the vector in~(\ref{eq:pvec}) for the expurgated ensemble. 
Since that is very tedious, the main challenge in this direction is finding a simpler way of characterizing the performance. Also, we have observed that higher size stopping sets dominate when $w < d_v$. 
An explicit proof for this could be insightful.

\let\oldbibitem\bibitem
\renewcommand{\bibitem}{\setlength{\itemsep}{0pt}\oldbibitem}
%
%
%


\phantomsection
\addcontentsline{toc}{chapter}{REFERENCES}

\renewcommand{\bibname}{{\normalsize\rm REFERENCES}}



%
%
%

\makeatletter
\xpatchcmd*{\@chapter}{\thechapter .}{\thechapter}{}{}
\makeatother
\addtocontents{toc}{\protect\renewcommand\protect\cftchapaftersnum{ }}

\begin{appendices}
\titleformat{\chapter}{\centering\normalsize}{APPENDIX \thechapter}{0em}{\vskip .5\baselineskip\centering}
\renewcommand{\appendixname}{APPENDIX}

%
%
%


\phantomsection

\chapter{\uppercase{Cooley-Tukey Formula}}
\label{sec:cooley_tukey}
\addtocontents{lof}{\protect\addvspace{-10pt}}%

\section{Discussion}
\label{sec:fft}

In this section, we will discuss the details of the Cooley-Tukey fast Fourier transform~\cite{{Blahut-1985,Cooley-moc65}} and derive the Kronecker product formulation of the same as given in Lemma~\ref{lem:Fab}. 

Consider two vectors $\vecnot{u}$ and $\vecnot{v}$ such that $\vecnot{u}$ is the Fourier transform of $\vecnot{v}$ and let $\ell=ab$ be the length of the vectors, where $a$ and $b$ are positive integers. The Fourier transform is given by
\begin{equation*}
u_{i} = [ F_{\ell} \vecnot{v}]_i = \sum_{j=0}^{\ell-1}\omega_{\ell}^{ij}v_{j},
\end{equation*}
where the matrix $F_{\ell}$ is defined by $[F_{\ell}]_{i,j} \triangleq \omega_{\ell}^{ij}$.
Now, express each of the indices with a coarse index and vernier index as
\begin{equation*}
j = j' + bj'' \hspace{3mm} \text{;} \hspace{3mm} i = ai' + i''
\end{equation*}
where $i',j' = 0,1,\ldots,b-1$ and $i'',j'' = 0,1,\ldots,a-1$. By making these substitutions we get
\begin{equation*}
u_{ai'+i''} = \sum_{j''=0}^{a-1} \sum_{j'=0}^{b-1} \omega_{\ell}^{(j'+bj'')(ai'+i'')} v_{j'+bj''} .
\end{equation*}
Now define $\gamma = \omega_{\ell}^{b}$ and $\beta = \omega_{\ell}^{a}$ so that they have multiplicative orders $a$ and $b$, respectively, in $\mathbb{F}_q$. Since $\omega_{\ell}$ has a multiplicative order of $\ell=ab$, $\omega_{\ell}^{abj''i'} = 1$ in the expansion of the above formula. Rearranging the expression gives us the following convenient form of the Cooley-Tukey fast Fourier transform.
\begin{equation*}
u_{a i' + i''} = \sum_{j'=0}^{b-1}\beta^{j'i'}\left[\omega_{\ell}^{j'i''}\left(\sum_{j''=0}^{a-1}\gamma^{j''i''}v_{j'+b j''}\right)\right].
\end{equation*}
This form allows us to fragment the Fourier transform into a sequence of operations which reduces the overall complexity as we will see next. The indices $i$ and $j$ are expressed in two different combinations of their coarse and vernier indices to arrive at this convenient form of the transform.

\begin{figure}
\begin{center}

$\begin{bmatrix}
v_{0}\\v_{1}\\v_{2}\\v_{3}\\v_{4}\\v_{5}
\end{bmatrix}
\xrightarrow{(i)}
\begin{bmatrix}
v_{0}'&v_{2}'&v_{4}'\\v_{1}'&v_{3}'&v_{5}'
\end{bmatrix}
\xrightarrow{(ii)}
\begin{bmatrix}
u_{0}'&u_{1}'&u_{2}'\\u_{3}'&u_{4}'&u_{5}'
\end{bmatrix}
\xrightarrow{(iii)}
\begin{bmatrix}
u_{0}\\u_{1}\\u_{2}\\u_{3}\\u_{4}\\u_{5}
\end{bmatrix}
$
\caption{\label{fig:cooley-tukey}Sequence of operations in the Cooley-Tukey fast Fourier transform for the case $\ell=6, a=3, b=2$.}

\end{center}
\end{figure}

Next, we will see the sequence of operations in computing the transform.
The above expression of the Cooley-Tukey FFT indicates that the computation is closely related to a two-dimensional (2-D) Fourier transform.
The input vector $\vecnot{v}$, of length $\ell$, is rearranged column-wise into a 2-D matrix of dimensions $b \times a$.
Fig.~\ref{fig:cooley-tukey} shows an example demonstrating the following sequence of operations.
\begin{itemize}

\item[$(i)$] Firstly, the inner summation is the 1-D FFT,
\begin{equation*}
v_{j'+b i''}' = \sum_{j''=0}^{a-1} \gamma^{j''i''} v_{j'+b j''} ,
\end{equation*}
of each row of this matrix; for each value of $j'$, compute a length-$a$ Fourier transform of the vector $\vecnot{v}_{j'} = (v_{j'+b j''})$ that outputs the vector $\vecnot{v}_{j'}' = (v_{j'+bi''}')$ which are the set of summations for $i''=0,1,\ldots,a-1$.
Hence this intermediate output vector can be indexed with $i=bi''+j'$ so that $i''=\lfloor i/b \rfloor$ and $j' = i \bmod b$.

\item[$(ii)$] Then, all elements are multiplied by $\omega_{\ell}^{j'i''}$ and reshuffled to give
\begin{equation*}
u_{aj'+i''}' = \omega_{\ell}^{j'i''} v_{j'+ bi''}' ,
\end{equation*}
where $j'$ and $i''$ vary across the rows and columns of the matrix, respectively.
The different expressions for $i$ and $j$ in terms of their coarse and vernier indices explain the need for reshuffling in this step. 
In a Fourier transform, all indices need to be involved in computing each output coefficient but step $(i)$ has only involved interleaved indices. Hence, in the next step, adjacent indices should be involved to complete the transform.

\item[$(iii)$] Finally, the outer summation is the 1-D FFT,
\begin{equation*}
u_{ai'+i''} = \sum_{j'=0}^{b-1} \beta^{j'i'} u_{aj'+i''}' ,
\end{equation*}
of each column of the resultant matrix obtained after the multiplication step; for each value of $i''$, compute a length-$b$ Fourier transform whose inputs are indexed by $j'$ and the output $\vecnot{u}$ is indexed by $i'=0,1,\ldots,b-1$, clearly indicating an output interleaved by $a$.
Hence, the (output) indexing implies that the output vector is to be read row-wise from the matrix after the last (row-FFT) step.

\end{itemize}
Na\"\i ve implementations of the $b$ length-$a$ transforms and the $a$ length-$b$ transforms would require a complexity of $O(ba^2)$ and $O(ab^2)$, respectively.
Therefore, the total complexity of the Fourier transform is now reduced from $O(\ell^2)$ to $O(\ell(a+b))$.

\section{Proof of Lemma~\ref{lem:Fab}}
\label{sec:lem-Fab}

\begin{proof}[\nopunct]

Consider $\vecnot{v} = (v_0,v_1,\ldots,v_{ab-1})^{T}$ and $\vecnot{u} = (u_0,u_1,\ldots,u_{ab-1})^{T}$ to be the input and output vectors of the transform, respectively.
We follow the sequence of operations described above to translate the summations into equivalent matrix operations.
\begin{itemize}

\item[$(i)$] First, we perform $b$ length-$a$ Fourier transforms on $b$ interleaved blocks as the matrix-vector product
\begin{equation*}
\vecnot{v}' = (F_{a} \otimes I_{b}) \vecnot{v} ,
\end{equation*}
where $A \otimes B$ denotes the Kronecker product of matrix $A$ with matrix $B$ as given in Definition~\ref{def:Kronecker}.
It is important to note that both the input and output vectors have no shuffling in the indices of their elements.

\item[$(ii)$] Next, we multiply each element of $\vecnot{v}'$, indexed by $i=bi''+j'$, by the twiddle factor $\omega_{ab}^{(\lfloor i/b \rfloor) (i \bmod b)}$. If we construct a diagonal matrix $D_{a,b}$ with these factors as its main-diagonal elements, then this step can be expressed as
\begin{equation*}
\vecnot{u}' = D_{a,b} \vecnot{v}' .
\end{equation*}

\item[$(iii)$] Finally, we perform the $a$ length-$b$ Fourier transforms on $a$ adjacent blocks of $\vecnot{u}'$ as the matrix-vector product
\begin{equation*}
\vecnot{u} = (I_{a} \otimes F_{b}) \vecnot{u}' .
\end{equation*}

\end{itemize}
Since this output vector has its indices shuffled, we also need to deinterleave it. Hence, the final expression for the transform is given by
\begin{equation*}
\vecnot{u} = S_{b,a} (I_{a} \otimes F_{b}) D_{a,b} (F_{a} \otimes I_{b}) \vecnot{v} ,
\end{equation*}
where $S_{b,a}$ is the perfect-shuffle permutation matrix introduced in Definition~\ref{def:shuffle}.  \qedhere
\end{proof}

\section{Proof of Lemma~\ref{lem:generalFFT}}
\label{sec:lem-generalFFT}

\begin{proof}[\nopunct]
From Lemma~\ref{lem:Fab} we have, for $N=a \times b$,
\begin{IEEEeqnarray*}{rCl}
F_{ab} & = & S_{b,a} (I_{a} \otimes F_{b}) D_{a,b} (F_{a} \otimes I_{b}) \\
       & = & (F_{b} \otimes I_{a}) S_{b,a} D_{a,b} (F_{a} \otimes I_{b}) \\
       & = & [ (S_{1,b} D_{b,1} \otimes I_{a}) (F_{b} \otimes I_{a}) ] \times [ (S_{b,a} D_{a,b} \otimes I_1) (F_{a} \otimes I_{b}) ] .
\end{IEEEeqnarray*}
Now, let us see the extension for $N=a \times bc$.
\begin{IEEEeqnarray*}{rCl}
F_{N} & = & (F_{bc} \otimes I_{a}) S_{bc,a} D_{a,bc} (F_{a} \otimes I_{bc}) \\
      & = & [ (F_{c} \otimes I_{b}) S_{c,b} D_{b,c} (F_{b} \otimes I_{c}) \otimes I_{a}) (F_{b} \otimes I_{a}) ] \times [ (S_{bc,a} D_{a,bc} \otimes I_1) (F_{a} \otimes I_{bc}) ] \\
      & = & [ F_{c} \otimes I_{b} \otimes I_{a} ] \times [ S_{c,b} D_{b,c} (F_{b} \otimes I_{c}) \otimes I_{a} ] \times [ (S_{bc,a} D_{a,bc} \otimes I_1) (F_{a} \otimes I_{bc}) ] \\
      & = & [ (S_{1,c} D_{c,1} \otimes I_{ba}) (F_{c} \otimes I_{ba}) ] \times [ (S_{c,b} D_{b,c} \otimes I_{a}) (F_{b} \otimes I_{c} \otimes I_{a}) ] \\
      &   & \times [ (S_{bc,a} D_{a,bc} \otimes I_1) (F_{a} \otimes I_{bc}) ] \\
      & = & [ (S_{N/abc,c} D_{c,N/abc} \otimes I_{abc/c}) (F_{c} \otimes I_{N/c}) ] \times [ (S_{N/ab,b} D_{b,N/ab} \otimes I_{ab/b}) (F_{b} \otimes I_{N/b}) ] \\
      &   & \times [ (S_{N/a,a} D_{a,N/a} \otimes I_{a/a}) (F_{a} \otimes I_{N/a}) ] .
\end{IEEEeqnarray*}
We see a pattern in the recurison which can be generalized for length $N=\prod_{m=1}^{n} \ell_{m}$. Define $p_j =  \prod_{m=1}^{j} \ell_{j}$ and
\[U_{m} =(S_{N/p_{m},\ell_{m}}D_{\ell_{m},N/p_{m}}\otimes I_{p_{m}/\ell_{m}})(F_{\ell_{m}}\otimes I_{N/\ell_{m}}). \]
Then, the Fourier transform can be expressed as
\begin{equation*}
F_N = U_{n} U_{n-1} \cdots U_1 .   \qedhere
\end{equation*}
\end{proof}

%
%
%


\chapter{\uppercase{Channel Polarization}}
\label{sec:channel_polarize}

In this section, we prove the polarization theorem, stated in Section~\ref{sec:erasure_code_design} Theorem~\ref{thm:polarization}, for the cyclic polar code construction.
First, we prove Lemma~\ref{lem:preserve_eps} so that the result can be used to prove the theorem.

\section{Proof of Lemma~\ref{lem:preserve_eps}}
\label{sec:lem-preserve_eps}

\begin{proof}[\nopunct]
Observing that every term in the summation of~\eqref{eq:epsilon0} is positive, we have the following:
\begin{itemize}

\item[$(i)$]%
\begin{IEEEeqnarray*}{rCl}
\frac{1}{\ell} \sum_{j=0}^{\ell - 1} \psi(\ell, j, \epsilon') & = & \frac{1}{\ell} \sum_{j=0}^{\ell - 1} \sum_{i=0}^{(\ell-1)-j} \binom{\ell}{i}(1-\epsilon')^{i} (\epsilon')^{\ell-i} \\
  & = & \sum_{i=0}^{\ell-1} \frac{(\ell-i)}{\ell} \binom{\ell}{i}(1-\epsilon')^{i} (\epsilon')^{\ell-i} \\
  & = & \epsilon' \sum_{i=0}^{\ell-1} \binom{\ell-1}{i}(1-\epsilon')^{i} (\epsilon')^{\ell-1-i} \\
  & = & \epsilon'   
\end{IEEEeqnarray*}

\item[$(ii)$] Given $\epsilon' \in (0,1)$, we have
\begin{equation*}
\psi(\ell, \ell-1, \epsilon') = \binom{\ell}{0} (1-\epsilon')^{0} (\epsilon')^{\ell} = (\epsilon')^{\ell} < \epsilon'
\end{equation*}     
and
\begin{equation*}
\psi(\ell, 0, \epsilon') = 1 - \binom{\ell}{\ell} (1-\epsilon')^{\ell} (\epsilon')^{0} =  1 - (1-\epsilon')^{\ell} .
\end{equation*}
Now, consider $\psi(\ell, 0, \epsilon') - \epsilon'$. We have
\begin{IEEEeqnarray*}{rCl}
\psi(\ell, 0, \epsilon') - \epsilon' & = & 1 - (1-\epsilon')^{\ell} - \epsilon' \\
                                     & = & (1-\epsilon') - (1-\epsilon')^{\ell} \\
                                     & > & 0 .
\end{IEEEeqnarray*}
Hence, $\psi(\ell, \ell-1, \epsilon') < \epsilon ' < \psi(\ell, 0, \epsilon') $. \qedhere
\end{itemize}
\end{proof}

\section{Proof of Theorem~\ref{thm:polarization}}
\label{sec:thm-polarization}

\begin{proof}[\nopunct]

We use the same strategy as \arikan~used in~\cite{Arikan-it09} but we slightly generalize the channel evolution tree and the mathematical framework to re-formulate the problem in our scenario.
The primary requirement for this is that, at every stage of the transform, each channel splits into multiple channels in our case whereas in the original polar code construction, each channel split into exactly two channels at every stage.

The root node of the tree is associated with the underlying QEC $W$.
At level $1$, $W$ evolves into $\ell_1$ channels, namely $W_{\ell_1}^{(1)}, W_{\ell_1}^{(2)}, \ldots, W_{\ell_1}^{(\ell_1)}$.
We have $\ell_1$ nodes corresponding to $\ell_1$ channels at level $1$.
At level $2$, every channel from level $1$ gives birth to $\ell_2$ channels.
Hence, we have the channels $W_{\ell_1 \ell_2}^{(1)}, W_{\ell_1 \ell_2}^{(2)}, \ldots, W_{\ell_1 \ell_2}^{(\ell_1 \ell_2)}$, and so on.
The $i^{\rm th}$ channel from the top at level $n$ will be denoted by $W_{\ell_1 \ell_2 \cdots \ell_n}^{(i)}$.

Since this is not a binary tree, the channels have to be indexed by $\ell_m$-ary symbols $s_m$'s, for $m=1,2,\ldots$.
Define $L_m \triangleq \{0,1,\ldots,\ell_m-1\}$.
The root node is indexed with a null sequence.
The nodes at level $1$ are indexed with symbol $s_1 \in L_1$.
Given a node at level $m$ with the symbol sequence $s_1 s_2 \cdots s_m$, the child nodes at the next level will have indices $s_1 s_2 \cdots s_m 0, s_1 s_2 \cdots s_m 1, \ldots$, $s_1 s_2 \cdots s_m (\ell_m-1)$.
According to this labeling, the channel $W_{\ell_1 \ell_2 \cdots \ell_m}^{(i)}$ is situated at the node $s_1 s_2 \cdots s_m$ with $i = 1 + \sum_{j=1}^{m} s_j \ell_{j}^{m-j}$.
Alternatively, we denote this channel as $W_{s_1 s_2 \cdots s_m}$.

We redefine the random tree process $\{ K_m; m \geq 0 \}$.
The process begins at the root node with $K_0 = W$.
At level $1$, the process takes the value $K_1 = W_{s_1}$, where all values for $s_1$ are equally likely.
In general, if $K_m = W_{s_1 s_2 \cdots s_m}$, then $K_{m+1} = W_{s_1 s_2 \cdots s_m j}$ for any $j \in L_{m+1}$ with probability $1/\ell_{m+1}$ each.
We need to associate the channel obtained as the value of the process at each stage with its reliability parameter, i.e. the Bhattacharyya parameter, in order to track the evolution of the erasure rates after each step of polarization.
Since the rate and reliability parameters have a complementary relation for the erasure channel, it is not necessary to also associate the rate parameter with the random tree process. 
Hence, we define the reliability random process $\{Z_m; m \geq 0\}$ as $Z_m = Z(K_m)$.

Now, consider the probability space $(\Omega,\mathcal{F},P)$ where $\Omega$ is the space of all sequences $(s_1,s_2,\ldots) \in L_1 \times L_2 \times \cdots$, $\mathcal{F}$ is the Borel field generated by the cylinder sets $S(s_1,s_2,\ldots,s_n) \triangleq \{ \omega \in \Omega : \omega_1 = s_1, \ldots, \omega_n = s_n \}, n\geq 1, s_m \in L_m$,  $P$ is the probability measure defined on $\mathcal{F}$ such that $P(S(s_1,\ldots,s_n)) = 1/\prod_{m=1}^{n} \ell_m$.
For each $n \geq 1$, we define $\mathcal{F}_n$ as the Borel field generated by the cylinder sets $S(s_1,s_2,\ldots,s_m), 1 \leq m \leq n, s_i \in L_i$. 
We define $\mathcal{F}_0$ as the trivial field consisting only of the null set and $\Omega$.
Clearly, $\mathcal{F}_0 \subset \mathcal{F}_1 \subset \cdots \subset \mathcal{F}$.

Then, we can define the random processes as follows.
For $\omega = (\omega_1,\omega_2,\ldots) \in \Omega$ and $n \geq 1$, define $K_{n}(\omega) = W_{s_1 s_2 \cdots s_n}$ and $Z_{n}(\omega) = Z(K_{n}(\omega))$.
For $n=0$, define $K_0 = W, Z_0 = Z(W)$.
Hence, for any fixed $n \geq 0$, the RVs $K_n$ and $Z_n$ are measurable with respect to $\mathcal{F}_n$.

\begin{lem}
\label{lem:Zmartingale}
The sequence of random variables and Borel fields $\{ Z_n, \mathcal{F}_n; n \geq 0 \}$ is a martingale, i.e.,
\begin{IEEEeqnarray}{Cl}
\label{eq:Fn}
& \mathcal{F}_n \subset \mathcal{F}_{n+1} \text{ and } Z_n \text{ is } \mathcal{F}_n\text{-measurable},\\
\label{eq:Z_n}
& E[|Z_n|] < \infty,\\
\label{eq:Zmart}
& Z_n = E[Z_{n+1} | \mathcal{F}_n].
\end{IEEEeqnarray}
\begin{proof}
Condition~\eqref{eq:Fn} is satisfied just by construction and~\eqref{eq:Z_n} is given by the fact that $0 \leq Z_n \leq 1$.
To prove~\eqref{eq:Zmart}, consider a cylinder set $S(s_1,s_2,\ldots,s_n) \in \mathcal{F}_n$ and set $Z(W_{s_1 \cdots s_n}) = \epsilon', Z(W_{s_1 \cdots s_n j}) = \psi(\ell_n, j, \epsilon')$ in the result of Lemma~\ref{lem:preserve_eps} to write
\begin{IEEEeqnarray*}{rCl}
E[Z_{n+1} | S(s_1,s_2,\ldots ,s_n)] & = & \frac{1}{\ell_{n+1}} \sum_{j=0}^{\ell_{n+1}-1} Z(W_{s_1 \cdots s_n j}) \\
                                    & = & Z(W_{s_1 \cdots s_n})
\end{IEEEeqnarray*}
Since $Z(W_{s_1 \cdots s_n})$ is the value of $Z_n$ on $S(s_1,s_2,\ldots ,s_n)$,~\eqref{eq:Zmart} follows.
This completes the proof that $\{ Z_n, \mathcal{F}_n \}$ is a martingale.
\end{proof}
\end{lem}

\begin{lem}
\label{lem:Zinfty}
The sequence $\{Z_n; n \geq 0\}$ converges a.e. to a random variable $Z_{\infty}$ such that
\begin{IEEEeqnarray}{Cl}
\label{eq:EZinf}
& E[Z_{\infty}] = Z_0, \\
\label{eq:ZinfBinary}
& Z_{\infty} \in \{0,1\} \text{ a.e.}
\end{IEEEeqnarray} 
\begin{proof}
Since $\{ Z_n, \mathcal{F}_n \}$ is uniformly integrable,~\eqref{eq:EZinf} follows from standard convergence results about such martingales (see, e.g.,~\cite[Theorem 9.4.6]{Chung-book74}).
From Lemma~\ref{lem:preserve_eps}, we see that the individual channel erasure rates $\psi(\ell,j,\epsilon')$ polarize away from the input channel erasure rate $\epsilon'$, while the mean is preserved to be $\epsilon'$.
Since the erasure rate $\psi$ is bounded in $[0,1]$, the polarization will recur until it reaches either of the fixed points in the set $\{0,1\}$.
\eqref{eq:ZinfBinary} follows automatically.
\end{proof}
\end{lem}
From the above results, we have 
\[ E[Z_{\infty}] = 1 \cdot P(Z_{\infty}=1) + 0 \cdot P(Z_{\infty}=0) = Z_0 . \]
Conditioning that we start with a channel of erasure rate $Z_0 = \epsilon$, the theorem follows.
This completes the proof of Theorem~\ref{thm:polarization}.   
\end{proof}

%
%
%


\chapter{\uppercase{Forney's Decoder for Small Blocks}}
\label{sec:Forney}

In 1965, Forney described a simplified algorithm for the decoding of RS and BCH codes~\cite{Forney-it65}.
The algorithm is suitable for errors and erasures decoding.
In this description, we focus on the case where:
\begin{itemize}
\item The locations of errors and/or erasures are known and are given by the erasure locator polynomial,
\begin{equation*}
\Lambda \left(x\right) = \prod_{l=1}^{\nu}(1-X_{l}x)
\end{equation*}
where, $X_{l}$ denotes the location of the $l$-th erasure and $\nu$ is the actual number of erasures.

\item Syndromes can be computed based on known values in the codeword spectrum.
\end{itemize}

For the QEC, the erased positions are known at the receiver and hence, the erasure locator polynomial $\Lambda\left(x\right)$ can be easily computed.
To verify the second condition, consider the Fourier transform pair $u(x)$ and $v(x)$ defined by
\begin{equation*}
u(x)=\sum_{i=0}^{\ell-1}u_{i}x^{i}=\sum_{i=0}^{\ell-1}v(\omega_{\ell}^{i})x^{i}
\end{equation*}
and
\begin{equation*}
v(x)=\sum_{i=0}^{\ell-1}v_{i}x^{i}=\sum_{i=0}^{\ell-1}\left(\ell^{-1}u(\omega_{\ell}^{-i})\right)x^{i}.
\end{equation*}
These are GFFT and inverse GFFT equations associated with the cyclic polar code construction.

Now, we assume that the information polynomial $u(x)$ has $r$ consecutive known values (not necessarily zeroes) starting from index $b$.
Thus, the value $u_{i}=v(\omega_{\ell}^{i})$ is known for
\begin{IEEEeqnarray}{rCl}
i\in \mathcal{B} & = & \{b+j \bmod (\ell-1) \, | \, j\in \mathbb{Z}, 0\leq j\leq r-1\}, \;\;
\label{eq:information}
\end{IEEEeqnarray}
 where $b\in \{0,1,\ldots,\ell-1\}$.
 These known values are available at both the transmitter and receiver and allow us to satisfy second condition above.

\subsection*{The Decoder}
Assume that the information polynomial $u(x)$ is encoded into $v(x)$ and transmitted via $\ell$ consecutive uses of QEC($\epsilon$).
Let the received polynomial be $y(x)=v(x)+e(x)$, where $e(x)$ is a ``error'' polynomial that changes the coefficients of $y(x)$ to be zero at all erasure locations.
To compute the syndromes $S_j = e(\omega_{\ell}^j)$, we note that $e(\omega_{\ell}^{j})=y(\omega_{\ell}^{j})-u_{j}$ for $j\in \mathcal{B}$.
These are computable at the receiver because $y(x)$ is known (except for erasures) and $u_j$ is known for $j\in \mathcal{B}$.
Now, we restrict the discussion to the case of $b=0$, which implies $\mathcal{B}=\{0,1,2,\ldots,r-1\}$. 

Assume the $\nu$ erasures occurred at positions $i_{l}$ for $l=1,2,3,\ldots,\nu$ and proceed as follows.
Let the erasure and syndrome polynomials be
\begin{IEEEeqnarray*}{rCl.s.rCl}
e(x) & = & \sum_{l=1}^{\nu}e_{i_{l}}x^{i_{l}} = \sum_{l=1}^{\nu}\left(-v_{i_{l}}\right)x^{i_{l}} \\
S(x) & = & \sum_{j=0}^{r-1}S_{j}x^{j},
\end{IEEEeqnarray*}
where
\[ S_{j}=e(\omega^{j}) = \sum_{l=1}^{\nu}e_{i_{l}}X_{l}^{j} \]
and $X_{l} \triangleq \omega_{\ell}^{i_{l}}$ is the location of the $l$-th erasure.
The erasure evaluator polynomial is defined as
\begin{IEEEeqnarray*}{rClsrl}
\Omega &(x)& = S(x)\Lambda(x) 			& (mod & x^{r} & ) \\
		  & = & \left[\sum_{j=0}^{r-1}\left(\sum_{l=1}^{\nu}e_{i_{l}}X_{l}^{j}\right)x^{j}\right]\left[\prod_{m=1}^{\nu}\left(1-X_{m}x\right)\right]		& (mod & x^{r} & ) \\		  
		  & = & \sum_{l=1}^{\nu}e_{i_{l}} \sum_{j=0}^{r-1}(X_{l}x)^{j} \prod_{m=1}^{\nu}(1-X_{m}x) & (mod & x^{r} & ) \\
		  & = & \sum_{l=1}^{\nu}e_{i_{l}}\underbrace{(1-X_{l}x)\sum_{j=0}^{r-1}(X_{l}x)^{j}}_{=1-(X_{l}x)^{r}} \prod_{m\neq l}^{\nu}(1-X_{m}x). \quad		& (mod & x^{r} & )		  
\end{IEEEeqnarray*}
Since $(X_{l}x)^{r} \bmod x^{r} =0$, we find that
\begin{IEEEeqnarray*}{rCl}
\Omega(x) & = \sum_{l=1}^{\nu}e_{i_{l}} \prod_{m\neq l}^{\nu}(1-X_{m}x)
\end{IEEEeqnarray*}
for $\nu-1 < r$.
Substituting $x=X_{k}^{-1}$ we get
\begin{equation*}
e_{i_{k}} = -\frac{X_{k} \Omega\left(X_{k}^{-1}\right)}{\Lambda^{'}\left(X_{k}^{-1}\right)},
\end{equation*}
where $i_{k}$ is the index of the $k$-th erasure.
Once the erased values are obtained, $v(x)$ and $u(x)$ can be obtained from the erasure polynomial.


%
%
%


\chapter{\uppercase{Capacity of QSCE}}
\label{sec:qsce_capacity}

Let the input alphabet be $\mathcal{X}=\{0,1,2,\ldots,q-1\}$, where each element is a representation of a unique $q$-ary symbol, with a probability distribution $p(X) = \left( p_0,\ldots,p_{q-1} \right)$.
Then, the output alphabet will be $\mathcal{Y}=\mathcal{X}\cup\{?\}$. 
%

The capacity of the channel is defined as,
\begin{equation}
C = \underset{p(X)}{\text{max }} I(X;Y) = \underset{p(X)}{\text{max }} (H(Y) - H(Y|X))
\label{eq:capacity}
\end{equation}
where, $X$ and $Y$ are random variables representing the input and output alphabet, respectively. Since, we have a symmetric channel, $H(Y|X)$ is independent of $p(X)$ and is given as
\begin{equation}
\label{eq:H_Y|X}
H(Y|X) = -\biggr[(1-\epsilon-\beta) \text{ log}_{q}(1-\epsilon-\beta) + \alpha \text{ log}_{q}(\alpha) + \beta \text{ log}_{q}\biggr(\frac{\beta}{q-1}\biggr) \biggr] .
\end{equation}

The probabilities for the output symbols in $\mathcal{Y}$ are,
\begin{IEEEeqnarray*}{rCl}
P(Y=i) & = & (p_i)(1-\epsilon-\beta) + \sum_{\substack{j=0 \\ j\neq i}}^{q-1} (p_j) \biggr(\frac{\beta}{q-1}\biggr)\\
       & = & (p_i)(1-\epsilon-\beta) + (1-p_i) \biggr(\frac{\beta}{q-1}\biggr),\\
       \\
P(Y=?) & = & \sum_{j=0}^{q-1} (p_j) (\epsilon) = \epsilon       
\end{IEEEeqnarray*}
for $i = 0,1,2,\ldots,q-1$.
Hence,
\begin{IEEEeqnarray*}{rCl}
H(Y) & = & -\sum_{y \in \mathcal{Y}} p(y) \text{ log}_{q} p(y) \\
     & = & -\biggr[ \sum_{i=0}^{q-1} P(Y=i) \text{ log}_{q}(P(Y=i)) + \epsilon \text{ log}_{q} \epsilon \biggr] .
\end{IEEEeqnarray*}
Since $H(Y|X)$ is independent of $p(X)$, $H(Y)$ has to be maximized in order to maximize capacity. Therefore,
\begin{equation*}
\frac{\partial H(Y)}{\partial p_{i}} = 0
\end{equation*}
which implies
\allowdisplaybreaks
\begin{IEEEeqnarray*}{CrCl}
&
\biggr(1-\epsilon-\beta-\frac{\beta}{q-1} \biggr) \biggr[1 + \text{ log}_{q} \biggr(p_i(1-\epsilon-\beta) + (1-p_i) \biggr(\frac{\beta}{q-1}\biggr) \biggr) \biggr] & = & 0 \\
\Rightarrow &
1 + \text{ log}_{q} \biggr(p_i(1-\epsilon-\beta) + (1-p_i) \biggr(\frac{\beta}{q-1}\biggr) \biggr) & = & 0 \\
\Rightarrow &
p_i(1-\epsilon-\beta) + (1-p_i) \biggr(\frac{\beta}{q-1}\biggr) & = & \frac{1}{q} .
\end{IEEEeqnarray*}
Now, sum the other $(q-1)$ equations for $j\neq i$ and equate that to ($q-1$) times the left hand side of the above equation since both of their values evaluate to $\left(\frac{q-1}{q}\right)$:
\begin{IEEEeqnarray*}{CrCl}
\Rightarrow &
(q-1)\biggr[p_i(1-\epsilon-\beta) + (1-p_i) &\biggr(&\frac{\beta}{q-1}\biggr) \biggr] \\
       && = & \sum_{j\neq i} \biggr(p_i(1-\epsilon-\beta) + (1-p_i) \biggr(\frac{\beta}{q-1} \biggr) \biggr)\\ 
\Rightarrow &
(q-1)p_{i}(1-\epsilon)+\beta - qp_{i}\beta & = & (1-p_i)(1-\epsilon)+\beta - (1-p_i)\biggr(\frac{q\beta}{q-1} \biggr)\\
\Rightarrow &
qp_{i}(1-\epsilon) - \biggr(qp_{i} \beta + \frac{qp_{i}\beta}{q-1} \biggr) & = & (1-\epsilon) - \frac{q\beta}{q-1} \\
\Rightarrow & (qp_i - 1) \biggr(1 -\epsilon-\frac{q\beta}{q-1} \biggr) & = & 0 .
\end{IEEEeqnarray*}
Thus, $p_i = \frac{1}{q}$ maximizes $H(Y)$ to give
\begin{IEEEeqnarray}{rCl}
H(Y) & = & -\biggr[(1-\epsilon-\beta+\beta) \text{ log}_{q}\biggr(\frac{1-\epsilon-\beta+\beta}{q} \biggr) + \epsilon \text{ log}_{q}\epsilon \biggr] \nonumber \\
	 & = & (1-\epsilon) + h_{q}(\epsilon),
\label{eq:H_Y}	 
\end{IEEEeqnarray}
where
\begin{equation*}
h_{q}(\epsilon) = -[\epsilon \text{ log}_{q}(\epsilon) + (1-\epsilon) \text{ log}_{q}(1-\epsilon)] .
\end{equation*}

Substituting values of $H(Y|X)$ and $H(Y)$ obtained in (\ref{eq:H_Y|X}) and (\ref{eq:H_Y}), respectively, into (\ref{eq:capacity}), we get the capacity of QSCE as
\begin{equation}
C = (1-\epsilon) + (1-\epsilon) \text{ log}_{q}\biggr(\frac{1-\epsilon-\beta}{1-\epsilon} \biggr) - \beta  \text{ log}_{q}\biggr(\frac{1-\alpha-\beta}{\beta} \biggr) - \beta \text{ log}_{q}(q-1) .
\end{equation}

\pagebreak{}

\end{appendices}

\end{document}